\documentclass[a4paper,fleqn,usenatbib]{mnras}

\usepackage{newtxtext,newtxmath}

\usepackage[T1]{fontenc}
\usepackage{ae,aecompl}
\usepackage[utf8]{inputenc}

\usepackage{graphicx}	
\usepackage{amsmath}	
\usepackage{longtable}

\usepackage{color}
\usepackage{booktabs}
\usepackage{siunitx}
\usepackage{multicol}
\usepackage{multirow}
\usepackage{float}
\usepackage{graphicx,subcaption}
\usepackage{tabulary,booktabs}
\usepackage{multirow}

\usepackage[dvipsnames]{xcolor}
\usepackage{pdflscape}
\usepackage{tabularx}
\usepackage{xltabular}

\title[CALIFA AGN]{AGN in the CALIFA survey: X-ray detection of nuclear sources}

\author[Osorio-Clavijo et. al. ]{
N. Osorio-Clavijo,$^{1}$ \thanks{e-mail: n.osorio@irya.unam.mx}
O. Gonzalez-Martín,$^{1}$ 
S.F. Sánchez,$^{2}$
M. Guainazzi,$^{3}$
\newauthor
I. Cruz-González$^{2}$ 
\\
$^{1}$Instituto de Radioastronomia and Astrofisica (IRyA-UNAM), 3-72 (Xangari), 8701, Morelia, Mexico \\
$^{2}$Instituto de Astronomía, Universidad Nacional Autónoma de México, A. P. 70-264, C.P. 04510, México, D.F., Mexico\\
$^{3}$ESA European Space Research and Technology Centre (ESTEC), Keplerlaan 1, 2201 AZ, Noordwijk ,The Netherlands \\
}

\date{Accepted 2023 March 28. Received 2023 March 22; in original form 2022 October 14}

\pubyear{2023}

\begin{document}
\label{firstpage}
\pagerange{\pageref{firstpage}--\pageref{lastpage}}
\maketitle
\begin{abstract}
A complete demographic of active galactic nuclei (AGN) is essential to understand the evolution of the Universe. Optical surveys estimate the population of AGN in the local Universe to be of $\sim$ 4\%. However, these results could be biased towards bright sources, not affected by the host galaxy attenuation. An alternative method for detecting these objects is through the X-ray emission. In this work, we aim to complement the AGN population of the optical CALIFA survey (941 sources), by using X-ray data from {\it Chandra}, which provides the best spatial resolution to date, essential to isolate the nuclear emission from the host galaxy. We study a total of 138 sources with available data. We find 34 new {\it bonafide} AGN and 23 AGN candidates, which could increase the AGN population to 7-10\% among the CALIFA {survey}. X-rays are particularly useful for {low-luminosity AGN} since they are excluded by the criterion of large equivalent width of {the $\rm{H\alpha}$ emission line when} applied to optical selections. Indeed, placing such a restrictive criteria might cause a loss of up to 70\% of AGN sources. X-ray detected sources are preferentially located in the right side of the {[$\ion{O}{III}$]/H${\beta}$} versus {[$\ion{N}{II}$]/H${\alpha}$} diagram, suggesting that this diagram might be the most reliable at classifying AGN sources. Our results support the idea that multi-wavelength studies are the best way to obtain a complete AGN population.

\end{abstract}

\section{Introduction}

It has been widely accepted that most galaxies (at least those with a well-formed bulge) host a super massive black-hole (SMBH), sometimes fed by an accretion disk, releasing energies that can go up to $\rm{L_{bol}\sim 10^{48} \ erg \ s^{-1}}$ \citep{Netzer-15}. {These are known as active galactic nuclei (AGN)}. Observational results suggest that AGN play a role in star formation and galaxy evolution, with the most accepted paradigm being that AGN may quench the star formation by heating \citep{Bower-06} or removing \citep[][]{fabian12, Trussler-20} the molecular gas available for star formation, ultimately also changing the galaxy morphology \citep{hopkins+2010}. {Thus, complete samples of AGN are crucial for understanding their role in galaxy evolution.}

{Our structural model of an AGN comprises} an accretion disk, both broad and narrow emission line regions (BLR and NLR, respectively), a dusty toroidal structure called the torus, and in some cases a jet. Most of the variety of the observational properties can be explained with different observation angles and whether or not the line-of-sight (LOS) is impeded by the torus, obscuring the innermost regions. Indeed the distinction between Type I (with both broad and narrow emission lines in the optical spectrum) and Type II (with only narrow emission lines) can be explained under this scenario \citep[Unified model][]{Antonucci-93, Urry-95}. {AGN can emit throughout the complete electromagnetic spectrum}. The big blue bump at optical/UV wavelengths is associated with the accretion disk \citep{Wills+85}, while the Compton-hump at X-rays can be associated with reflection occurring in the torus \citep{Murphy+09} or the disk \citep{Fabian+00}. The optical emission lines can be associated with the BLR and NLR and their profiles give information about the physical characteristics of these regions \citep[see the review by][]{Netzer-15}. However, the AGN detection is not trivial; due to their location at the center of galaxies, their detection strongly depends on the intrinsic luminosity but also on the amount of obscuration provided by the host galaxies. Particularly at optical wavelengths, one of the most common methods to detect and classify AGN  is through the use of diagnostic diagrams \citep[BPT diagrams,][]{baldwin81, veilleux87}, which calculate ratios between different narrow emission lines.
These diagrams can potentially identify star-forming (SF) dominated galaxies, and differentiate them from galaxies dominated by nuclear activity, shock excitation \citep{Kewley-13}, post-AGB stars \citep{singh:2013} and low ionization nuclear emission regions \citep[LINERs,][]{heckman80}.
The interpretation of these diagrams is still under debate, particularly because several of the processes mentioned above can locate a galaxy in the same region of the BPT \citep[][and references therein]{rich11, singh:2013, Belfiore-16, lacerda20, Comerford-22}. 
Additionally, the presence of dust in the host galaxy can attenuate the nuclear emission, even preventing us from {detecting} any trace of nuclear activity  \citep[][and references therein]{Moran-02, Trump-09, Agostino-19, Comerford-22}.

Therefore, spatially resolved spectroscopy \citep[e.g.,][]{singh:2013, Belfiore-16, sanchez18,Wylezalek-18} is crucial to study all the processes occurring in galaxies and particularly, for the proper characterization of their nuclei. Indeed, several works have been {performed} aiming to characterize these regions and have been proven successful at detecting AGN. For instance, \cite{lacerda20} found 34 objects in the The Calar Alto Legacy Integral Field Area survey \citep[CALIFA][]{sanchez:2012} to host an AGN. 

They followed a method proposed by \citet{sanchez18} which consists of two main criteria: the object should reside in the AGN region in all BPT diagrams and additionally should have an $\rm{EW(H\alpha)}>$ 3 \AA. This optical classification method is particularly useful to find bright sources where the nuclear emission is not strongly affected by the host galaxy. {In a similar work, \cite{Comerford-22} found 10 AGN in the MaNGA survey, which had been previously classified as SF dominated or LINERs, thanks to the higher spatial resolution. Therefore, spatially resolved spectroscopy is essential for a correct classification of the sources}.
However, different works have shown that this method fails at finding low-luminosity AGN \citep[LLAGN][]{heckman80} where the attenuation by the host galaxy is highly significant, or galaxies with active star formation processes and AGN residing in low-mass galaxies \citep[][and references therein]{Goulding-09,Kewley-13, trump+2015,Cann-19}. 

Another way to detect AGN is through the X-rays, which are dominated by the AGN emission above 2 keV (known as hard X-rays) with a continuum described as a power-law {of the form $A(E) = K E^{-\Gamma} $, where $\Gamma$ can range between 0.7-3, with a mean of $\Gamma = 1.8$ in the local Universe \citep{Nandra+94, Bianchi+09}}. These objects are seen as point-like sources \citep{Netzer-15}, and constitute the bulk of the Cosmic X-ray Background \citep{Comastri-95, Ueda-03}. The brightest {AGN} can even outshine the galaxy \citep{Azadi-17} and therefore are less affected by host-galaxy attenuation \citep{Brandt-15}. The spectral and imaging identification of these objects strongly depend on the telescope resolution, spectral range, and sensitivity. This is particularly important for LLAGN, where the imaging identification can be difficult, but also because there are other energetic processes associated with stellar processes that can emit X-rays, such as X-ray binaries (XRBs) {and} Ultra-Luminous X-ray sources (ULXs), and the distinction between these objects and nuclear activity is crucial. In fact, several works have shown that these processes have a luminosity function that differs from that of AGN \citep[see the review by][and references therein]{Fabbiano+06}. Several studies have attempted to properly characterize the X-ray emission of galaxies, aiming to detect AGN sources. In particular, \cite{Roberts-00} use data from the {\it ROSAT} satellite, finding that 45 out of the 83 galaxies in their parent sample host nuclear sources associated with an AGN. Among the sample, a significant amount of sources (33\%) host LLAGN with luminosities as low as $\rm{L_{X} = 10^{38} \ erg \ s^{-1}}$. {More recently, with the exceptional spatial resolution provided by {\it Chandra}, several works have estimated the population of AGN in different types of galaxies. For instance, \cite{Zhang+09} study a total of 187 local galaxies, with a distance below 15 Mpc. They find that 46\% of them present point-like sources at the center associated with black holes, most of them with low Eddington rates. In addition, \cite{She+17} find a similar fraction of active galaxies in X-rays which had been previously classified as HII galaxies in optical. In a very recent work by \cite{Williams-22}, they find a similar fraction of active galaxies, with a significant amount of them in the low-luminosity regime.} 
Heavily obscured objects (the so-called Compton-thick - CT-AGN) are also difficult to be detected at X-rays, as the material from the inner parts of the AGN can suppress the intrinsic emission \citep{Brightman-08, Georgantopoulos-10, Comastri-11,Azadi-17, Ricci-17}, and observations covering above 10 keV are necessary to properly characterize the spectra \citep{Ricci-11, Netzer-15, Ramos-17}. 

Thus, the use of more than one wavelength to identify AGN sources leads to a more complete statistics on the AGN population. In fact, several works have proven to be successful at identifying AGN using optical and X-ray wavelengths for local \citep{LaMassa-09, Vasudevan-09, Yan-11, Pouliasis-19} and high-redshift objects \citep{Malizia-12, Azadi-17, Agostino-19}. It is clear that the fraction of AGN sources is lower when using a single wavelength \citep{Torbaniuk-21, Birchall-22, Comerford-22}.

Our aim is to identify AGN objects in the CALIFA survey, which provides a census of the local galaxies with $z<0.1$ using spatially resolved spectroscopy, through their X-rays properties, in order to complement the AGN census in the nearby Universe as reported through optical wavelengths by \cite{lacerda20}. Furthermore, it allows us to compare both optical and X-ray wavelengths to understand how reliable each method is at detecting these sources, and what information is missing when only one of them is used. In this work we use data from the \textit{Chandra} X-ray satellite \citep{chandra} because it provides the best spatial resolution in X-rays to date (0.49 arcsec/px), crucial to properly separate our sources from emission due to gas or stellar sources in host galaxy. Moreover, we also perform a spectral analysis to characterize these sources and determine their nuclear nature and whether or not these X-ray sources can be classified as AGN.

This work is divided as follows: in Section\,\ref{sec:optical-reduction} we present the parent CALIFA sample and {a brief description of the data products} employed for the analysis. In Section\,\ref{sec:x-ray_data} we present the X-ray data reduction and imaging analysis, in Section\,\ref{sec:spec-analysis} we present the spectral analysis, while in Section\,\ref{sec:results} we present the results obtained, and in Section\,\ref{sec:discussion} we discuss them, to finally summarize and conclude our work in Section\,\ref{sec:conclusions}. Throughout the analysis, we assume a cosmology of $\rm{H_0 = 70 \ km \ s^{-1} \ Mpc^{-1}}$, $\rm{q_0 = 0}$ and $\rm{\Omega_{\lambda_0} = 0.73}$.

\section{Optical data selection and processing}
\label{sec:optical-reduction}

We selected our sources from the extended CALIFA sample \citep[eCALIFA\footnote{Throughout this text we will refer as CALIFA to the last available internal release of data of this project, i.e., the eCALIFA compilation.}][]{sanchez16, galbany18}. CALIFA is a survey of nearby galaxies observed at the Calar Alto 3.5m telescope, using the {Potsdam MultiAperture Spectrophotometer \citep[PMAS][]{roth:2005} in the PPaK configuration \citep{kelz06}, obtained in the low-resolution mode (V500)}. The sources were drawn from the 7th release of the Sloan Digital Sky Survey \citep[SDSS DR7][]{walcher:2014} with most of their optical extent falling in the field-of-view (FoV) of the instrument and a spatial resolution of {$\sim$1\,kpc ($\sim$2.5\arcsec per spaxel)}. The final sample extends in a wide range stellar masses, {and therefore} a wide range of black hole masses ($\rm{M_{BH}}$), morphological type, and color, with redshifts $0.005 \leq z \leq 0.3$. Note, however, that restrictions on the galaxy mass are imposed in the CALIFA sample since it is known that the evolution of dwarf galaxies (i.e., M$< 10^{7}\rm{M_{\odot}}$) is different from that of their more massive peers. 
The Integral Field Unit technique (IFU) of the PMAS/PPaK allows the possibility of studying both spectroscopic and spatially resolved properties of all galaxies. Altogether, the CALIFA sample {used in this work comprises 941 sources observed in the spectral range [3750-7500]\AA \ with good quality observations in the V500 setup \citep[][and references therein]{sanchez16}. }
{The data have been reduced using version 2.2 of the pipeline \citep{sanchez16}. This code comprises all the usual reduction steps for fiber-fed IFS \citep{sanchez06a}, including fiber tracing, extraction, wavelength calibration, fiber-to-fiber corrections, flux calibration, spatial re-sampling and registration, and differential atmospheric refraction correction. As a result of the reduction, a three-dimensional cube is created, in which the spatial information is registered in the {\it x} and {\it y} axis, and the spectral one in the {\it z} axis. As reported in \citet{sanchez16}, the astrometry accuracy of the final datacubes is $\sim$0.5$\arcsec$, the precision of the absolute photometric calibration is $\sim$8\% ($\sim$5\% relative from blue to red), and the average 3$\sigma$ depth of the spectra {is} $r\sim$23.6 mag/arcsec$^2$. Further details on the data reduction and the quality of the data can be found in \cite{sanchez:2012} \citep[see also][]{husemann13,rgb15,sanchez16}. 

The Pipe3D pipeline \citep{sanchez+2016a, sanchez+2016b, Lacerda22} has been applied to the CALIFA data providing a number of database files where measurements performed in the data cubes are recorded. For the purpose of this analysis we use flux ratios of different emission lines, including the {[$\ion{N}{II}$]$\lambda 6584$/H${\alpha}$}, {[$\ion{O}{III}$]$\lambda 5007$/H${\beta}$}, {[$\ion{S}{II}$]}$\lambda 6717$/H${\alpha}$ and {[$\ion{O}{I}$]$\lambda 6301$/H${\alpha}$}, measured in three different regions of the galaxy: the inner 2.5$\arcsec$ x 2.5$\arcsec$ ({hereafter} {\it center}), effective radius, which is the radius at which half of the luminosity of the galaxy is contained ({hereafter} $R_e$) and the average across the entire galaxy ({hereafter} $All$). Table\,\ref{tab:flux} provides these values, as well as the equivalent width (EW) of the H${\alpha}$ emission line for our compiled sample when detected in the three regions mentioned above. Upper limits correspond to the 3$\sigma$ limit of the corresponding flux.}
\section{X-ray data selection and processing}
\label{sec:x-ray_data}
\begin{table}
\footnotesize
\begin{tabular}[htp]{llllllll}
\hline \\
          Name &  Obsid &         RA &        DEC & z &  Dist. & H. T. &  Exp. T. \\
          & & (deg) & (deg) & & (Mpc) & & (ks)  \\
          \hline
          (1) & (2) & (3) & (4) & (5) & (6) & (7) &(8)  \\
    \hline \hline
  NGC7803   &  6978 & 0.333 & 13.111 & 0.018 & 76.7 & Sb & 28.17 \\
  NGC0023   & 10401 & 2.472   &  25.923 &   0.016 &  51.5 &   Sbc  &   19.98 \\
  NGC0192   & 8171  &  9.806   &  0.864  &   0.014 &  59.0 &   S0a &    19.42 \\
  NGC0197   & 8171   & 9.828  &   0.892   &  0.011 &  58.9 &   E7  &    19.42 \\
  NGC0214   &  9098   & 10.367  &  25.499 &   0.015 &  51.1 &   Sb &     5.04\\ 
  NGC0384    &  2147   & 16.854   & 32.292 &   0.014 &  60.7 &   I  &     44.98\\
  NGC0495    &  10536  & 20.733  &  33.471 &   0.014 &  69.9 &   E4 &     18.64\\
  NGC0499    &  10536  & 20.798   & 33.46  &   0.015 &  66.8 &   Sbc &    18.64\\
  NGC0496    &  10536  & 20.798  &  33.529 &   0.021 &  63.4 &   E2 &     18.64\\
  NGC0504    &  317    & 20.866  &  33.204  &  0.014 &   64.7 &  Sb  &    27.19\\
  NGC0507    &  317   &  20.916  &  33.256 &   0.016 &  69.1 &   Sb &     27.19\\
 \hline \hline \\

\end{tabular}
\caption{{First 10 rows of the }compiled sample of 138 sources with {\it Chandra} data available from the extended CALIFA sample. (1) Name of the source. (2) Observation ID in the {\it Chandra} database. (3) Right ascension in degrees. (4) Declination in degrees. (5) redshift. (6) Distance in Mpc. (7) Hubble type. (8) Exposure time of the observation measured in kiloseconds. The rest of the table is available as supplementary online material.}
\label{tab:sample}
\end{table} 

 {We searched in the HEASARC\footnote{https://heasarc.gsfc.nasa.gov} archive for \textit{Chandra} observations of the 941 sources with the following criteria}: we select sources that use the ACIS instrument without {grating, the sources} should be located within 30$\arcsec$ from the central coordinates and the observations should have at least 5 ksec of net exposure time to impose a minimum data quality. {We obtained a total of 139 sources with at least one observation. Note that in order to {make sure that we select the same source in both bands}, we use the coordinates from the CALIFA datacubes, which have an astrometry accuracy of 0.5 arcsec . We obtained spectra of the central 3 arcsec for these observations \citep[in this way, we avoid possible contamination by ULXs, see][]{Walton+11, Lehmer+20} and, for each galaxy, we keep the observation that maximizes the signal-to-noise ratio (SNR)}. We further discard M\,87 from the analysis due to the contamination of the jet at X-rays \citep{Prieto16}. Thus, we explore here the 138 targets (compiled sample) with the maximum number of counts observed with \emph{Chandra}. This naturally selects observations close to the focal plane of the telescope and the longest exposures. Observational properties of the sources are indicated in Table\,\ref{tab:sample}.

\subsection{X-ray data reduction}
\label{sec:xray-data-red}
All X-ray data were reduced using the \emph{CXC Chandra Interactive Analysis of Observations} (CIAO\footnote{\url{http://asc.harvard.edu/ciao}}) software version 3.1, following standard procedures \citep[e.g.,][]{Gonzalez-09}. The exposure time was processed to exclude background flares, using the task {\sc lc\_clean.sl}\footnote{\url{http://cxc.harvard.edu/ciao/download/scripts/}} in source-free sky regions of the same observation.

{We selected a circular region of 3 arcsec centred at the source coordinates. This region corresponds to the central source extraction and is used to study the existence and properties of the AGN. We also selected an annular region with an inner radius of 3 arcsec and an outer radius that varied from 5 to 30 arcsec such that its SNR is maximized. This region corresponds to the extended emission extraction and is used to detect any possible emission associated with the host galaxy or the NLR of the AGN. Additionally, we extracted the background for both the central source and extended emission, by selecting circles around the outer radius of the annular extraction with position angles from 0 to $\pm 90$ degrees in steps of 10 degrees. We impose a distance between the background region and the extended emission in order to avoid any possible contamination for both the central and extended regions.  }

We extracted the spectra of the circle, annulus and background, as well as both ARF and RMF files, using the {\sc dmextract}, {\sc mkwarf} and {\sc mkacisrmf} tools. Furthermore, we produced binned spectra of both the nuclear and extended emission, with the {\sc ftgrouppha} tool, using the optimal binning scheme by \cite{Kaastra-16}.

\begin{figure}
\centering
    \includegraphics[scale = 0.35, trim = 50 0 0 0]{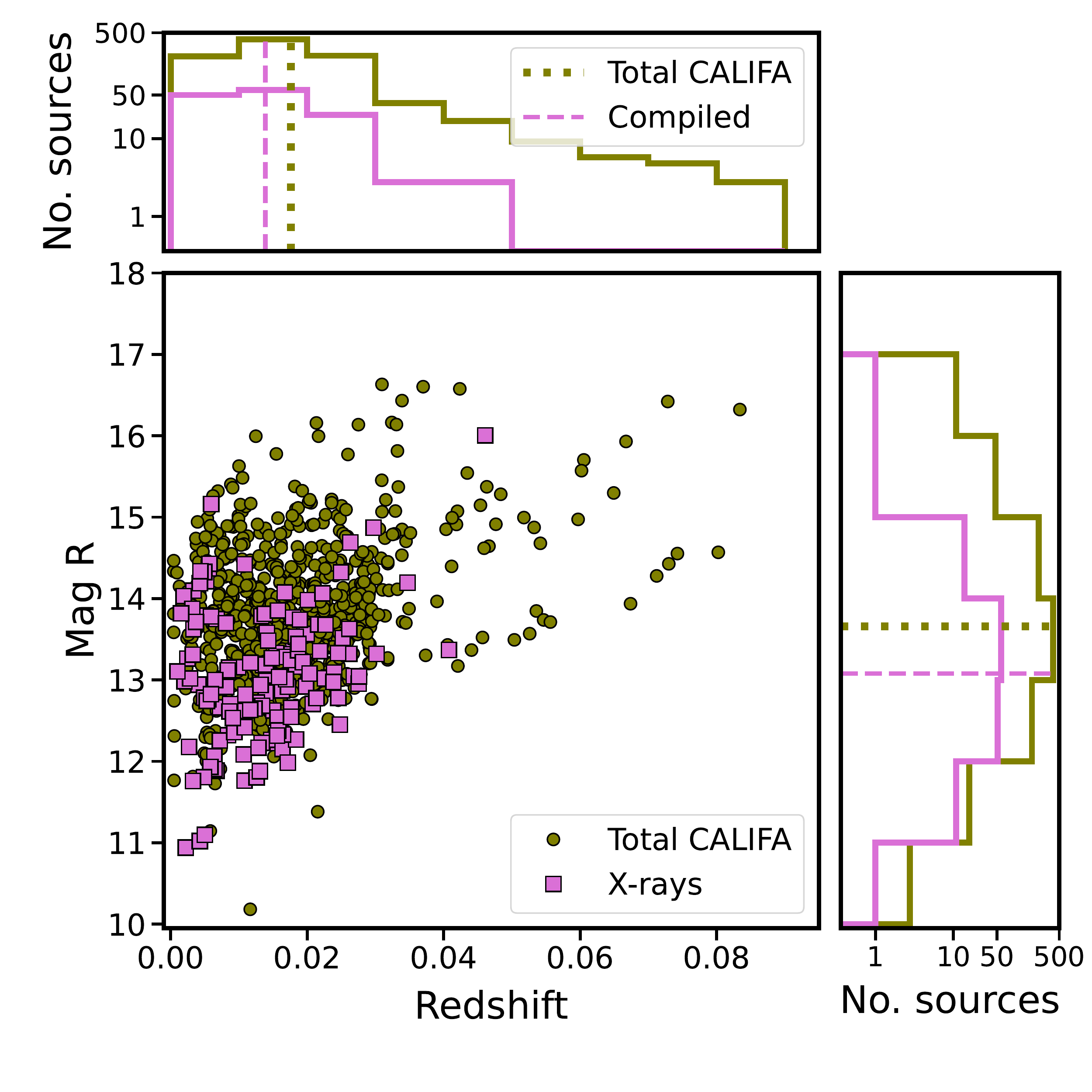}
    \caption{{Magnitude in the R filter versus distance represented in redshift for the CALIFA sample (green circles) and our compiled sample (pink squares). The top and side histograms represent the distribution of redshift and magnitude, respectively, for both the CALIFA sample (green histograms) and our compiled sample (pink). {Note that the axes are presented in logarithmic scale for clarity}. The dotted green and dashed pink lines in both panels represent the mean value in each case for both samples (see text). }}
    \label{fig:redshift}
\end{figure}

{We explore possible biases in the compiled sample compared to the total CALIFA sample.} The {green circles and pink squares} in Fig.\,\ref{fig:redshift} show the absolute magnitude of all the objects in the CALIFA and compiled samples, respectively, measured in the $R$ filter compared to the cosmological distances at which they are located. Top and side histograms show the distributions of distance and magnitude, respectively, of both samples. The mean magnitude of the CALIFA sample is $\langle R \rangle = 13.76 \ \rm{mag}$, while for the compiled sample is $\langle R \rangle= 13.08 \ \rm{mag}$. As for the redshift, the mean value of the CALIFA sample is $\langle z \rangle = 0.019$, while for the compiled sample it is $\langle z \rangle = 0.013$. Moreover, when performing the Kolmogórov-Smirnov (KS) test, we find that for the redshift, the p-value$=0.012$, showing slight differences between both samples, although not very significant, while for the magnitude the p-value$=4\times10^{-12}$, showing much more significant differences between both samples. Therefore, our compiled sample is slightly biased towards the nearest and brightest objects, compared to the total CALIFA sample. This can be naturally explained as for a certain luminosity, sources become fainter with distance, and are harder to detect. Despite any bias, we are losing sources with shorter exposure times than 5 ks, which are discarded based on our selection criteria, although for such short exposure times, we expect for most sources to be undetected. Furthermore, we discuss possible biases on the different samples are analyzed in Sec.\,\ref{sec:bias}. 

\subsection{Structural analysis of the X-ray emission}
\label{sec:morphology}

We aim to separate both the extended and nuclear X-ray emission. To do this, we start by fitting the full band image of all the 138 sources with the {\tt Gaussian2D} model available in {\tt python}. This model has as free parameters: amplitude $a$, the central coordinates ($x$, $y$), dispersion in both directions ($dispx$, $dispy$) and the position angle ($\theta$). 
\begin{figure*}
\includegraphics[width = 1\textwidth,clip,trim=0 0 20 0]{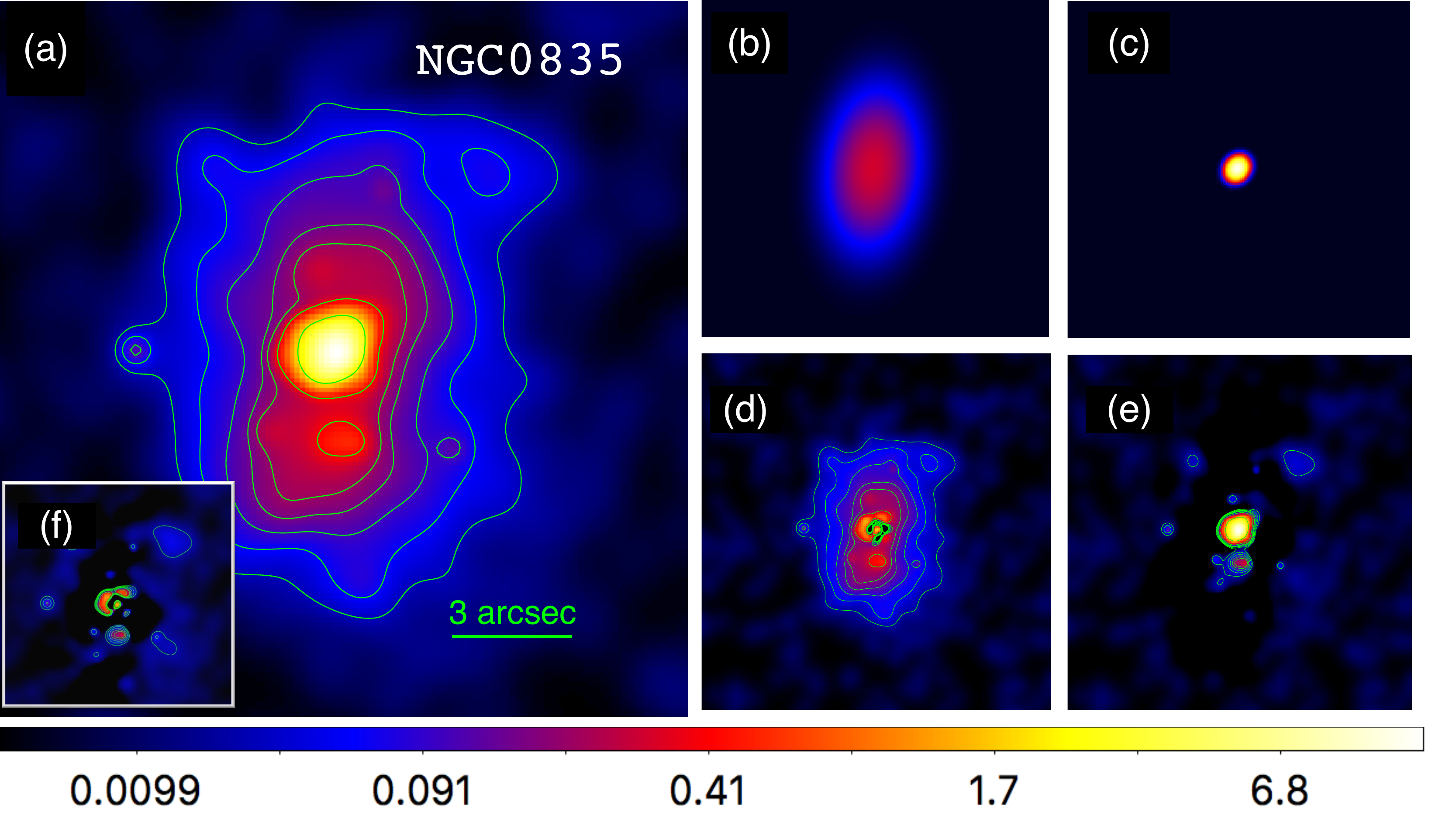}
\caption{{Gaussian fit for the extended emission in NGC\,0835. (a) is the total emission, (b) is the Gaussian profile of the extended emission, (c) is the Gaussian profile of the nuclear emission, (d) corresponds to the extended emission (i.e., (a)-(c)), (e) is the nuclear emission (i.e., (a)-(b), while (f) are the residuals of the image, once accounting for both nuclear and extended emission. The bar at the bottom corresponds to the flux range in the total emission image. Cases in which there is nuclear detection are expected to have a point-like morphology (deconvolved by the PSF of the instrument). Values in the color-bar correspond to counts/s/pixel}.}
\label{fig:NGC0835}
\end{figure*}
We first fit a Gaussian profile for the central source, allowing the amplitude to vary between zero and the maximum value at the central position, which corresponds to the brightest pixel at the center of coordinates. The $x$ and $y$ positions are allowed to vary $\pm 7$ pixels ($\sim 1.2\arcsec$ ), while the dispersion in both directions is allowed to vary from 0 to 4.3 pixels (i.e., $\sim 1\arcsec$). We minimize the values for the Gaussian profile using a least-square minimization routine included in {\tt astropy} (SLSQPLSQFitter). Once we obtain the best-fit for the nuclear Gaussian, we subtract it from the original image, in order to obtain a first estimate of the extended emission. 
Then, to fit the extended Gaussian profile, we allow the $x$ and $y$ positions to vary $10$ pixels ($\sim 1.5\arcsec$), to account for the cases in which the extended emission is not in the same position of the nucleus. As for the dispersion in both directions, we visually inspect each source and depending on the extent of the diffuse emission, we allow this value to vary between 150 and 300 pixels ($\rm{\sim}$18-37$\arcsec$). We then minimize the parameters of the extended Gaussian profile. 

Once again, we subtract the modelled extended Gaussian profile from the original image. In this way, we obtain the nuclear emission. We then re-fit the nuclear emission to a Gaussian profile, subtract the corresponding nuclear Gaussian from the original image, obtaining the extended emission, and we perform a final fit on the extended emission. Note that the double loop on each component of the image allows us to properly decompose both nuclear and extended emission of the sources. In this way, we minimize the contribution of one component on the other.

We finally obtain the best-fit parameters of the nuclear and extended Gaussian profiles, and the residuals of the fit. Our best-fit parameters are those that minimize the residuals. Figure\, \ref{fig:NGC0835} shows the outcome of the process for NGC\,0835 as an example. The left large panel (a) is the original 0.5-10.0 keV image, while (b) and (c) are the extended and nuclear Gaussians best-fit components, respectively, (d) and (e) are the extended and nuclear emission, respectively, and (f) are the residuals of the fit.

We identify a detection of either nuclear or extended emission if the amplitude of the corresponding Gaussian profile is larger than the value of the SNR of the image at the 3-$\sigma$ level. This value is determined in a source-free region in a box of 5x5 pixels. Nuclear regions with detection are referred to as detected sources {in} this manuscript. {The analysis of the extended X-ray emission is the focus of a subsequent paper}. 

\section{X-ray spectral modelling}
\label{sec:spec-analysis}

After determining the sources with nuclear detection (i.e., those for which the amplitude of the nuclear Gaussian is larger than the noise) we select those with more than 50 counts in the 0.5-7.0 keV spectra. {Note that although we are able to obtain images of the full energy range, the effective area of the {\it Chandra} optics significantly decreases above 7 keV. Therefore, for the spectroscopic analysis we restrict the energy band to 7 keV. We model the spectra using a forward-folding approach in order to recover their physical properties.} 

{All models used in this section are summarized in Tab.\,\ref{tab:models}, and are motivated by previous works fitting the intrinsic continuum of AGN \citep{Gonzalez-09, HernandezGarcia-13, HernandezGarcia-15, OsorioClavijo-20, OsorioClavijo-22}, but also by scenarios {unrelated} to AGN such as diffuse hot gas (perhaps from inter-cluster emission) among others \citep{GonzalezMartin-06, Yoshino-09, Ota-14}. We already account for the Galactic absorption in all the models, using the {\sc nh} tool within {\sc ftools} \citep[retrieved from NED\footnote{\url{https://ned.ipac.caltech.edu}} and assuming the HI maps of][]{Kalberla-05}. 

For model $\rm{M_1}$ we use a simple absorbed power-law, which is the simplest approximation to the AGN intrinsic continuum at X-rays. The parameters of this model are the intrinsic absorption (material in the LOS of the AGN), photon index and normalization. We allow these parameters to vary in the range presented in cols. (2), (3) and (4), respectively. 

Model $\rm{M_2}$ reproduces the spectrum of a thermal diffuse gas ({\tt apec} within {\tt xspec}). This model is used to represent warm thermal emission from optically-thin material associated with heating, either by the AGN or any other source (e.g., stellar processes, shocks, etc). The free parameters of this model are the temperature, metal abundances and normalization. The ranges of variation are listed in cols. (6), (7) and (8). Note that in all cases, the metal abundance is fixed to the Solar one.\footnote{Due to the quality of the data, the spectral fits are not sensitive to the metal abundances.}

The third and fourth models are combinations of the previous two. On the one hand, $\rm{M_3}$ represents the partial-covering scenario, which has been widely used in the modelling of bonafide AGN \citep{Gonzalez-09, Ricci-17, OsorioClavijo-22}. In this scenario, the intrinsic absorption is not due to an uniform medium but rather is non-homogeneously distributed (in clumps/clouds or hollow cones). Note that this model can be seen as a combination of two power-laws with the addition of a constant, known as the covering factor (see column (4) in Tab.\,\ref{tab:models}). Altogether, the free parameters of $\rm{M_3}$ are the ones listed in cols. (2)-(4) of Tab.\,\ref{tab:models}.
As for $\rm{M_4}$ it is a combination of models 1 and 2 (i.e., $\rm{M_4 =M_1 + M_2}$) and it is associated with a scenario in which the AGN is surrounded by thermal material from the host galaxy and the free parameters are all those listed in the table.
\begin{table}
\centering
\begin{tabular}{l|ccc|cc}
\hline
           &   \multicolumn{3}{c}{powerlaw}   & \multicolumn{2}{c}{apec}  \\
        
         & $\rm{\log(N_{H,int})}$ & $\rm{\Gamma}$ &  $\rm{C_f}$ & $\rm{kT}$ & $\rm{abund}$  \\
(1) & (2) & (3) & (4) & (5) & (6)  \\              
\hline \hline           
$\rm{M_1}$ & 22-25            & 0.7-3                   & -          & -         & -                            \\
$\rm{M_2}$ & -                & -                       & -          & 0-1.5     & 1            \\
$\rm{M_3}$ & 22-25            & 0.7-3                    & 0-1        &  -         &  -                             \\
$\rm{M_4}$ & 22-25            & 0.7-3                    & 0-1        & 0-1.5     & 1                      
\end{tabular}
\caption{{Models used in the spectroscopic analysis. In all models we already account for the Galactic absorption (see text).  Columns (2) - (6) represent the parameters used in each model version. The parameters with a dash symbol are not used in the model version. The range of variability is represented in each column. Note that each model has a normalization associated with it. For this parameter, we allow it to vary in the whole range.}}
\label{tab:models}
\end{table}
Models $\rm{M_1}$ and $\rm{M_2}$ are defined as {\it simple} models as because they are constituted by one component, while $\rm{M_3}$ and $\rm{M_4}$ are defined as {\it complex} models because they are combinations of the previous ones.

Note that we do not use more sophisticated models of AGN, first due to the quality of the data which might cause an over-fitting of the data, and second because the spectral range recovered by {\it Chandra} is narrow and does not allow the use of such complex models. Nonetheless, our models allow us to determine the existence of AGN. We also tested models including a blackbody component, but they failed to reproduce any of the sources in our spectroscopic sample. In total, three out of the four models used in this work account for the AGN scenario ($\rm{M_1}$, $\rm{M_3}$ and $\rm{M_4}$) while one model accounts for a non-AGN scenario ($\rm{M_2}$). In the following section, we explain how we compare {\it simple} models to {\it complex} ones and ultimately, how we choose the best-fit model of the sources.}

\subsection{X-ray spectral fitting}
\label{sec:spectral-fitting}

For the spectral fitting, we use the X-ray spectral analysis software {\tt xspec} v.12.10.0c. Due to the binning technique used in this work, we use the Cash statistic ($C$-statistic) throughout the analysis \citep{Kaastra-16}. We start by fitting the data to the simple models ($\rm{M_1 \ , \ M_2}$), described in the preceding section and keep the $\rm{C/dof}$ values and the number of bins. Following \cite{Buchner-14}, we estimate the Bayesian Information Criterion (BIC):
\begin{multline}
    \rm{BIC} =\rm{C - m*ln(n)}
\end{multline}

\noindent where $C$ is the $C$-stat of the fit, $\rm{m}$ are the free parameters in the fit and $\rm{n}$ is the number of data points (pha bins of the spectra). The model with the smallest BIC is the best-fit one if the difference is of at least 10 when compared to the other.

If the difference between both BIC values is smaller than 10, then the two simple models produce statistically equivalent spectral fits. We then fit the data to the complex models with the following procedure:
\begin{enumerate}
    \item If there is only one best-fit simple model, we fit the data with complex model including it(i.e., $\rm{M_1}$-$\rm{M_3}$, $\rm{M_1}$-$\rm{M_4}$,  
    $\rm{M_2}$-$\rm{M_4}$.

    \item  In order to test if the {\it complex} model is statistically significant compared to the {\it simple} one, we use the following criterion \citep[in agreement with][]{Brightman-14}. For each free parameter, we require a minimum of $\delta C = 2.71$ when comparing the simple and complex models. {This is equivalent to using the f-test within {\tt xspec}.} 
    \item If this condition is fulfilled, then the complex model is better than its corresponding simple one. Note that, by construction, $\rm{M_3}$ can be compared to $\rm{M_1}$, and $\rm{M_4}$ can be compared to both $\rm{M_1}$ and $\rm{M_2}$. 
 {However, if both simple models are statistically equivalent, we fit the data to the corresponding complex models. In addition to using the f-test accordingly, we use the BIC value between the complex models, in order to determine if there is one preferred model over the others.}
\end{enumerate}
Note that this methodology allows the possibility for more than one model to be able to reproduce the observed spectra of the sources. Once we have the best-fit model (or models) for each source, we calculate the parameters associated with it (or them) and their $1\sigma$ errors using the command {\sc error} within {\tt xspec}.

\section{Results}
\label{sec:results}

In this section, we present the results obtained in the aforementioned morphological (Section\,\ref{sec:morph-results}) and spectral (Section\,\ref{sec:spectral-results}) analyses.

\subsection{Nuclear morphology}
\label{sec:morph-results}

{We find that 66 out of the 138 sources in the compiled sample present nuclear detection. We refer to these sources as {the} detected sample and it corresponds to $\sim 50\%$ of the compiled {sample}. Table\,\ref{tab:Gaussian_fits} shows the Gaussian parameters for the nuclei of the detected sample. To indicate that a source has nuclear detection, the amplitude of the nuclear Gaussian must be larger than the noise of the image. Moreover, the ratio between semi-axes is consistent with a point-like sources in all of the objects when accounting for the errors. Note that although we have made a thorough decomposition of the nucleus and diffuse extended emission, there might still be some contamination of thermal emission in the LOS, mostly in the soft X-ray band (i.e., below 2 keV). While for most sources this contamination is negligible, around 10\% of the sample (six sources) have $>50\%$ of contribution from the extended emission in the nuclear extraction. Three of these sources are spectroscopically analyzed ({namely NGC\,2639, NGC\,3842 and UGC\,11958}), while the remaining three do not have enough SNR in their spectra. Therefore, although the sources with nuclear detection are AGN candidates, their true nature, based on the spectroscopic analysis, is further discussed in the next section.   }

\begin{table}
\centering
\begin{tabular}{p{1.3cm}rrrrrrr}
\toprule
Name &  Amp. &   b & $b/a$ &  Noise & Cont (\%) \\
& {$\rm{cts/s/px}$} &  ($\rm{arcsec})$ &  & (cts/s/px) \\
(1) & (2) & (3) & (4) &(5) & (6) \\
\midrule \midrule
  NGC0023   &      0.7$\pm$0.1 &      0.4$\pm$0.1  &    1*            &    0.039 &  0.33        \\
  NGC0192   &      0.08$\pm$0.01 &      0.5*            &    1*            &    0.089 &  0.14    \\
  NGC0214   &      1.7$\pm$0.3 &      0.3$\pm$0.2  &    1.1$\pm$0.2 &    0.009 &  0.04           \\
  NGC0499   &      0.16$\pm$0.03 &      0.5$\pm$0.1  &    1.1$\pm$0.1 &    0.049 &  0.27         \\
  NGC0507   &      0.5$\pm$0.6 &      0.2*            &    1*            &    0.058 &  0.27       \\
  NGC0508   &      0.07$\pm$0.01 &      0.4$\pm$0.3  &    0.9$\pm$0.3 &    0.01 &  0.032          \\
  NGC0741   &      0.4$\pm$0.1 &      0.6*            &    1*            &    0.09 &  0.31       \\
  NGC0833   &      0.8$\pm$0.5 &      0.1*            &    1*            &    0.094 &  0.27\\
  NGC0835   &      6.2$\pm$7.1 &      0.4$\pm$0.2  &    0.9$\pm$0.2 &    1.636 &  0.015           \\
  UGC01859  &      0.3$\pm$0.1 &      0.2*            &    1*            &    0.008 &  0.27      \\
\end{tabular}
\caption{{First 10 rows of the best-fit values of the Gaussian fit for the nuclear source. Col. (1) is the name of the source, (2) is the amplitude of the nuclear Gaussian, (3) is the semi-major axis of the Gaussian, in arcsec. (4) is the ratio between the semi-major and semi-minor axes, (5) is the noise of the image, calculated in a source-free region and (6) is the contamination from the extended emission. Values with an $*$ symbol have errors below $0.005$. The rest of the table is available as supplementary online material.}}
\label{tab:Gaussian_fits}
\end{table}

\subsection{Spectral fits}
\label{sec:spectral-results}

\begin{table*}
\begin{tabular}{l|ll|ll|lll|lll}
\toprule
 Name &   \multicolumn{2}{c}{PL ($\rm{M_1}$)} &  \multicolumn{2}{c}{APEC ($\rm{M_2}$)} &  \multicolumn{3}{c}{2PL ($\rm{M_3}$)} &  \multicolumn{3}{c}{A+PL ($\rm{M_4}$)}    \\

\midrule

         & $\rm{C/dof}$ & $\rm{BIC}$ & $\rm{C/dof}$ & $\rm{BIC}$ &  $\rm{C/dof}$ & $\rm{BIC}$ &  $\rm{f_{1}}$ & $\rm{C/dof}$ & $\rm{BIC}$ &  $\rm{f_{1}}$/$\rm{f_{2}}$  \\
         (1) & (2) & (3) & (4) & (5) & (6) &(7) & (8) & (9) & (10) & (11)   \\
\toprule

  NGC0023   &   95.42/46             &    83.75  & 86.01/47           &   78.23   &  95.43/45              &    79.86    &   x              &  49.21/44$\clubsuit$    &    29.75   &      $\checkmark$/$\checkmark$    \\
  NGC0192   &   65.49/44             &    53.94  & 90.86/45           &   83.16   &  44.01/43$\clubsuit$     &    28.61  &   $\checkmark$   &  64.21/42               &   44.96    &      x/$\checkmark$              \\
  NGC0499   &   58.09/44             &    46.54  & 55.41/45           &   47.71   &  58.10/43              &    42.7     &   x              &  45.83/42$\clubsuit$    &    26.57   &      $\checkmark$/$\checkmark$    \\
  NGC0507   &   99.51/45             &    87.9   & 30.08/46$\clubsuit$  &   22.33  &   99.51/44              &    84.03  &   x             &   30.08/43             &     10.72    &      $\checkmark$/x             \\
  NGC0741   &   310.20/50            &    298.29 & 91.20/51           &   83.25   &  305.01/49             &    289.12   &   $\checkmark$    &   53.22/48$\clubsuit$    &    33.37 &      $\checkmark$/$\checkmark$   \\
  NGC0833   &   315.49/52            &    303.47 & 889.70/53          &   881.68  &  58.07/51$\clubsuit$     &    42.04  &   $\checkmark$   &   80.84/50              &    60.8    &      $\checkmark$/$\checkmark$   \\
  NGC0835   &   989.69/60            &    977.26 & 3023.78/61         &   3015.5  &  144.52/59             &    127.94   &   $\checkmark$    &   87.55/58$\clubsuit$    &    66.84 &      $\checkmark$/$\checkmark$   \\
  NGC1060   &   76.47/39             &    65.26  & 62.78/40           &   55.31   &  75.80/38              &    60.85    &   x              &  41.57/37$\clubsuit$    &    22.88   &      $\checkmark$/$\checkmark$    \\
  NGC1167   &   37.94/44$\clubsuit$    &    26.39 &  50.60/45           &   42.9   &   35.71/43              &    20.31  &   x             &   35.78/42             &     16.53    &      x/$\checkmark$             \\
  NGC1277   &   63.91/56             &    51.68  & 95.28/57           &   87.12   &  63.91/55              &    47.6     &   x              &  34.83/54$\clubsuit$    &    14.44   &      $\checkmark$/$\checkmark$    \\

\bottomrule
\end{tabular}
\caption{{First 10 rows of the statistical results of the spectral fits for the nuclear extraction. Column (1) is the name of the source. Columns (2), (4), (6) and (9) are the $\rm{C/dof}$ all the models used in the analysis respectively, while Columns (3), (5), (7) and (10) are the Bayesian Information Criterion (BIC) values for each of the models, and Columns (8) and (11) are the f-tests values for the comparison between models $\rm{M_1-M_3}$ and $\rm{M_1-M_4}$/$\rm{M_2-M_4}$, respectively. The club-suit ($\clubsuit$) symbol in each of the $\rm{C/dof}$ columns, represents the preferred model (models) for each source. The rest of the table is available as supplementary online material. }}

\label{tab:spectral-fits}

\end{table*}

{Out of the 66 detected sources, 48 have enough quality (i.e. they have enough SNR, see Sec.\,\ref{sec:spec-analysis}) to perform a spectral fitting (hereafter spectroscopic sample). We fit them to the models described in Section\,\ref{sec:spectral-fitting} and show the $\rm{C/dof}$ for all of them in Table\,\ref{tab:spectral-fits}. The club-suit symbol ($\rm{\clubsuit}$) next to the $\rm{C/dof}$ value for each model highlights the best-fitted model for the corresponding source. 
Out of the 48 sources, 15 prefer the $\rm{M_1}$ model, six prefer $\rm{M_2}$, eight prefer $\rm{M_3}$ and 20 prefer $\rm{M_4}$. Note that one source (NGC\,5675) prefers both $\rm{M_3}$ and $\rm{M_4}$ equally.  

Table\,\ref{tab:tab_params} shows the best-fit parameters for the spectroscopic sample. In general terms, the photon index has a mean value of $\langle\rm{\Gamma}\rangle = 1.9$, which is in agreement with the mean value for the local {Universe} \citep{Nandra+94, Bianchi+09}. The {LOS} column density has a mean value of $\langle\rm{log(N_H)}\rangle = 22.2 \ cm^{-2}$, and the temperature has a mean value of $\langle\rm{kT}\rangle = 0.8\,keV$. {Note that all average values are determined for the best-fit model parameters.} 
In models with a non-negligible absorption, all but nine sources are low obscured (i.e., the obscuration has a value $\rm{N_H < 1\times 10^{22} \ cm^{-2}}$). Additionally, the temperature in either model has average values consistent with strong thermal processes \citep{Strickland-02}, with five of them presenting temperatures above $\rm{kT > 1 \ keV}$.}

As it was mentioned in Sec.\,\ref{sec:spec-analysis}, the simplest approximation to the AGN spectra at X-rays is a power-law. Thus, objects containing a power-law component with photon index consistent with the AGN nature (i.e., $\rm{\Gamma \lesssim 3}$), can be referred to as \textit{bonafide} AGN. This is the case for most objects in the sample with the exception of six (namely NGC\,0507,  NGC\,2639,  NGC\,5953, NGC\,6166N01, UGC\,12127 and  NGC\,7619) which are best fitted to pure-thermal models and could be in principle rejected as AGN candidates. However, these objects could also be low-luminosity sources dominated by the extended emission, or even highly obscured sources. In either case, longer exposure times are needed for the intrinsic continuum to be detected. Therefore, we treat these sources as AGN candidates, along with the rest of the detected sample, for which better SNR is needed to perform a spectral analysis. Thus, the fraction of AGN might raise when the spectroscopic sample is enlarged (such as with spectra from {\it XMM}-Newton and/or {\it NuSTAR}). {Three sources (NGC\,2639, NGC\,3842 and UGC\,11958) have a significant amount of contamination from the extended emission in the morphological decomposition. While the spectrum of NGC\,2639 is described by a thermal component, NGC\,3842 presents in the spectrum with both power-law and thermal components, and UGC\,11958 presents a second power-law in the form of partial-covering. All three scenarios can explain the significant contribution of the extended emission in the nuclear region. }

\subsection{X-ray selection bias}
\label{sec:bias}

{After performing the morphological and spectroscopic analyses, we check if the sources within the detected (see Sec.\,\ref{sec:morphology}) or spectroscopic (see Sec.\,\ref{sec:spec-analysis}) samples have larger exposure times and count-rate with respect to the X-ray compiled sample.
Fig.\,\ref{fig:exptime} shows the distribution of net exposure times in the compiled X-ray sample {(green histogram)}. Around 95\% of the sample (130 sources) have exposure times below 100 ks. The objects with nuclear detection  have a similar distribution with the majority of them having exposure times shorter than 100\,ks. Additionally, sources with enough SNR for the spectral modelling follow the same trend. Therefore, although there is a bi-modality in the distribution of sources with respect to their exposure times, we do not find a clear bias on the distribution of detected sources when compared to that of non-detected sources. Note that from the analysis in sections\,\ref{sec:morphology} and \ref{sec:spec-analysis}, around half of the sample with exposure times below 100 ks have nuclear detection (62 sources), while only 32\% (43 sources) have enough SNR to perform a spectral analysis. As for sources with exposure times above 100 ks, those which are detected can also be spectroscopically analyzed.
\begin{figure}
\centering
    \includegraphics[scale=0.55]{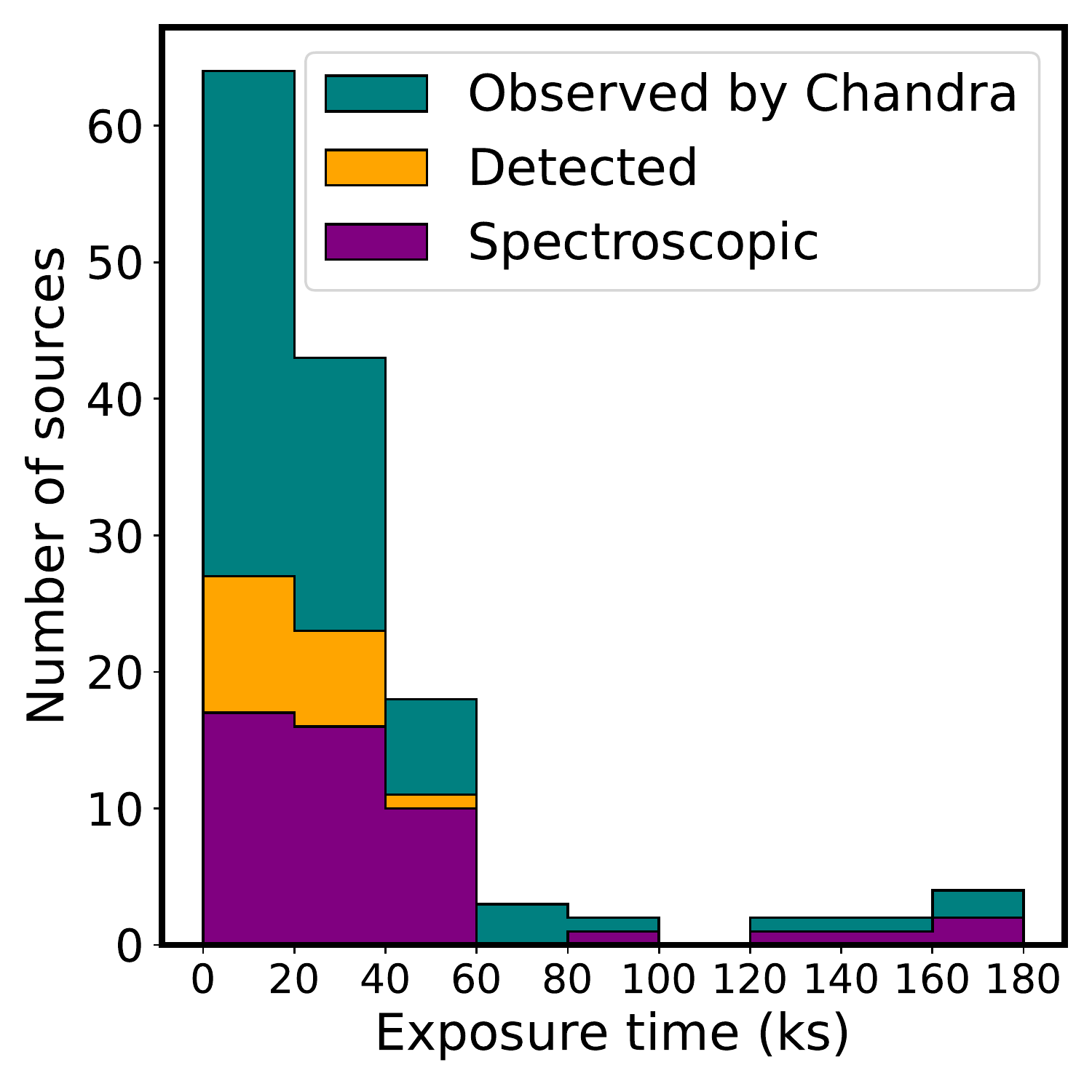}
    \caption{{Histogram of the exposure times for the X-ray observations in our sample. The green, orange and purple  histograms correspond to the compiled sample (138 sources), detected sample (66 sources) and spectroscopic sample (48 sources). For detected and spectroscopic samples, see Sec.\,\ref{sec:morphology} and Sec.\,\ref{sec:spec-analysis}, respectively.}}
    \label{fig:exptime}
\end{figure}

\begin{figure}
\centering
    \includegraphics[width = 0.9\columnwidth]{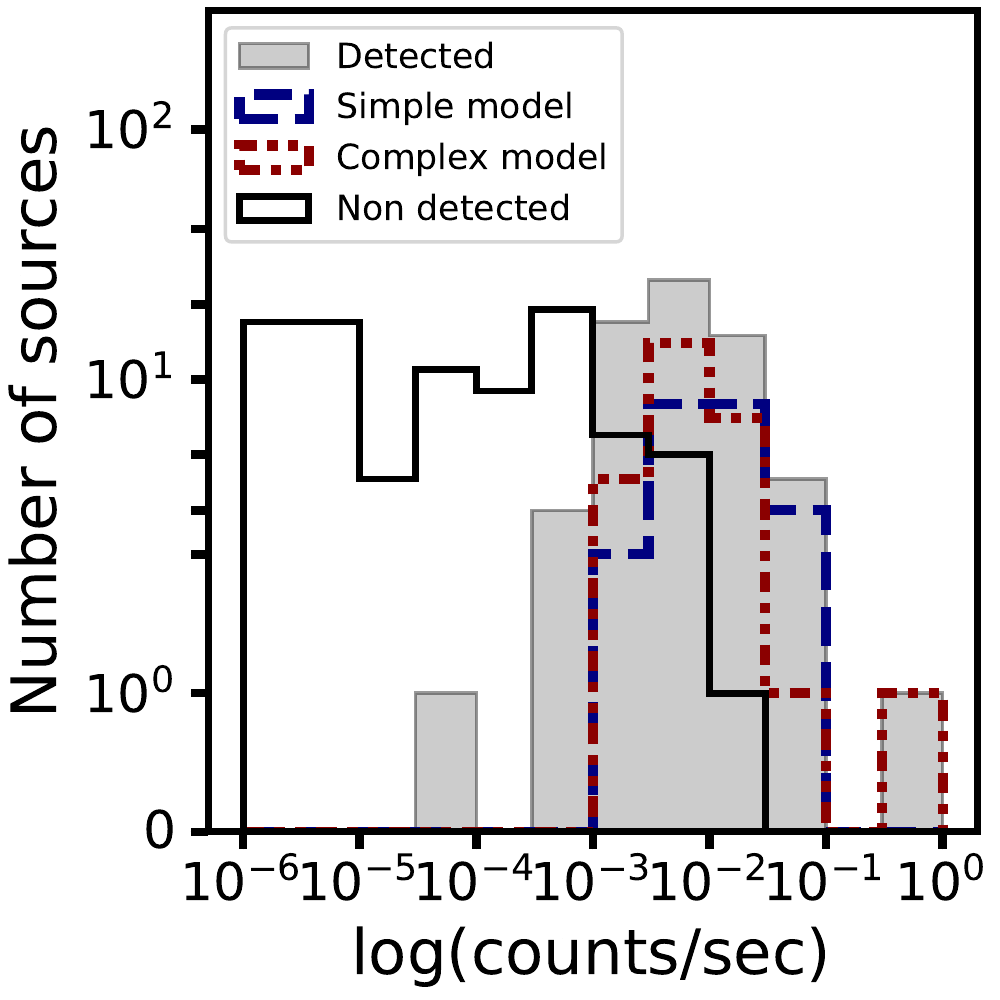}
    \caption{{Distribution of count rate in the sample. The gray histogram is the distribution of sources for which there is detection in the nuclear 3 arcsec extraction, while the blue and red histograms correspond to the sources which are best fitted to simple and complex models, respectively (see Sec.\,\ref{sec:spec-analysis}). Non-detected sources are represented with the blank histogram.}}
    \label{fig:hist-crate}
\end{figure}
On the other hand, Fig.\,\ref{fig:hist-crate} shows the distribution of count rate the different samples analyzed here. Around 52\% of the compiled sample (72 objects, blank histogram) are not detected in the central 3 arcsec extraction, while the remaining $\sim 48\%$ (66 objects) have a count rate between $10^{-5}$ and 1 counts $\rm{s^{-1}}$ (detected sample, see gray histogram in Fig.\,\ref{fig:hist-crate}). Moreover, spectral models are applied for sources with enough SNR in the spectra (48 objects, see Sec.\,\ref{sec:spectral-results}), out of which 27 are best fitted to a simple model (dashed blue histogram) and 21 are best fitted to a complex model (red dotted histogram) and have count-rates above $10^{-3} \ \rm{counts \ s^{-1}}$.}

\subsection{BPT diagrams}

\label{sec:BPT}

Once we select the AGN candidates according to the morphological (see Section\,\ref{sec:morphology}) and the spectral (see Section\,\ref{sec:spec-analysis}) X-ray analyses, we place these objects in the three classical optical diagnostic diagrams: {[$\ion{O}{III}$]/H${\beta}$} versus {[$\ion{N}{II}$]/H${\alpha}$}, {[$\ion{O}{III}$]/H${\beta}$} versus {[$\ion{S}{II}$]}/H${\alpha}$ and {[$\ion{O}{III}$]/H${\beta}$} versus {[$\ion{O}{I}$]/H${\alpha}$}. As mentioned in Sec.\,\ref{sec:optical-reduction}, the Pipe3D {pipeline} provides line flux measurements for three regions: {\it center}, $R_e$ and the average {across the entire galaxy}. From the 138 sources in our compiled sample, 119 have constrained measurements of all the lines required for the diagnostic diagrams explored in this section, and the $\rm{EW\left(H\alpha\right)}$ at the {\it center} CALIFA region. On the other hand 121 have complete measurements at the $R_e$. 

The color-coded area in both panels of Fig.\,\ref{fig:BPT} represent the CALIFA sample, while the white border circles and black border squares are the compiled and detected samples, respectively. The {\it bonafide} AGN are represented with a black cross symbol, while the non-AGN are marked with a black `x' symbol. The color-coded (and color-bar at the top of both panels) represents the $\rm{\log|EW(H\alpha)|}$. The yellow circles are the sources optically classified AGN in common with \cite {lacerda20} (namely NGC\,0833, NGC\,2639, NGC\,5216, NGC\,5675, NGC\,5929, NGC\,6251, NGC\,3861, UGC\,03995 and UGC\,1859). The dot-dashed line in all panels is the \cite{kewley01} demarcation line between {ionization} from AGN and star formation. The dashed line in the {[$\ion{O}{III}$]/H${\beta}$} versus {[$\ion{N}{II}$]/H${\alpha}$} panels corresponds to the one reported by \citet{kauff03} to distinguish from AGN and star forming processes. The dotted line in both {[$\ion{O}{III}$]/H${\beta}$} versus {[$\ion{S}{II}$]}/H${\alpha}$ and {[$\ion{O}{III}$]/H${\beta}$} versus {[$\ion{O}{I}$]/H${\alpha}$} panels is the demarcation line between AGN and LINER ionization \citet{kewley06}. 

Note that the region between the \cite{kewley01} and \cite{kauff03} demarcation lines is also known as mixed or composite region \citep[although this strongly depends on the aperture/resolution of the data, e.g.,][]{mast:2014,davies+2016a,sanchez20}.

{In sections\,\ref{sec:cen} and \ref{sec:re} we explore the distribution of sources in the {diagrams} when the line flux ratios are measured at the $center$ and $R_e$ CALIFA regions. Note that, irrespective of the region, we only report sources with constrained measurements of the flux line ratios. {The results found using the values at $R_e$ and those using the average galactic value ($All$) do not present significant differences. This is expected as the values at $R_e$ are already recovering the main characteristics of the galaxies \citep{sanchez20}. Figure\,\ref{fig:BPTre} shows the distribution of sources in the BPT diagrams for the $All$ region.} 
\subsubsection{Line flux ratios measured at the Center }
\label{sec:cen}

We evaluate the distribution of the objects in the BPT diagrams for all the samples. We start with the {[$\ion{O}{III}$]/H${\beta}$} versus {[$\ion{N}{II}$]/H${\alpha}$} diagram (Fig.\,\ref{fig:BPT} left panel). The distribution of objects in the CALIFA sample for this diagram has a $v$ shape with two main branches (left and right), although the sources are distributed throughout all three regions (AGN/LINER, composite and SF). Our compiled sample also reproduces this distribution. On the other hand, note that the detected sample is mostly located at the right branch of the diagram. 

We start by quantifying the difference between the different samples, by performing a set of KS tests (results reported in Tab.\,\ref{tab:ks}). The p-value of the KS test between the CALIFA and compiled samples suggest no significant differences, while in the case of the CALIFA and detected samples, the p-value suggests that there are significant differences (see Tab.\,\ref{tab:ks}, Column (3)), which is also the case when comparing the CALIFA and spectroscopic samples. Thus, although the compiled sample is representative of the CALIFA sample, there is a lack of X-ray detection on the left branch of this diagram. The numbers reported for the {[$\ion{O}{III}$]/H${\beta}$} versus {[$\ion{N}{II}$]/H${\alpha}$} diagram (first three rows named as NII region in the panel (a) of Table\,\ref{tab:BPT-numbers}) reinforce this result. 

Indeed, around half of the CALIFA sample (426 sources) are located in the SF region of this diagram, while 24\% (213 sources) are located in the composite region and 28\% (248 sources) are located in the AGN region. As for the compiled sample, 45 sources (36\%) are in the SF, 36 sources (28\%) are in the composite region and 36\% are in the AGN regions. The majority of sources in the detected sample are located in either composite (29\%, 18 sources) or AGN (55\%, 34 sources) regions, with only $\rm{\sim}$16\% (10 sources) in the SF region. Interestingly, most of the sources in the non-detected sample are located in the SF region (55\% corresponding to 35 sources), while 27\% (17 sources) are located in the composite region and 18\%  (12 sources) are located in the AGN region. {This is an expected bias as X-ray observations are frequently requested for objects that a priori are expected to have an strong X-ray source. This is the case of already known AGNs, galaxy clusters and mergers and post-mergers.} 
As for the {\it bonafide} AGN, 15\% (six sources) are located in the SF region, while 28\% (11 sources) are located in the composite region, and 57\% (22 sources) are located in the AGN region.

When analysing the {[$\ion{O}{III}$]/H${\beta}$} versus {[$\ion{S}{II}$]}/H${\alpha}$ diagram, we find that the distribution of both the CALIFA and compiled samples are very similar. {This is reinforced when using KS test between these samples (see Column (4), first row of Tab.\,\ref{tab:ks}. This result is also reproduced when comparing the CALIFA {sample with both the detected and spectroscopic samples} (see second and third rows of Col. (4) in the same table).} Figure\,\ref{fig:BPT} and Table\,\ref{tab:BPT-numbers} (panel a) shows that $\sim 60\%$ of the sources in both the CALIFA and compiled samples fall in the SF region (560 and 75, respectively), less than 10\% fall in the AGN region (52 and 10, respectively), while in both cases around 30\% of each sample fall in the LINER region (265 and 39). For the detected sample, around half of the sources (27) are located in the SF region, while around 13\% (8) are located in the AGN region and 42\% (25) are located in the LINER region. {However, when applying the KS test between the CALIFA and detected sample, they do not seem to have statistical significances (see column (4), second row of Tab.\,\ref{tab:ks}.) The behaviour of the non-detected sample is similar to both the CALIFA and compiled samples, with most of the sources located in the SF region (48), while two are located in the AGN region and the remaining 14 sources are located in the LINER region. As for the {\it bonafide} AGN, 17 are located in the SF region, four are located in the AGN region and 17 are located in the LINER one. }
\begin{table}
\begin{tabular}{ccccc}
\multicolumn{1}{c}{Sample} & EW & \multicolumn{1}{c}{OIII vs NII} & \multicolumn{1}{c}{OIII vs SII} & \multicolumn{1}{c}{OIII vs OI}            \\
(1) & (2) & (3) & (4) & (5) \\
\hline \hline
Compiled                 & 0.02/2e-06 & 0.04/0.03        & 0.02/0.08  & 0.03/2E-04 \\ 
Detected                 & 0.14/6e-06  & 2E-05/5E-5 & 0.01/0.02  & 0.033/0.002   \\
Spectroscopic                      & -             & 6E-04/3E-06 & 0.02/0.01 & 0.04/4E-4   \\
\hline \hline
\end{tabular}

\caption{{KS tests performed between the CALIFA, compiled, detected and spectroscopic samples for the BPT diagram diagnostic ratios and the distribution of the $\rm{H\alpha}$ equivalent width. Column (1) is the sample compared to CALIFA, Columns (2-5) are the KS tests in each BPT diagram or EW($\rm{H\alpha}$). The values correspond to the KS in the {\it center}/$R_e$ region. }}
\label{tab:ks}
\end{table}
In the {[$\ion{O}{III}$]/H${\beta}$} versus {[$\ion{O}{I}$]/H${\alpha}$} none of the KS tests performed, suggest a difference in the distribution of the samples in the $center$ measurement. 

\begin{landscape}
\begin{figure}

     \includegraphics[width = 0.9
\columnwidth]{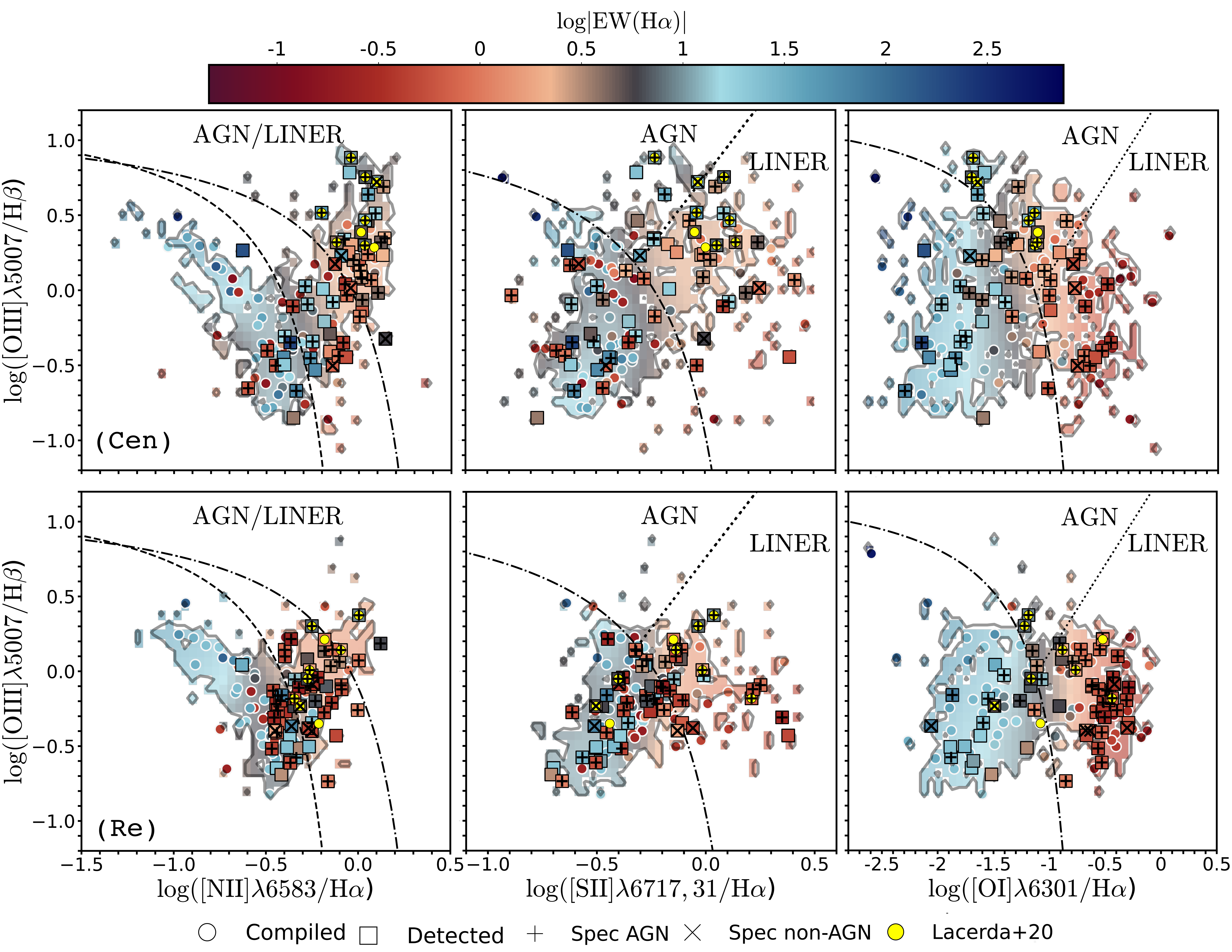}
    \caption{{Top panel: diagnostic diagrams for the CALIFA measurements at the {\it center} region. Bottom panel: diagnostic diagrams for the CALIFA measurements at the {\it $R_e$} region. In all the diagrams, the color-coded image at the back corresponds to the CALIFA, while the white border circles, black border squares, black crosses and $\times$ symbols correspond to our compiled detected, spectroscopic AGN, and non-AGN samples, respectively, and the yellow circles are the sources in common with \citet{lacerda20}, respectively. The dot-dashed line in all panels is the \citet{kewley01} demarcation line between AGN and star formation ionization, the dashed line in the {[$\ion{O}{III}$]/H${\beta}$} versus {[$\ion{N}{II}$]/H${\alpha}$} panels is the one reported by \citet{kauff03}, and the dotted line in both {[$\ion{O}{III}$]/H${\beta}$} versus {[$\ion{S}{II}$]}/H${\alpha}$ and {[$\ion{O}{III}$]/H${\beta}$} versus {[$\ion{O}{I}$]/H${\alpha}$} panels is the demarcation line between AGN and LINER ionization \citep{kewley06}.}}
    \label{fig:BPT}
\end{figure}
\end{landscape}
Moreover, half of the sources for the CALIFA sample fall in the SF region (549), while around 16\% (135) fall in the AGN region and the remaining 21\% (182) fall in the LINER region (Table\,\ref{tab:BPT-numbers}, panel a). The total compiled sample is similarly distributed, with 54\% of the sources (66) falling in the SF region, while 17\% (21) fall in the AGN region and the remaining 29\% (34) fall in the LINER region. Non-detected and detected samples are similarly distributed, with 53\% and 57\% of the samples located in the SF region, 22\% and 13\% of the samples falling in the AGN region and 24\% and 32\% of the samples located in the LINER region, respectively. 
In the case of the {\it bonafide} AGN, 53\% are located in the SF region, 23\% are located in the AGN region and the remaining 23\% are located in the LINER region. As for the non-AGN, they are located evenly in all three regions of the {diagram}. 
Finally, Fig.\,\ref{fig:eq} shows the distribution of the $\rm{EW\left(H\alpha\right)}$ for the CALIFA, compiled and detected samples for sources with constrained values of flux ratios. The CALIFA sample follows a bimodal distribution with a mean value of $0.71\pm0.69 \ \rm{dex}$ when measured at the {\it center} (top panel). As for the compiled sample, the distribution has a mean value of $0.56\pm 0.83  \ \rm{dex}$. Finally, for the detected sample, the EW has a mean value of $0.57\pm 0.73  \ \rm{dex}$. Additionally, the p-value when comparing the distribution of EW in the CALIFA and compiled samples is $0.02$ and $2\times10^{-6}$ for the {\it center} {and {\it $R_e$} region (Col. 2 of the table) {which suggests that the compiled sample is distributed statistically different regardless of the region when compared to the CALIFA sample. Additionally, note that the X-ray samples tend to have lower values of $\rm{EW\left(H\alpha\right)}$}, although all of the distributions have a large dispersion, i.e., span a wide set of values.} 

\begin{table*}
    \renewcommand{\tabcolsep}{0.055cm}
    \begin{tabular}{lccccccc|ccccccc}
    \toprule
    & & \multicolumn{6}{c}{(a)} & \multicolumn{6}{c}{(b)} \\
    \toprule
     & &  CALIFA & Compiled &  Det. & Non det. & AGN& non-AGN & CALIFA & Compiled &  Det. & Non det. & AGN & non-AGN  \\
    &  (1) & (2) & (3) & (4) & (5) & (6) &  (7) & (2) & (3) & (4) & (5) & (6)  & (7)\\
    \toprule \toprule
    \multirow{4}{*}{{\rotatebox[origin=c]{90}{NII}}} & SF & 426 (0.48) & 45 (0.36) & 10 (0.16) & 35 (0.55) & 6 (0.15) & 0 (0) & 595 (0.67) & 72 (0.56) & 24 (0.39) & 48 (0.73) & 12 (0.3) & 2 (0.33)     \\
     & Mix & 213 (0.24) & 35 (0.28) & 18 (0.29) & 17 (0.27) & 11 (0.28) & 1 (0.17) & 245 (0.28) & 51 (0.40) & 34 (0.55) & 17 (0.26) & 24 (0.6) & 4 (0.67)\\
    & AGN & 248 (0.28) & 46 (0.36) & 34 (0.55) & 12 (0.18) & 22 (0.57) & 5 (0.83) &  41 (0.05) & 5 (0.04) & 4 (0.06) & 1 (0.01) & 4 (0.1) & 0 (0) \\
    
    & AGN+EW & 83 (0.1) & 21 (0.17) & 18 (0.29) & 3 (0.05) & 12 (0.31) & 3 (0.5) & 11 (0.01) & 2 (0.01) & 2 (0.04) & 0 (0) & 2 (0.05) & 0 (0) \\ 
   \hline
   \multirow{3}{*}{{\rotatebox[origin=c]{90}{SII}}} & SF & 560 (0.64) & 75 (0.60) & 27 (0.46) & 48 (0.74) & 17 (0.44) & 2 (0.33) & 691 (0.79) & 90 (0.71) & 39 (0.64) & 51 (0.78) & 24 (0.6) & 4 (0.8)           \\
     & AGN & 52 (0.06) & 10 (0.09) & 8 (0.13) & 2 (0.03) & 4 (0.12) & 2 (0.33) & 10 (0.01) & 1 (0.01) & 0 (0) & 1 (0.02) & 0 (0) & 0 (0) \\
    & LINER & 265 (0.30) & 39 (0.31) & 25 (0.42) & 14 (0.22) & 17 (0.44) & 2 (0.33) &  172 (0.20) & 35 (0.28) & 22 (0.36) & 13 (0.20) & 16 (0.4) & 1 (0.2)  \\
    \hline
    \multirow{3}{*}{{\rotatebox[origin=c]{90}{OI}}} & SF &  549 (0.63) & 66 (0.54) &  31 (0.53) & 35 (0.57)  & 20 (0.53) & 1 (0.2) & 504 (0.57) & 48 (0.37) & 19 (0.31) & 29 (0.43) & 8 (0.2) & 2 (0.33)         \\
     & AGN & 135 (0.16) & 21 (0.17) & 13 (0.22) & 8 (0.13) & 9 (0.23) & 1 (0.2) & 31 (0.04) & 6 (0.05) & 4 (0.07) & 2 (0.03) & 4 (0.1) & 0 (0) \\
    & LINER & 182 (0.21) & 34 (0.29) & 14 (0.24) & 20 (0.32) & 9 (0.23) & 3 (0.5) & 346 (0.39) & 75 (0.58) & 38 (0.62) & 37 (0.54) & 28 (0.7) & 4 (0.67) \\
    \bottomrule \bottomrule
    \end{tabular}
   \caption{Number of sources for each sample in each region of the BPT diagrams. Rows 1-4 correspond to the  {[$\ion{O}{III}$]/H${\beta}$} versus {[$\ion{N}{II}$]/H${\alpha}$} diagram, rows 5-7 correspond to the {[$\ion{O}{III}$]/H${\beta}$} versus {[$\ion{S}{II}$]}/H${\alpha}$ diagram and rows 8-10 correspond to the {[$\ion{O}{III}$]/H${\beta}$} versus {[$\ion{O}{I}$]/H${\alpha}$} diagram. Col. (1) is the region (i.e., SF, composite, AGN and/or LINER) depending on the diagram. Cols. (2-7) are the CALIFA, compiled, detected, non-detected {\it bonafide} AGN, and non-AGN samples, respectively. The values in each column correspond to the number of sources, showing also between parenthesis the fraction of the corresponding sample. (a) corresponds to the measurements at the {\it center} and (b) corresponds to the measurements at the $R_e$. No the sources in each region). Note that the fourth row in the {[$\ion{O}{III}$]/H${\beta}$} versus {[$\ion{N}{II}$]/H${\alpha}$} region, corresponds to values within the AGN region which also have $\rm{EW(H\alpha)}$ above 3 \AA.}
    \label{tab:BPT-numbers}

\end{table*}

\subsubsection{Line flux ratios measured at the Effective radius}
\label{sec:re}


{In the case of the [$\ion{O}{III}$]/H${\beta}$ versus [$\ion{N}{II}$]/H${\alpha}$ diagram (panel b in Fig.\,\ref{fig:BPT}), we find that the CALIFA and compiled samples do not have significant differences (as can be seen in Column (3), second row of Tab.\,\ref{tab:ks}). On the contrary, the CALIFA and detected samples are statistically different (see Column (3), third row of the same table), and the same happens between the CALIFA and sepctroscopic samples. However, as for the number of sources in each region, most of the sources in the CALIFA, compiled and non-detected samples fall in the SF region (595, 72 and 48 sources), respectively. The second most populated region is the composite, where there are 245 sources from the CALIFA sample, 51 sources from the compiled sample, 34 sources from the detected sample, 17 sources from the non-detected sample and 24 sources from the {\it bonafide} AGN sample. Therefore, there seems to be a preference towards the SF/composite region in all the samples. This will be further discussed in Section\,\ref{sec:discussion}.

When analyzing the [$\ion{O}{III}$]/H${\beta}$ versus [$\ion{S}{II}$]/H${\alpha}$ diagram, the fraction of sources is similar to that at the $center$ (see panel (b) of Tab.\,\ref{tab:BPT-numbers}, and there are no significant differences as seen through the KS tests between the samples (see Col. 4 in Tab.\,\ref{tab:ks}). The vast majority of sources fall in the SF region (691, 90, 39, 51, 24 and four sources from the CALIFA, compiled, detected, non-detected, {\it bonafide} AGN and non-AGN, respectively), followed by the LINER region (172, 35, 22, 13, 16 and one sources, respectively) and only few sources fall in the AGN region. Interestingly, the detected sample does not have any sources in the AGN region.

As for the [$\ion{O}{III}$]/H${\beta}$ versus [$\ion{O}{I}$]/H${\alpha}$ diagram, 57\% of the CALIFA sample is located in the SF region, while 4\% is located in the AGN region and 39\% is located in the AGN region. The compiled, detected and non-detected samples follow a similar behaviour, with 37\%, 31\% and 43\% of the sources in the SF region, respectively, while 5\%, 7\% and 3\% are located in the AGN region, respectively, and 58\%, 62\% and 54\% located in the LINER region. As for the {\it bonafide} AGN sample, $\sim20\%$ fall in the SF region, 10\% in the AGN region and 70\% fall in the LINER region, while for the non-AGN, the fraction slightly changes as no source populates the AGN region. Note that in the case of the $R_e$ measurements, the KS test suggest significant differences between the CALIFA sample and the compiled/detected/spectroscopic samples samples.

Finally, from the distribution of $\rm{EW\left(H\alpha\right)}$ (Fig.\,\ref{fig:eq}, it is possible to see the bimodality of the CALIFA, compiled and detected samples. The mean of the CALIFA sample is $0.75\pm0.72 \ \rm{dex}$. As for the compiled sample, the distribution has a mean value of $0.43\pm0.83  \ \rm{dex}$. Finally, for the detected sample, the EW has a mean value of $0.33\pm0.70 \ \rm{dex}$. Note that in this region, there is a larger fraction of sources with $\rm{EW\left(H\alpha\right)}< 0.5 \ \rm{dex}$ in both the compiled and detected samples, in contrast with the fraction of sources in the CALIFA sample.}

\begin{figure}

     \includegraphics[width = 0.9
\columnwidth]{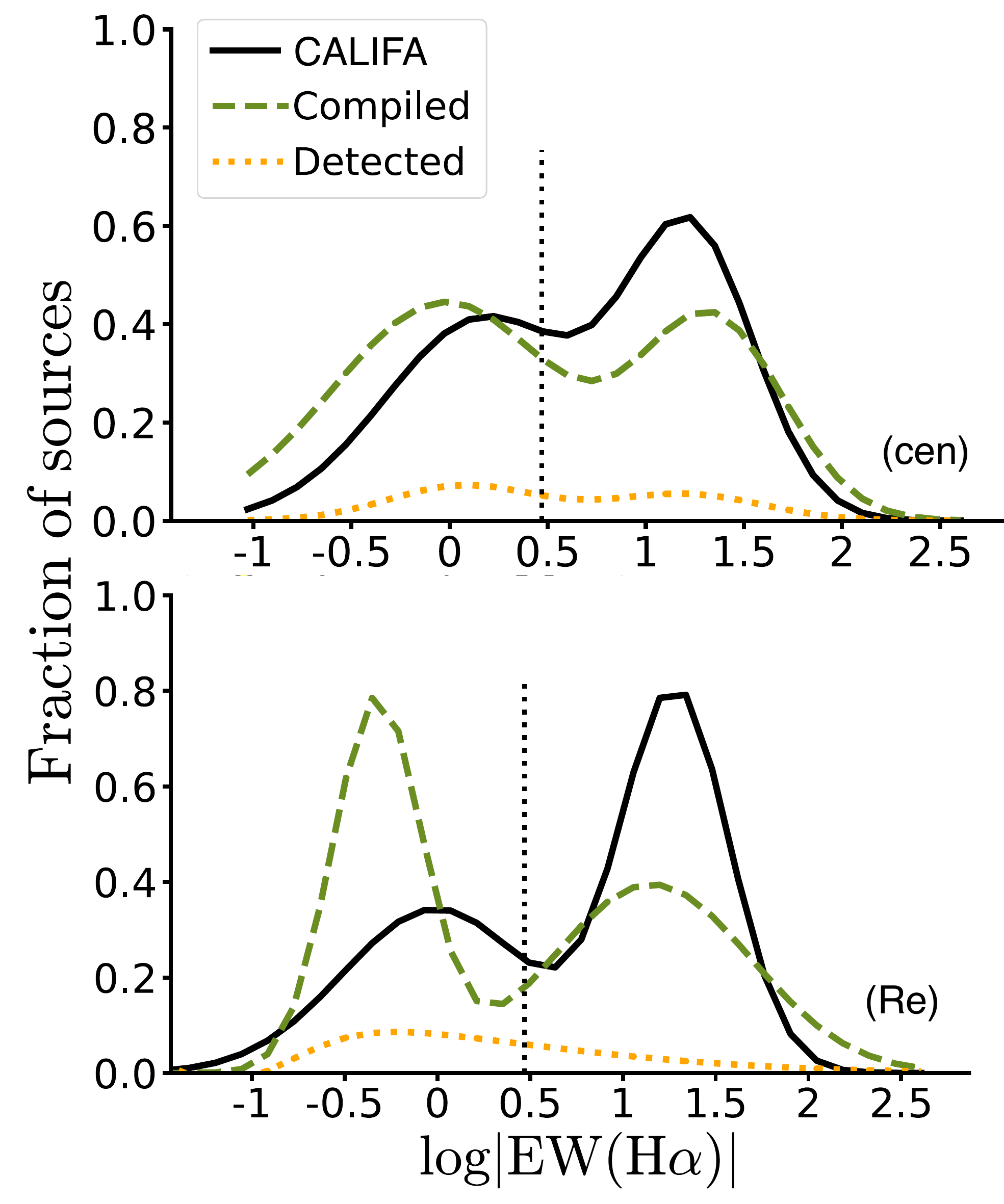}
    \caption{{Top: Histogram distribution for $\rm{EW\left(H\alpha\right)}$ measured at the $center$ region. Bottom: Histogram distribution for $\rm{EW\left(H\alpha\right)}$ measured at the $R_e$ region. The black, green and orange lines correspond to the distribution of $\rm{EW(H\alpha)}$ for the CALIFA, the X-ray compiled and detected samples, respectively, and the dotted line corresponds to the limit of $\rm{EW(H\alpha) = 3}$\AA \ imposed by \citet{lacerda20} for AGN sources. Both compiled and detected samples are normalized such that the integral over the range is 1. All the distributions are weighted by the total number of sources in each sample.}}
    \label{fig:eq}
\end{figure}

\section{Discussion}
\label{sec:discussion}

From the results presented in the previous section, we find that from the 138 sources studied in this analysis (i.e., $\sim 15\%$ of the CALIFA) 66 present a clear nuclear emission at X-rays ($\sim 48\%$ of the X-ray sample and $\sim 7\%$ of the CALIFA sample) out of which 42 (30\% of the X-ray sample and 5\% of the CALIFA sample) are {\it bonafide} AGN from the spectroscopic analysis. 
In this section we analyze how many AGN are in the CALIFA sample (Section\,\ref{sec:dis1}), what are the differences between sources found at X-rays in contrast to those found at optical by analyzing their location in the BPT diagrams (Section\,\ref{sec:dis2}), and ultimately what are the benefits of studying AGN at each wavelength (Section\,\ref{sec:dis3}).

\subsection{AGN rate in the CALIFA}\label{sec:dis1}

{From the X-ray spectroscopic analysis, we find that 42 out of the 48 sources in the spectroscopic sample ($\sim 88\%$) can be classified as {\it bonafide} AGN, while six ($\sim 12\%$) can be ruled out as {\it bonafide} AGN. However, these six objects and the rest of detected nuclear sources are treated as AGN candidates. 

{From the 42 AGN sources classified in this analysis, six are already reported in the work by \cite{lacerda20} using optical wavelengths. Therefore, our work adds at least 36 extra {\it bonafide} AGN to the sample of AGN in the CALIFA survey, for a total of 70 {\it bonafine} AGN, which raises the percentage of nuclear activity from 4\% to $\sim$7\%. }

\citet{lacerda20} found a total of 34 AGN in the CALIFA using optical diagnostics. We find that among the nine objects classified as AGN by \cite{lacerda20} in common with our X-ray sample {(namely NGC\,0833, NGC\,2639, NGC\,5215, NGC\,5675, NGC\,5929, NGC\,6251, NGC\,3861, UGC\,3995 and UGC\,01859)}, all but one (NGC\,3861) have clear X-ray emission based on our analysis. Interestingly, NGC\,3861 presents among the shortest exposure times (see Table\,\ref{tab:sample}) which may be the cause for its non-detection. Moreover, it is classified as a Type-II AGN, and therefore it might be intrinsically obscured, thus more difficult to detect. Indeed, \cite{Terashima-15} show that this object is Compton-thin obscured in X-rays. Therefore, short exposure times may also explain why we do not detect an AGN in the sources fitted to the pure-thermal scenario.
This reinforces the commonly accepted scenario in which most of the X-ray detected sources are {\it bonafide} AGN. 

{We find in our work that six sources are not fitted to a model associated with the AGN scenario. These six soures are NGC\,0507, NGC\,2639,  NGC\,5953, NGC\,6166N1, UGC\,12127 and NGC\,7619. NGC\,2639 is the only object in common with \cite{lacerda20} that is not classified as {\it bonafide} AGN in our work. We find that the best-fit model only accounts for thermal soft X-ray emission. This object is classified as a Type-II AGN in their work, which might suggest that is highly obscured and cannot be detected with the {\it Chandra} capabilities. This was also found by \citet{Gonzalez-09, Williams-22}. Additionally, \cite{Sebastian-20} finds radio emission at the core of the galaxy. Therefore, the lack of bright X-ray sources can be due to the fact that the spectral range covered by {\it Chandra} ({only up to $\sim7$ keV}) might be preventing us from detecting highly obscured X-ray AGN sources \citep[see also][]{LaMassa-09, Yan-11, Azadi-17} Thus, information above 10\,keV is crucial to ultimately ascertain the nature of this object.}

Nonetheless, other physical mechanisms can produce X-ray emission. For instance, galactic-scale outflows can produce X-ray emission in the soft X-ray band (0.5-2.0\,keV). Indeed, out of the 17 galaxies in \cite{LopezCoba-19} which present galactic scale outflows seen at optical wavelengths, we have detected X-ray emission in two objects with {\it Chandra} (NGC\,4676A and NGC\,6286). Unfortunately, neither of these objects have enough SNR in the nuclear region to perform a detailed spectroscopic analysis. It should be noted, though, that these outflows can also be detected in the extended X-ray emission when they are prominent and/or very extended. {From a very preliminary analysis, we note that in the case of NGC\,4676A, there is a significant extended component along the major axis of the X-ray emission, which might be consistent with the outflow. In fact, the nuclear emission has a significant contribution from the extended one ($\sim 47\%$). This is not the case for NGC\,6286, which is not detected in the {\it Chandra} observation. However, this will be further explored in the following paper.} Sources within the spectroscopic sample which are best-fitted with a ionized thermal component could also be candidates for outflows. This will be investigated in Paper II, aiming to understand the extended emission of these sources. X-ray binaries can also produce hard X-ray emission. In fact, their luminosities can even be up to $\rm{10^{38} \ erg \ s^{1}}$\citep[see the review by][]{Fabbiano+06}, or higher in the case of ULXs \citep{Walton+22}. Therefore, accounting for the intrinsic X-ray luminosity is an additional way to ensure the detection of AGN sources. If the central source has a luminosity above this value, a stellar process cannot account for such luminosity, and the source is most likely to be an AGN.}

Thus, we explore the nature of the selected AGN when using both criteria (optical vs. X-rays) based on well known correlations for $\rm{L_{(2-10 \ keV)}}$ and {$\rm{L_{[\ion{O}{III}]}}$} \citep{Bassani+99}. To do this, we estimate the luminosity of the best-fit model for the objects in the spectroscopic sample, by adding the {\tt clumin} component as a convolution component within {\sc xspec}. The intrinsic luminosity is computed from the {unabsorbed} power-law component when the best-fit model includes it, and luminosity of the thermal component otherwise. The X-ray and \ion{O}{III}\footnote{The $\rm{L_{[\ion{O}{III}]}}$ luminosities are corrected by dust attenuation, assuming a Balmer decrement $\rm{H\alpha/H\beta = 2.86}$ and an extinction law \citep[e.g.,][]{cardelli:1989}.} luminosities are reported in Columns (12) and (13) of Tab.\,\ref{tab:tab_params}, respectively. Note that although we fit the spectra in the 0.5-7 keV energy band, we prefer to calculate the luminosity in the 2-10 keV energy band, as it can  be directly compared to the optical luminosity given the well-known correlations. We use convolution model {\tt clumin}, which can escalate the luminosity up to 10 keV. Figure\,\ref{fig:LxversusLOIII} shows this relation for {\it bonafide} AGN (i.e. objects modelled as a power-law, circles) and non-AGN (i.e. modelled with a thermal component, squares). We compare our results with previously found relationships between both quantities for larger samples of AGN \citep[e.g.,][see caption]{Panessa-06, Gonzalez-09, Georgantopoulos-10, Berney-15, Williams-22}. Although most of the sources follow the expected correlation, there are some sources that are located outside of it. {This can be explained under different scenarios: high LOS obscuration, variability, the dust-correction, which might be poorly estimated for objects in which the $\rm{H\beta}$ line is not constrained, and even calibration issues. This can either cause an an underestimation in the X-ray luminosity or an overestimation of the [$\ion{O}{III}$] luminosity.}

Moreover, four of the non-AGN sources also fall along this relation, while the remaining two are clearly located outside this trend, among which is NGC\,2639, optically classified as AGN. However, these six sources do not seem to follow a particular trend, opposite to what is seen with the detected sample. Indeed, four of the non-AGN sources have around the same X-ray luminosities (when accounting for errors) and different optical luminosities. However, obscuration may play an important role in these sources, for which the quality of the data might be insufficient to provide a correct estimate of the obscuration and therefore of the intrinsic X-ray luminosity.
The sources in common with \citet{lacerda20} (yellow dots) have among the largest X-ray and optical luminosities, except for NGC\,2639. This result suggests that BPT diagrams are useful to select luminous/strong AGN, but they are likely to fail at detecting most of the LLAGN. Indeed, \citet{Georgantopoulos-09} propose that for AGN with low X-ray luminosity and high {[$\ion{O}{III}$]} luminosity, the latter might be significantly contaminated by star-formation. Similarly, \cite{Yan-11} use a sample of 146 objects at redshifts $0.3 <z<0.8$, out of which $\rm{\sim}$ 40\% are classified as star-forming galaxies but their X-ray luminosity coincides with a scenario of both SF and nuclear activity, where optical emission is dominated by SF and X-ray emission is dominated by the AGN.

\begin{figure}
    \includegraphics[width=\columnwidth]{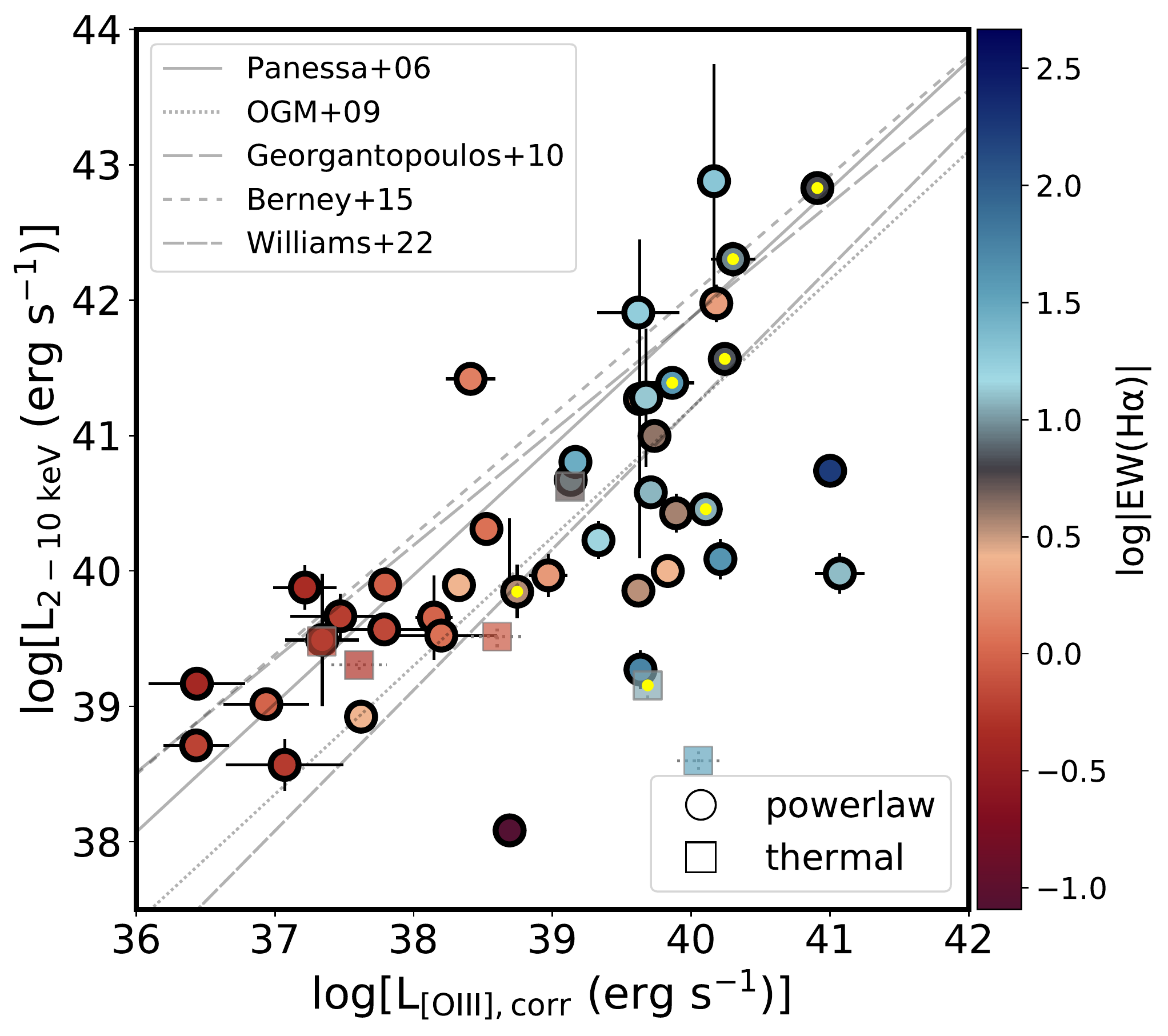}
    \caption{X-ray luminosity in the 2-10 keV band compared to the {[$\ion{O}{III}$]} optical emission line for the spectroscopic sample. The objects are color-coded depending on their $\rm{EW(H\alpha)}$. Squares are object best-fitted to thermal models ($\rm{M_2}$) , while circles are objects best-fitted to power-law models ($\rm{M_1}$, $\rm{M_3}$, and/or $\rm{M_4}$). The yellow dots represent sources in common with \citet{lacerda20}. The grey solid, dotted, long-dashed and dashed lines correspond to the correlations predicted by \citet{Panessa-06, Gonzalez-09, Georgantopoulos-10, Berney-15} and \citet{Williams-22}, respectively.}
    \label{fig:LxversusLOIII}
\end{figure}

All put together, 42 out of 138 X-ray studied sources are {\it bonafide} AGN according to our analysis (i.e. 30\%), which implies an AGN population of 5\% among the CALIFA sample (studying X-ray data only). If all the AGN candidates are treated as AGN, this percentage will increase up to 48\% among the X-ray sample and 7\% among the CALIFA, for a total of 11\% of AGN sources when accounting for both optical and X-ray information. In a subsequent paper, data from \textit{XMM}-Newton and {\it NuSTAR} will help to confirm the nature of these AGN candidates and to study a larger portion of the CALIFA at X-rays.
If the behaviour of the AGN candidates is similar to the spectroscopic one, we could expect up to $\sim 52$ AGN in the CALIFA,  {in addition to} the 34 sources reported by \cite{lacerda20}, which together would represent $\sim 10\%$ of the CALIFA galaxies with possible nuclear activity. This might imply more than twice the AGN population compared to the optical selection \citep[4\% in CALIFA according to][]{lacerda20}. This low percentage of 4\% has also been reported in the MaNGA survey \citep{sanchez18}, using optical information only. Indeed, \cite{Comerford20} report a fraction of $\sim$7\% when using multi-wavelength data in that survey, which is consistent with our result when combining optical and X-ray results.

Previous X-ray {censuses} of AGN have shown that the fraction of AGN in the nearby Universe may be even larger. For instance, in the work by \cite{Zhang+09} where they analyze a sample of 187 nearby sources, they find that $\sim 46\%$ of sources (86) present evidence of nuclear activity. They show that these sources tend to have lower Eddington rates and obscuration, which might explain why the optical fraction is lower. {In addition}, a similar fraction is reported by \cite{She+17} ($\sim 44\%$), most of them optically classified as SF galaxies. Indeed, our lower fraction of AGN sources relies on the incompleteness that our selection criteria impose on the sample, and on the intrinsic properties with which the objects in the CALIFA sample were selected.}

\subsection{Location of X-ray AGN in the BPT diagrams}\label{sec:dis2}

As indicated before, one of our main goals is to compare both optical and X-ray classification criteria. Several authors have shown that spatially resolved spectroscopy is essential to minimize the contamination from the host galaxy when exploring AGN \citep{singh:2013, Belfiore-16, sanchez18, Wylezalek-18}. Indeed, this can be seen in figures\,\ref{fig:BPT} (bottom panel) and \ref{fig:BPTre}, which show the position of the sources in the BPT diagrams when the line flux ratios are measured at $R_e$ and the average value across the entire galaxy ($All$), respectively. Sources tend to group towards the center of the diagrams and are less scattered in contrast to what is seen when the line flux ratios are measured at the 2.5"x2.5" {\it center} aperture (Fig.\,\ref{fig:BPT}, top panel). This effect is also known as host galaxy dilution and can play a significant role in hiding LLAGN emission or sources hosted by galaxies with large amounts of dust \citep{sanchez18, Wylezalek-18}. Therefore, the $R_e$ measurements do not seem to be as reliable to find AGN sources as the $center$ measurements. 

Considering our sample of X-ray AGN candidates and {\it bonafide} AGN, the {[$\ion{O}{III}$]/H${\beta}$} versus {[$\ion{N}{II}$]/H${\alpha}$} diagram is the most reliable {in} separating AGN candidates because the X-ray AGN candidates are clearly segregated from the main sample (see Sec.\,\ref{sec:BPT}). On the contrary the X-ray sources seem to be located throughout all the range of values, not showing any clear difference with the distribution of the remaining sources in the other two diagrams. \cite{Azadi-17} find that high-redshift X-ray/optical AGN are preferentially located in the right branch of the {[$\ion{O}{III}$]/H${\beta}$} versus {[$\ion{S}{II}$]}/H${\alpha}$ diagram, while our local AGN are located mostly in the right branch of the {[$\ion{O}{III}$]/H${\beta}$} versus {[$\ion{N}{II}$]/H${\alpha}$} diagram. However, their BPT diagrams might be similar to those when the line flux ratios are measured at $R_e$ (see above). This is probably due to the fact that high redshift AGN are not only more severely affected by host galaxy emission (due to poorer spatial resolution) but also the redshift might cause source confusion at optical wavelengths \citep{Azadi-17} and they might also miss heavily absorbed weak AGN at X-rays \citep{Malizia-12}. 

\subsection{Completeness of AGN selection criteria}\label{sec:dis3}

In order to compare the selection at both wavelengths, and to determine which type of sources we select in each one, we determine how many objects we may lose at optical wavelengths when the different criteria are used for the BPT diagrams (i.e. completeness). In addition, we also explore how many non-AGN objects we are considering in our statistics when we relax each selection criterion as well (i.e. contamination). We only explore the {[$\ion{O}{III}$]/H${\beta}$} versus {[$\ion{N}{II}$]/H${\alpha}$} diagram since it is the one providing the best separation of X-ray AGN candidates (see Section\,\ref{sec:dis2}). 

For both the AGN/non-AGN and detected/non-detected samples, when not applying any restriction, we obtain a total of {42} {\it bonafide} AGN (i.e., we are 100\% complete) and 62 AGN candidates, although we are contaminating  the sample with $\sim 13-51\%$ of the non-AGN targets. This is clearly a too simplistic approach, in which it is essentially assumed that the location in the diagnostic diagrams presents no useful information regarding the presence or lack of AGN. When applying the \citet{kauff03} demarcation line, which is the less restrictive criterion to select AGN, we are {lose} 15\% of the {\it bonafide} AGN (sources that are now located at the so-called SF region of the diagram as seen in Table\,\ref{tab:BPT-numbers}). 

Additionally, by applying this criterion, the contamination decreases to a range between 13-36\% by counting non-AGN and non-detected sources, respectively. {These results can be compared to those obtained by \cite{Williams-22} in which they find that 89 objects located in the AGN/LINER region of the BPT diagram have X-ray detection (which corresponds to 60\% of their sample). Therefore, they would be 89\% complete, and would have $\sim 11\%$ of contamination.} Furthermore, when applying the \citet{kewley06} line (a more restrictive selection procedure), we lose 43\% of the {\it bonafide} AGN, 35\% of the AGN candidates, and we reduce the contamination to 18-26\%. Therefore, we lose a significant fraction of the {\it bonafide} AGN and AGN candidates, and the contamination does not decrease significantly. {This would be similar in the work by \cite{Williams-22}, where there are only 13 AGN sources and the remaining 68 are LINERs. Therefore, it would correspond to $\sim 14\%$ of completeness}. Finally, if we impose as an additional criterion a limit on the EW of the $\rm{H\alpha}$ line, following \citet{lacerda20} (i.e., considering sources with EW($\rm{H\alpha}$) > 3 \AA \ only), we miss 69\% of the {\it bonafide} AGN and 76\% of the AGN candidates, while the contamination is $16-20\%$ for both samples, respectively. We then miss most of the AGN and the decontamination does not improve. Thus, imposing such a restrictive criterion on the EW might be causing us to miss the weakest AGN at optical wavelengths, just detecting the brightest AGN in this spectral regime. In Fig.\,\ref{fig:LxversusLOIII} we color-code objects according to the EW($\rm{H\alpha}$) to explore possible {patterns} with respect to this parameter. It is clear that most of the bonafide AGN at X-rays show low EW($\rm{H\alpha}$).

It is worth noticing that using either the {[$\ion{O}{III}$]/H${\beta}$} versus {[$\ion{S}{II}$]}/H${\alpha}$ or the {[$\ion{O}{III}$]/H${\beta}$} versus {[$\ion{O}{I}$]/H${\alpha}$} diagram, may cause a much more significant loss, up to 50\% AGN. This is relevant since it is usually considered that AGN selected through line ratios more sensitive to harder ionizing spectra (e.g., {[$\ion{S}{II}$]}/H${\alpha}$ and in particular {[$\ion{O}{I}$]/H${\alpha}$}) are more reliable than those selected when using {[$\ion{N}{II}$]/H${\alpha}$}. Although it should be confirmed with a larger sample, our results clearly indicate the contrary. Therefore, the strict criteria that combines the use of the Kewley demarcation line in the {[$\ion{O}{III}$]/H${\beta}$} versus {[$\ion{N}{II}$]/H${\alpha}$} diagram and the $\rm{EW(H\alpha)}$ are specially useful to select luminous/strong AGN, but it will clearly fail at detecting LLAGN. 

Our results support the idea that the BPT diagrams, although useful at distinguishing between SF and other ionization processes, have long been misused for AGN detection \citep{sanchez:2012, singh:2013, canodiaz:2016, LopezCoba-19, lacerda20}. This is the case for the demarcation line proposed by \cite{kewley01}, which should be used to differentiate those objects harbouring ionization processes not associated with SF. However, as most of the optically-detected AGN sources are located above this line, it has been improperly used to differentiate between pure SF and pure AGN ionization. Additionally, our work reinforces the idea that a single spectral range fails at detecting all families of AGN, as each of them presents its own bias.

\section{Summary and conclusions}
\label{sec:conclusions}

The AGN population in the CALIFA was previously reported to be 4\% by \cite{lacerda20} using optical emission line ratios and the diagnostic diagrams. In this work we have performed an analysis aiming to complement the fraction of AGN sources in the CALIFA (941 sources), using X-ray data from {\it Chandra} for the 138 sources with X-ray observations, and to compare the differences of identifying AGN sources at both spectral ranges. Here we summarize our main results:

\begin{itemize}
    \item The fraction of \textit{bonafide} AGN sources in the CALIFA survey is $7\%$ when accounting for optical and X-ray spectroscopic analyses. However, this fraction could rise up to $10\%$ if AGN candidates are treated as {\it bonafide}. 
    \item X-ray sources are mostly distributed along the right branch in the {[$\ion{O}{III}$]/H${\beta}$} versus {[$\ion{N}{II}$]/H${\alpha}$}, while there is no preferential location for these sources in neither {[$\ion{O}{III}$]/H${\beta}$} versus {[$\ion{S}{II}$]}/H${\alpha}$ or {[$\ion{O}{III}$]/H${\beta}$} versus {[$\ion{O}{I}$]/H${\alpha}$}. This suggests that {[$\ion{O}{III}$]/H${\beta}$} versus {[$\ion{N}{II}$]/H${\alpha}$} is the most reliable diagram to detect AGN sources.
    \item Using only optical data might cause a significant loss of AGN sources that could be obscured by the host galaxy, sources which could be hosted by a high SFR galaxy and low-luminosity AGN. The best way to get a complete statistics on the AGN population is by using multi-wavelength information.
     \item Objects classified as AGN at both wavelengths have among the largest X-ray and optical luminosities, suggesting that a strict criterion in the EW might be biased {towards} brighter sources.
    \item Most of our {\it bonafide} AGN follow the expected correlation between $\rm{L(2-10 \ kev)}$ and $\rm{L}$[$\ion{O}{III}$] which reinforces the AGN nature of these sources.

\end{itemize}

Finally, sources detected at optical wavelengths but missing at X-rays could be intrinsically obscured, unreachable by the capabilities of {\it Chandra}. Observations from more sensitive instruments (e.g. {\it XMM}-Newton) or at higher energies (e.g. {\it NuSTAR}) might help to disentangle this effect.

\section*{Acknowledgements}
We thank the anonymous referee for her/his comments which significantly improved this work. This research has made use of data obtained from the Chandra Data Archive and the Chandra Source Catalog, and software provided by the Chandra X-ray Center (CXC) in the application packages CIAO and Sherpa. We acknowledge support from ESA through the Science Faculty - Funding reference ESA-SCI-SC-LE-083. NOC and OGM acknowledge support from DGAPA-UNAM grant IN105720. NOC would like to thank CONACyT scholarship No. 897887. ICG acknowledges support from DGAPA-UNAM grant IN119123.

\section*{Data Availability}
The X-ray data underlying this article are available in \url{https://heasarc.gsfc.nasa.gov/}. The datasets were derived from sources in the public domain. The optical data underlying this article are available \url{https://califa.caha.es/}.

\bibliographystyle{mnras}
\bibliography{bib}

\label{lastpage}
\appendix

\onecolumn
\section{Best-fit parameters of the spectral fits}
We present here the best-fit parameters of the spectral fits performed in Sec.\,\ref{sec:spectral-fitting}.

\begin{minipage}{\linewidth}

\small
\renewcommand{\tabcolsep}{0.07cm}
\captionof{table}{Best-fit values for the spectral analysis of the spectroscopic sample. In all model versions, we only list for each object, the best-fit values, i.e., the values of the model(s) that best fit the data. Column (1) are the names of the sources, Colum (2) is the AGN classification according to the spectral analysis. Columns (3), (6), (9) and (12) are the column densities (obscuration in the los) in units of $\rm{cm^{-2}}$, while Columns (4), (7), (10) and (14) are the photon indices of the power-law, Columns (5), (11) and (14) are the temperatures of the plasma or neutral material and Column (8) is the covering fraction of the source. Note that we do not list model $\rm{M_3}$ as none of the sources are best fitted to this model.  }
\begin{tabular}{l|c|ll|l|lll|lll|cc}
\toprule
 Name & Type &  \multicolumn{2}{c}{$\rm{M_1}$} &  \multicolumn{1}{c}{$\rm{M_2}$} &  \multicolumn{3}{c}{$\rm{M_3}$} &  \multicolumn{3}{c}{$\rm{M_4}$} & $\rm{log\left(L_{2-10 \ keV}\right)}$ & $\rm{log\left(L_{\ion{O}{III}}\right)}$\\ 
 \midrule
& & $\rm{\log(N_H)}$ & $\Gamma$ & $\rm{kT}$ & $\rm{\log(N_H)}$ & $\Gamma$ & Cfrac & $\rm{\log(N_H)}$ & $\Gamma$ & $\rm{kT}$\\  
& & ($\rm{cm^{-2}}$) & & ($\rm{keV}$) & ($\rm{cm^{-2}}$) & & & ($\rm{cm^{-2}}$) & & ($\rm{keV}$) & $\rm{erg \ s^{-1}}$ & $\rm{erg \ s^{-1}}$ \\
(1) & (2) & (3) & (4) & (5) & (6) & (7) & (8) & (9) & (10) & (11) & (12) & (13)  \\ 

\midrule \midrule
NGC0023 & AGN & - & - & - & - & - & - & $<20.9$ & ${1.9}_{-0.4}^{+0.3}$ & ${0.9}_{-0.1}^{+0.1}$ &  $40.1\pm0.1$	&	$40.21\pm0.04$	 \\                                 
NGC0192 & AGN & - & - & - & ${24.3}_{-0.2}^{+0.2}$ & ${2.4}_{-0.3}^{+0.3}$ & $<0.003$ & - & - & - &  $42.9\pm0.8$	&	$40.2\pm0.1$	 \\                               
NGC0499 & AGN & - & - & - & - & - & - & $<21.4$ & $>1.6$ & ${1.2}_{-0.1}^{+0.1}$  &  $39.7\pm0.3$	&	$38.2\pm0.1$	 \\                                               
NGC0507 & Non-AGN & - & - & ${0.9}_{-0.1}^{+0.1}$ & - & - & - & - & - & -   &   $39.5\pm0.2$	&	$37.3\pm0.1$	 \\                                                    
NGC0741 & AGN & - & - & - & - & - & - & $<21.2$ & $>2.9$ & ${0.78}_{-0.03}^{+0.03}$  &  $39.6\pm0.1$	&	$37.8\pm0.4$	 \\                                            
NGC0833 & AGN & - & - & - & ${23.49}_{-0.04}^{+0.04}$ & ${2.2}_{-0.3}^{+0.3}$ & $<0.01$ & - & - & -  &  $41.6\pm0.1$	&	$40.2\pm0.1$	 \\                            
NGC0835 & AGN & - & - & - & - & - & - &${23.30}_{-0.03}^{+0.03}$ & $<0.8$ & ${0.84}_{-0.03}^{+0.03}$  &    $41.91\pm0.03$	&	$39.6\pm0.3$	 \\                         
NGC1060 & AGN & - & - & - & - & - & - & ${21.6}_{-0.6}^{+0.3}$ & $>2.4$ & ${0.9}_{-0.1}^{+0.1}$  &  $39.9\pm0.1$	&	$37.8\pm0.1$	 \\                            
NGC1167 & AGN & $<20.9$ & ${2.5}_{-0.3}^{+0.3}$ & - & - & - & - & - & - & -   &  $39.9\pm0.1$	&	$41.1\pm0.2$	 \\                                                   
NGC1277 & AGN & - & - & - & - & - & - & ${21.1}_{-0.6}^{+0.3}$ & ${1.6}_{-0.3}^{+0.3}$ & ${1.1}_{-0.1}^{+0.1}$ &     $40.01\pm0.09$	&	$<38.7$	 \\               
UGC03816 & AGN & - & - & - & ${24.1}_{-0.4}^{+0.4}$ & ${2.4}_{-0.3}^{+0.3}$ & ${<0.12}$ & - & - & -  &    $41.3\pm1.2$	&	$39.6\pm0.1$	 \\                          
UGC03995 & AGN & - & - & - & ${23.7}_{-0.1}^{+0.1}$ & ${2.2}_{-0.3}^{+0.3}$ & ${<0.2}$ & - & - & - &  $42.3\pm0.1$	&	$40.3\pm0.2$	 \\                            
NGC2445 & AGN & - & - & - & - & - & - & ${21.9}_{-0.3}^{+0.2}$ & $>2.4$ & ${0.9}_{-0.1}^{+0.1}$  & $39.3\pm0.1$	&	$39.6\pm0.1$	 \\                                  
NGC2484 & AGN &  $<20.9$ & ${2.0}_{-0.2}^{+0.3}$ & - & - & - & - & - & - & -   & $41.9\pm0.1$	&	$40.2\pm0.1$	 \\                                                   
NGC2623 & AGN & - & - & - & ${23.7}_{-0.4}^{+0.3}$ & $<1.2$ & ${0.2}_{-0.2}^{+0.7}$ & - & - & -  &  $41.3\pm0.5$	&	$39.67\pm0.04$	 \\                                
NGC2639 & Non-AGN & - & - & $>1.5$ & - & - & - & - & - & -  &   $39.1\pm0.1$	&	$39.6\pm0.1$	 \\                                                                    
NGC2787 &  AGN & ${21.1}_{-0.2}^{+0.1}$ & ${2.2}_{-0.2}^{+0.2}$ & - & - & - & - & - & - & -  & $38.9\pm0.1$	&	$37.6\pm0.1$	 \\                                     
PGC32873 & AGN & - & - & - & - & - & - & $<21.4$ & ${1.6}_{-0.3}^{+0.5}$ & ${0.9}_{-0.1}^{+0.1}$  &  $40.3\pm0.1$	&	$<38.7$	 \\                               
PGC033423 & AGN & - & - & - & - & - & - & ${21.4}_{-1.2}^{+0.4}$ & ${1.8}_{-0.4}^{+0.5}$ & ${0.7}_{-0.1}^{+0.1}$  &  $40.7\pm0.1$	&	$39.13\pm0.04$	 \\               
NGC3842 & AGN & - & - & - & - & - & - & ${21.6}_{-0.3}^{+0.3}$ & $>2.6$ & ${0.96}_{-0.03}^{+0.03}$  &  $39.7\pm0.2$	&	$37.5\pm0.4$	 \\                             
NGC3860 & AGN &${20.9}_{-0.5}^{+0.2}$ & ${1.6}_{-0.2}^{+0.2}$ & - & - & - & - & - & - & -   &  $40.6\pm0.1$	&	$39.7\pm0.1$	 \\                                     
NGC3945 & AGN & ${20.8}_{-0.7}^{+0.3}$ & ${2.4}_{-0.3}^{+0.2}$ & - & - & - & - & - & - & -  &   $39.5\pm0.1$	&	$38.1\pm0.4$	 \\                                    
NGC4291 & AGN & - & - & - & - & - & - & ${21.9}_{-0.9}^{+0.4}$ & $>1.4$ & ${0.7}_{-0.1}^{+0.1}$  &   $39.2\pm0.1$	&	$36.4\pm0.3$	 \\                               
NGC4486B & AGN & $<20.9$ & ${1.9}_{-0.2}^{+0.2}$ & - & - & - & - & - & - & -  &  $38.7\pm0.1$	&	$36.4\pm0.2$	 \\                                                   
NGC4676B & AGN & $<21.4$ & ${1.8}_{-0.4}^{+0.4}$& - & - & - & - & - & - & -  & $40.2\pm0.1$	&	$39.33\pm0.01$	 \\                                                     
NGC4874 & AGN & - & - & - & - & - & - & ${21.4}_{-0.6}^{+0.3}$ & $>2$ & ${1.0}_{-0.1}^{+0.1}$  &    $39.9\pm0.2$	&	$37.2\pm0.2$	 \\                               
NGC5216 & AGN & $<21.44$ & $>1.94$ & - & - & - & - & - & - & -  &  $39.9\pm0.2$	&	$38.7\pm0.1$	 \\                                                                 
NGC5395 & AGN & $<21.4$ & ${1.5}_{-0.3}^{+0.6}$& - & - & - & - & - & - & -  &  $39.9\pm0.2$	&	$38.9\pm0.1$	 \\                                                       
NGC5532 & AGN & - & - & - & - & - & - & ${21.3}_{-0.4}^{+0.4}$ & ${1.2}_{-0.1}^{+0.2}$ & ${0.79}_{-0.02}^{+0.02}$  &  $41.42\pm0.03$	&	$38.4\pm0.1$	 \\              
NGC5576 & AGN & $<21.6$ & $>1.8$ & - & - & - & - & - & - & - &  $38.6\pm0.2$	&	$37.1\pm0.4$	 \\                                                                    
NGC5614 & AGN & $<20.9$ & ${2.3}_{-0.2}^{+0.3}$ & - & - & - & - & - & - & -  &  $39.9\pm0.1$	&	$39.8\pm0.1$	 \\                                                    
NGC5623 & AGN &$<21.1$ & ${2.2}_{-0.3}^{+0.3}$ & - & - & - & - & - & - & -  &  $39.9\pm0.1$	&	$38.3\pm0.1$	 \\                                                     
NGC5675 & AGN & - & - & - & ${22.7}_{-0.2}^{+0.1}$ & $>1.9$ & ${0.03}_{-0.01}^{+0.06}$ & ${22.8}_{-0.2}^{+0.1}$ & $>2.3$ & $>1.4$ & $40.4\pm0.1$	&	$40.11\pm0.02$	 \\
NGC5845 & AGN & $<21.3$ & ${1.8}_{-0.3}^{+0.4}$ & - & - & - & - & - & - & -   &  $39.0\pm0.1$	&	$36.9\pm0.3$	 \\                                                   
NGC5929 & AGN & - & - & - &${23.3}_{-0.1}^{+0.1}$ & ${1.4}_{-0.3}^{+0.3}$ & ${0.05}_{-0.02}^{+0.03}$ & - & - & -  & $41.4\pm0.1$	&	$39.9\pm0.2$	 \\                
NGC5953 & Non-AGN & - & - & ${1.0}_{-0.1}^{+0.1}$ & - & - & - & - & - & -   &  $38.6\pm0.1$	&	$40.1\pm0.2$	 \\                                                     
ARP220 & AGN & - & - & - & - & - & - & ${21.7}_{-0.2}^{+0.2}$ & ${0.9}_{-0.2}^{+0.2}$ & ${0.9}_{-0.1}^{+0.1}$   & $40.81\pm0.03$	&	$39.2\pm0.1$	 \\                  
NGC6090 & AGN & - & - & - & - & - & - & ${21.9}_{-0.5}^{+0.2}$ & $>2.3$ & ${0.9}_{-0.1}^{+0.1}$  & $40.7\pm0.1$	&	$41.00\pm0.05$	 \\                                 
NGC6125 & AGN & - & - & - & - & - & - & $<20$ & $3 *$ & ${0.9}_{-0.1}^{+0.1}$  &  $39.5\pm0.5$	&	$37.3\pm0.3$	 \\                                                  
NGC6166N1 & Non-AGN &  - & - &$>1.5$ & - & - & - & - & - & -   &  $40.62\pm0.02$	&	$39.13\pm0.03$	 \\                                                                  
NGC6251 & AGN & - & - & - & - & - & - & ${21.2}_{-0.1}^{+0.1}$ & ${1.55}_{-0.03}^{+0.03}$ & ${0.29}_{-0.01}^{+0.02}$  & $42.83\pm0.01$	&	$40.91\pm0.04$	 \\            
NGC6278 & AGN  & ${21.2}_{-0.3}^{+0.2}$ & ${1.9}_{-0.2}^{+0.2}$ & - & - & - & - & - & - & -  & $40.3\pm0.1$	&	$38.5\pm0.1$	 \\                                     
NGC6338 & AGN & - & - & - & - & - & - & ${21.3}_{-0.2}^{+0.2}$ & $>2.9$ & ${1.10}_{-0.05}^{+0.04}$  &  $40.4\pm0.1$	&	$39.89\pm0.03$	 \\                             
UGC11958 & AGN & - & - & - & ${22.1}_{-0.2}^{+0.1}$ & ${2.4}_{-0.4}^{+0.4}$ & ${0.06}_{-0.03}^{+0.05}$& - & - & -  &  $40.9\pm0.1$	&	$39.74\pm0.03$	 \\              
UGC12127 & Non-AGN &  - & - &${0.8}_{-0.1}^{+0.1}$ & - & - & - & - & - & -  &  $39.5\pm0.1$	&	$38.6\pm0.2$	 \\                                                     
NGC7457 & AGN & $<20.8$ & $>2.8$ & - & - & - & - & - & - & -  &  $38.1\pm0.1$	&	$<38.7$	 \\                                                                   
NGC7619 & Non-AGN & - & - & ${0.97}_{-0.03}^{+0.02}$ & - & - & - & - & - & -  &  $39.30\pm0.03$	&	$37.6\pm0.2$	 \\                                                   
NGC7052 & AGN & - & - & - & - & - & - & $<21.19$ & ${2.35}_{-0.27}^{+0.31}$ & ${0.74}_{-0.10}^{+0.07}$  &   $39.8\pm0.1$	&	$39.619\pm0.001$	 \\                        
                               
\bottomrule
\end{tabular}
%
\label{tab:tab_params}

\end{minipage}
\section{BPT diagrams}
\label{Appendix-A}
\begin{figure}

\includegraphics[width = 1\columnwidth]{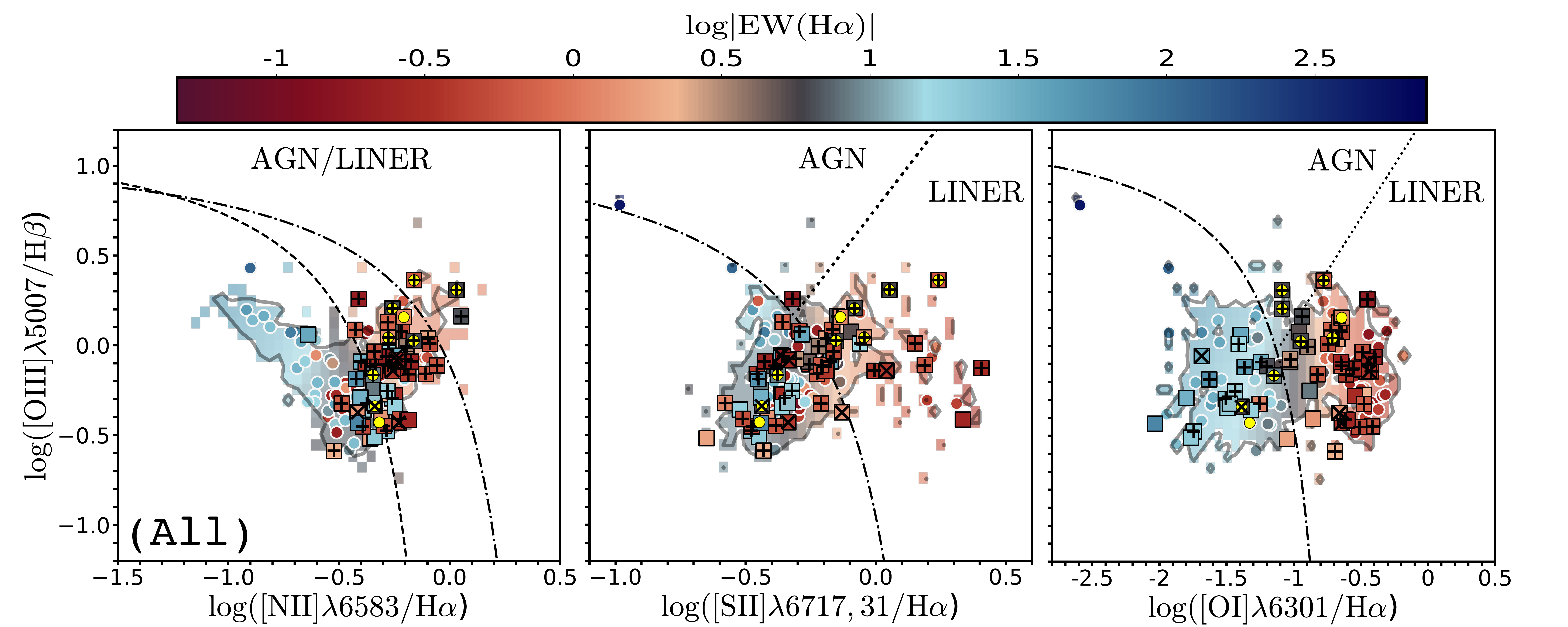}
\caption{{CALIFA measurements at the average galaxy value ($All$) region. The lines and symbols are as in Fig.\,\ref{fig:BPT}.}}
\label{fig:BPTre}
\end{figure}    



\section{Optical emission line flux ratios}
\label{Appendix-B}
\begin{table}
\scriptsize
\begin{tabular}{l|rrrrr|rrrrr}
Name & \multicolumn{5}{c}{{\it Center} } & \multicolumn{5}{c}{$R_e$}  \\
\hline
& {[$\ion{N}{II}$]/H${\alpha}$} & {[$\ion{O}{III}$]/H${\beta}$} & {[$\ion{S}{II}$]}/H${\alpha}$ & {[$\ion{O}{I}$]/H${\alpha}$} & EW(H${\alpha}$) &  {[$\ion{N}{II}$]/H${\alpha}$} & {[$\ion{O}{III}$]/H${\beta}$} & {[$\ion{S}{II}$]}/H${\alpha}$ & {[$\ion{O}{I}$]/H${\alpha}$} & EW(H${\alpha}$)  \\
(1) & (2) & (3) & (4) & (5) & (6) & (7) & (8) & (9) & (10) & (11) \\
\hline \hline
NGC7803 & $-0.311\pm0.047$ & $-0.349\pm0.137$ & $-0.375\pm0.047$ & $-2.303\pm0.164$ & $1.181\pm0.829$ & $-0.347\pm0.028$ & $-0.468\pm0.468$ & $-0.404\pm0.049$ & $-2.265\pm0.217$ & $1.486\pm1.311$  \\
NGC0023 & $-0.262\pm0.028$ & $-0.453\pm0.062$ & $-0.432\pm0.047$ & $-1.844\pm0.140$ & $1.610\pm0.964$ & $-0.250\pm0.045$ & $-0.346\pm0.346$ & $-0.356\pm0.081$ & $-1.593\pm0.253$ & $1.365\pm1.093$  \\
NGC0192 & $-0.291\pm0.040$ & $0.027\pm0.241$ & $-0.496\pm0.044$ & $-1.802\pm0.139$ & $1.289\pm0.477$ & $-0.183\pm0.086$ & $0.136\pm0.136$ & $-0.319\pm0.139$ & $-1.151\pm0.245$ & $0.581\pm0.269$    \\
NGC0197 & $-0.413\pm0.115$ & $-0.181\pm0.269$ & $-0.438\pm0.068$ & $<-1.240$ & $0.995\pm0.779$ & $-0.559\pm0.239$ & $-0.047\pm0.047$ & $-0.197\pm0.225$ & $-0.447\pm0.331$ & $0.808\pm0.852$                 \\
NGC0214 & $0.047\pm0.061$ & $0.462\pm0.206$ & $-0.309\pm0.093$ & $-1.447\pm0.267$ & $0.547\pm0.231$ & $-0.442\pm0.056$ & $-0.647\pm0.647$ & $-0.698\pm0.123$ & $-1.697\pm0.295$ & $1.335\pm0.950$    \\
NGC0384 & $-0.025\pm0.127$ & $0.150\pm0.281$ & $-0.429\pm0.177$ & $-1.141\pm0.288$ & $0.026\pm0.567$ & $-0.363\pm0.226$ & $-0.185\pm0.185$ & $-0.365\pm0.287$ & $-0.329\pm0.329$ & $-0.419\pm0.525$   \\
NGC0495 & $-0.401\pm0.202$ & $-0.390\pm0.504$ & $-0.732\pm0.061$ & $-0.677\pm0.327$ & $-0.352\pm0.760$ & $-0.562\pm0.293$ & $-0.245\pm0.245$ & $-0.504\pm0.134$ & $-0.731\pm0.255$ & $-0.383\pm0.236$ \\
NGC0499 & $0.016\pm0.101$ & $0.077\pm0.256$ & $-0.280\pm0.114$ & $-1.214\pm0.404$ & $-0.011\pm0.648$ & $-0.143\pm0.226$ & $-0.211\pm0.211$ & $-0.362\pm0.231$ & $-0.515\pm0.283$ & $-0.368\pm0.671$   \\
NGC0496 & $-0.467\pm0.015$ & $-0.781\pm0.139$ & $-0.569\pm0.025$ & $-1.900\pm0.194$ & $1.474\pm0.515$ & $-0.526\pm0.087$ & $-0.427\pm0.427$ & $-0.432\pm0.149$ & $-1.462\pm0.399$ & $1.370\pm1.228$  \\
NGC0504 & $-0.239\pm0.156$ & $0.174\pm0.153$ & $-0.488\pm0.065$ & $-0.353\pm0.185$ & $-0.598\pm0.930$ & $-0.267\pm0.147$ & $-0.505\pm0.505$ & $-0.138\pm0.222$ & $-0.402\pm0.225$ & $-0.429\pm0.519$  \\

    \end{tabular}
    \caption{{First 10 rows of the emission line flux ratios. Col. (1) is the name of the source. Cols. (2), (7) and (12) is the ratio between {[$\ion{N}{II}$]} and H${\alpha}$ for central CALIFA, $R_e$ and {\it complete galaxy}, respectively. Cols. (3), (8) and (13) is the ratio between {[$\ion{O}{III}$]} and H${\beta}$ for the {\it center}, $R_e$ and {\it complete galaxy}, respectively. Cols. (4), (9) and (14) is the ratio between {[$\ion{S}{II}$]} and H${\alpha}$ for the {\it center}, $R_e$ and {\it complete galaxy}, respectively. Cols (5), (10) and (15) is the ratio between {[$\ion{O}{I}$]} and H${\alpha}$ for the {\it center}, $R_e$ and {\it complete galaxy}, while Cols. (6), (11) and (16) is the $\rm{EW(H\alpha)}$ for {\it center}, $R_e$ and {\it complete galaxy}, respectively. All measurements are in logarithmic scale. The rest of the table is available as supplementary online material.}}
    \label{tab:flux}
\end{table}

\section{AGN in the CALIFA survey: Online supplementary material}
\onecolumn
\begin{table}
\scriptsize
\begin{tabular}[htp]{llllllll|llllllll}
\hline \\
          Name &  Obsid &         RA &        DEC & z &  Dist. & H. T. &  Exp. T. &  Name &  Obsid &         RA &        DEC & z &  Dist. & H. T. &  Exp. T.\\
          & & (deg) & (deg) & & (Mpc) & & (ks) & & & (deg) & (deg) & & (Mpc) & & (ks) \\
         
          \hline
          (1) & (2) & (3) & (4) & (5) & (6) & (7) &(8) &  (1) & (2) & (3) & (4) & (5) & (6) & (7) &(8)  \\
    \hline \hline

       NGC7803 &   6978 &    0.333 &   13.111 &  0.017657 &      76.7 &        Sb &      28.17  &        NGC3896 &  21091 &  177.235 &   48.674 &  0.003184 &      13.1 &        E4 &      10.07 \\
       NGC0023 &  10401 &    2.472 &   25.923 &  0.015682 &      51.5 &       Sbc &      19.98 &        NGC3945 &   6780 &  178.307 &   60.675 &  0.004277 &      23.2 &        Sc &      15.17 \\
       NGC0192 &   8171 &    9.806 &    0.864 &  0.013928 &      59.0 &       S0a &      19.42 &        NGC4059 &  12990 &  181.047 &   20.409 &  0.023845 &     107.2 &        Sc &       5.06 \\
       NGC0197 &   8171 &    9.828 &    0.892 &  0.010838 &      58.9 &        E7 &      19.42 &         IC3065 &   8076 &  183.802 &   14.433 &  0.003332 &      17.1 &        S0 &       5.17 \\
       NGC0214 &   9098 &   10.367 &   25.499 &  0.015056 &      51.1 &        Sb &       5.04 &        NGC4291 &  11778 &  185.074 &   75.370 &  0.005817 &      35.4 &        E5 &      30.16 \\
       NGC0384 &   2147 &   16.854 &   32.292 &  0.013994 &      60.7 &         I &      44.98 &        NGC4390 &  19425 &  186.461 &   10.459 &  0.003726 &      22.4 &        E4 &      15.56 \\
       NGC0495 &  10536 &   20.733 &   33.471 &  0.013516 &      69.9 &        E4 &      18.64 &      PGC040616 &   8128 &  186.491 &   10.053 &  0.003354 &      17.0 &        E3 &       5.16 \\
       NGC0499 &  10536 &   20.798 &   33.460 &  0.014726 &      66.8 &       Sbc &      18.64 &        NGC4470 &  12888 &  187.407 &    7.824 &  0.008145 &      18.8 &        E4 &     161.35 \\
       NGC0496 &  10536 &   20.798 &   33.529 &  0.020121 &      63.4 &        E2 &      18.64 &        NGC4479 &   8066 &  187.577 &   13.577 &  0.002864 &      18.3 &        E3 &       5.16 \\
       NGC0504 &    317 &   20.866 &   33.204 &  0.014030 &      64.7 &        Sb &      27.19 &       NGC4486 &   5827 &  187.633 &   12.490 &  0.005299 &      15.4 &        S0 &     158.27 \\
       NGC0507 &    317 &   20.916 &   33.256 &  0.016394 &      69.1 &        Sb &      27.19 &         IC3586 &   8083 &  189.228 &   12.520 &  0.005682 &      20.0 &        E7 &       5.16 \\
       NGC0508 &    317 &   20.919 &   33.280 &  0.018359 &      76.7 &        E6 &      27.19 &         IC3652 &   8079 &  190.244 &   11.184 &  0.001912 &      15.2 &        Sb &       5.16 \\
       NGC0548 &   7823 &   21.511 &   -1.226 &  0.017951 &      82.0 &        Sa &      65.68 &       NGC4676A &   2043 &  191.542 &   30.732 &  0.022241 &      94.5 &        Sa &      28.91 \\
       NGC0741 &  17198 &   29.088 &     5.629 &  0.018367 &      70.7 &        Sc &      92.62 &       NGC4676B &   2043 &   191.547 &    30.722 &  0.021610 &      94.4 &        Sc &      28.91 \\
       MCG-02-06-016 &   6106 &    30.229 &    -8.842 &  0.005946 &      23.0 &       Sbc &      35.77 &      PGC092948 &   8101 &   191.603 &    11.949 &  0.046083 &     186.3 &       S0a &       5.16 \\
       NGC0833$^{T2}$ &  15667 &    32.337 &   -10.133 &  0.012657 &      55.2 &       Scd &      59.11 &       NGC4841A &  20052 &   194.383 &    28.476 &  0.022594 &      90.5 &        E7 &      24.05 \\
       NGC0835 &  15667 &    32.352 &   -10.136 &  0.013331 &      34.0 &        Sa &      59.11 &        NGC4861 &  20992 &   194.751 &    34.845 &  0.001526 &       7.5 &         I &      59.23 \\
       PGC008502 &  18022 &    33.315 &    -7.663 &  0.001003 &      68.4 &       Sbc &      30.06 &        NGC4874 &  13996 &   194.899 &    27.959 &  0.023956 &      96.9 &        E7 &     124.68 \\
       NGC0890 &  19325 &    35.504 &    33.266 &  0.013076 &      37.0 &        Sc &      35.06 &        NGC5198 &   6786 &   202.547 &    46.670 &  0.008408 &      48.5 &        Sb &      14.97 \\
       UGC01859$^{T1}$ &  17064 &    36.185 &    42.623 &  0.019842 &     116.0 &        S0 &      10.04 &   NGC5216$^{T2}$ &  10568 &   203.029 &    62.700 &  0.009874 &      62.1 &       Sab &       5.47 \\
       IC0225 &  11351 &    36.617 &     1.160 &  0.005202 &      18.6 &        Sc &       7.56 &        NGC5218 &  10568 &   203.041 &    62.768 &  0.009717 &      51.4 &        Sa &       5.47 \\
       NGC0991 &   7861 &    38.886 &    -7.155 &  0.005150 &       8.8 &       Sbc &       5.11 &        NGC5358 &  14903 &   208.502 &    40.277 &  0.008112 &      34.5 &       Scd &      40.80 \\
       NGC1060 &  18713 &    40.813 &    32.424 &  0.017177 &      78.4 &       Sdm &      29.57 &        NGC5394 &  10395 &   209.640 &    37.453 &  0.011670 &      32.9 &        Sd &      16.08 \\
       NGC1132 &   3576 &    43.216 &    -1.275 &  0.023168 &      87.9 &       Sdm &      40.17 &        NGC5395 &  10395 &   209.658 &    37.424 &  0.011646 &      46.4 &        Sa &      16.08 \\
       NGC1129 &    908 &    43.614 &    41.579 &  0.017655 &      74.2 &        E4 &      48.46 &        NGC5426 &   4847 &   210.854 &    -6.069 &  0.008558 &      34.1 &       S0a &       9.74 \\
       PGC11179 &   4181 &    44.390 &     5.976 &  0.022694 &     110.5 &       Scd &      21.78 &        NGC5427 &   4847 &   210.858 &    -6.031 &  0.009128 &      33.8 &         I &       9.74 \\
       NGC1167 &  19313 &    45.426 &    35.205 &  0.016492 &      70.6 &       Sdm &      13.05 &        NGC5473 &  19322 &   211.180 &    54.892 &  0.006717 &      27.3 &        E1 &       9.95 \\
       NGC1259 &   9097 &    49.322 &    41.385 &  0.019364 &      71.7 &       Scd &      35.18 &        NGC5485 &  19375 &   211.797 &    55.001 &  0.006402 &      29.8 &        E4 &      10.07 \\
       NGC1270 &    502 &    49.742 &    41.470 &  0.016307 &      80.8 &         I &       5.38 &        NGC5532 &   3968 &   214.221 &    10.807 &  0.024809 &      69.5 &        E5 &      50.08 \\
       NGC1271 &  12037 &    49.797 &    41.353 &  0.019926 &      82.2 &        S0 &      85.76 &        NGC5546 &   7057 &   214.538 &     7.564 &  0.024562 &     104.0 &        Sa &       5.25 \\
       NGC1277 &   4952 &    49.965 &    41.573 &  0.016768 &      60.7 &       Sdm &     166.42 &        NGC5557 &  19324 &   214.607 &    36.493 &  0.010829 &      38.8 &        Sb &       8.95 \\
       NGC1281 &   4952 &    50.025 &    41.629 &  0.014107 &      93.3 &        Sb &     166.42 &        NGC5576 &  11781 &   215.266 &     3.271 &  0.005020 &      21.0 &        Sb &      30.05 \\
       PGC012562 &   4948 &    50.252 &    41.562 &  0.015724 &      68.9 &        E6 &     120.18 &        NGC5614 &  11679 &   216.032 &    34.858 &  0.012877 &      35.8 &       S0a &      14.75 \\
       UGC02698 &  17065 &    50.512 &    40.863 &  0.021366 &     111.1 &        E7 &       8.07 &        NGC5631 &  19376 &   216.639 &    56.582 &  0.006441 &      24.2 &        E5 &      10.07 \\
       NGC2315 &  12564 &   105.638 &    50.590 &  0.020354 &      89.9 &        E3 &      10.04 &        NGC5623 &   9895 &   216.786 &    33.252 &  0.011443 &      47.9 &        E1 &      31.03 \\
       UGC03816 &  16611 &   110.802 &    58.064 &  0.010948 &      61.6 &       Sab &      32.07 &        NGC5656 &  19673 &   217.606 &    35.321 &  0.010716 &      55.8 &       BCD &      22.80 \\
       UGC03995$^{T1}$ &  12869 &   116.038 &    29.247 &  0.015888 &      60.6 &        Sb &      10.96 &   NGC5675$^{T2}$ &   9135 &   218.166 &    36.302 &  0.013176 &      56.8 &        E5 &      36.22 \\
       NGC2445 &  14906 &   116.729 &    39.014 &  0.013291 &      62.3 &        Sb &      39.51 &        UGC9562 &  13930 &   222.810 &    35.542 &  0.004277 &      23.8 &        Sa &      31.04 \\
       NGC2484 &    858 &   119.617 &    37.786 &  0.040769 &     171.0 &       Sdm &       8.26 &        NGC5794 &  19531 &   223.973 &    49.726 &  0.013968 &      59.6 &       Sab &      34.60 \\
       UGC04132 &   7570 &   119.804 &    32.914 &  0.017605 &      75.7 &        S0 &      33.04 &        NGC5797 &  19531 &   224.100 &    49.696 &  0.013410 &      56.8 &       Sab &      34.60 \\
       NGC2513 &  19318 &   120.603 &     9.413 &  0.015655 &      70.3 &        E2 &      14.06 &        UGC9661 &  12952 &   225.515 &     1.841 &  0.004364 &      17.7 &        S0 &     144.90 \\
       NGC2553 &   7935 &   124.396 &    20.903 &  0.015502 &      67.4 &        E5 &      31.13 &        NGC5845 &   4009 &   226.503 &     1.633 &  0.004893 &      27.1 &        S0 &      30.79 \\
       NGC2558 &   7936 &   124.803 &    20.510 &  0.016720 &      81.3 &        E4 &      28.05 &        NGC5929$^{T1}$ &  20623 &   231.526 &    41.670 &  0.008364 &      38.5 &        E1 &      27.42 \\
      IC2341 &   7937 &   125.922 &    21.434 &  0.017060 &      74.8 &        Sc &      30.03 &        NGC5930 &  20623 &   231.533 &    41.676 &  0.008732 &      35.0 &        E4 &      27.42 \\
       UGC04414 &  10268 &   126.775 &    21.645 &  0.025220 &     112.0 &        E3 &      10.15 &        NGC5953 &   2930 &   233.635 &    15.193 &  0.007202 &      27.1 &        Sc &      10.16 \\
       NGC2595 &  10268 &   126.925 &    21.479 &  0.014301 &      68.6 &        Sc &      10.15 &        NGC5954 &   2930 &   233.645 &    15.200 &  0.006518 &      36.4 &        S0 &      10.16 \\
       UGC04461 &   1643 &   128.344 &    52.532 &  0.016729 &      69.8 &        E6 &       9.25 &         ARP220 &  16092 &   233.739 &    23.503 &  0.017995 &      77.6 &        E6 &     171.46 \\
       NGC2623 &   4059 &   129.600 &    25.754 &  0.018350 &      81.7 &        E4 &      20.81 &        NGC6027 &  11261 &   239.802 &    20.763 &  0.014831 &      68.8 &        E3 &      70.05 \\
       NGC2639$^{T2}$ &   5682 &   130.909 &    50.205 &  0.010686 &      47.7 &        E6 &       5.08 &       UGC10205 &  20434 &   241.668 &    30.099 &  0.021893 &     122.5 &        S0 &       7.06 \\
       NGC2780 &  11777 &   138.185 &    34.925 &  0.006737 &      55.5 &        E0 &      29.55 &        NGC6090 &   6859 &   242.920 &    52.457 &  0.030163 &     125.5 &        S0 &      14.98 \\
       NGC2748 &  11776 &   138.429 &    76.475 &  0.004926 &      19.2 &       Sdm &      30.05 &        NGC6125 &  10550 &   244.798 &    57.984 &  0.015764 &      69.0 &        Sb &      10.04 \\
       NGC2787 &   4689 &   139.827 &    69.203 &  0.002224 &      22.3 &       Sdm &      31.24 &   NGC6166N1 &    498 &   247.159 &    39.551 &  0.026819 &     132.9 &        E4 &      19.16 \\
       NGC2805 &  12984 &   140.086 &    64.103 &  0.005918 &      14.0 &       Sab &      10.06 &        NGC6251$^{T1}$ &   4130 &   248.133 &    82.537 &  0.024528 &      98.2 &        E3 &      49.17 \\
       NGC2906 &  19298 &   143.026 &     8.441 &  0.007231 &      38.8 &        E3 &      41.76 &     PGC2172338 &    887 &   249.372 &    40.880 &  0.026332 &     104.0 &       Scd &      74.32 \\
       UGC05187 &  19438 &   145.776 &    41.093 &  0.004929 &      20.9 &       BCD &      49.31 &        NGC6285 &  10566 &   254.599 &    58.955 &  0.018949 &      81.3 &        E3 &      14.19 \\
       MCG+08-19-17 &  19033 &   154.742 &    46.454 &  0.029742 &     128.3 &        E0 &      24.05 &        NGC6286 &  10566 &   254.630 &    58.936 &  0.018668 &      78.6 &        E6 &      14.19 \\
       NGC3353 &  13927 &   161.343 &    55.960 &  0.004087 &      18.9 &        Sb &      18.07 &        NGC6278 &   6789 &   255.209 &    23.011 &  0.009339 &      39.1 &        E1 &      16.68 \\
       IC2604 &   2042 &   162.354 &    32.772 &  0.005587 &      23.3 &       Sab &      19.76 &        NGC6338 &   4194 &   258.846 &    57.411 &  0.027541 &     128.8 &        S0 &      47.94 \\
       NGC3395 &   2042 &   162.458 &    32.983 &  0.005493 &      17.6 &        E5 &      19.76 &        NGC7236 &   6392 &   333.687 &    13.846 &  0.026167 &      91.2 &        S0 &      33.13 \\
       NGC3396 &   2042 &   162.480 &    32.990 &  0.005916 &      24.9 &       Scd &      19.76 &       UGC11958 &   6392 &   333.695 &    13.841 &  0.026178 &     112.6 &        E3 &      33.13 \\
       PGC32873 &  21377 &   164.067 &    42.333 &  0.024971 &     106.7 &        S0 &      58.07 &       UGC12127 &   2191 &   339.622 &    35.329 &  0.027611 &     118.2 &        E2 &      10.14 \\
PGC033423 &  12977 &   165.976 &    40.850 &  0.034718 &     147.9 &        E3 &      53.01 &        NGC7457 &  17007 &   345.249 &    30.144 &  0.002701 &      13.3 &        S0 &      45.27 \\
       NGC3600 &  19356 &   168.967 &    41.591 &  0.002462 &      16.3 &        E1 &       7.07 &         IC5309 &   2074 &   349.799 &     8.109 &  0.013793 &      53.0 &        E6 &      27.09 \\
       NGC3605 &   2073 &   169.194 &    18.017 &  0.002018 &      24.5 &        E1 &      39.00 &        NGC7611 &   3955 &   349.902 &     8.063 &  0.010849 &      42.6 &        S0 &      37.95 \\
       NGC3773 &  17071 &   174.554 &    12.112 &  0.003262 &      17.1 &        E0 &      10.10 &        NGC7619 &   3955 &   350.060 &     8.206 &  0.012598 &      53.8 &        S0 &      37.95 \\
       NGC3842 &   4189 &   176.009 &    19.949 &  0.020816 &      99.6 &       Sab &      48.11 &        NGC7623 &   2074 &   350.125 &     8.395 &  0.012278 &      51.0 &        E4 &      27.09 \\
       NGC3860 &    514 &   176.205 &    19.795 &  0.018738 &     101.3 &        S0 &      41.05 &        NGC7684 &   8616 &   352.633 &     0.081 &  0.017198 &      73.3 &       Sab &       8.95 \\
       NGC3861$^{T2}$ &    514 &   176.266 &    19.973 &  0.016911 &      85.3 &       Sbc &      41.05 &        NGC7716 &  11728 &   354.131 &     0.297 &  0.008635 &      32.4 &        Sb &      17.04 \\
       NGC3893 &  21091 &   177.159 &    48.710 &  0.003271 &      15.8 &        Sb &      10.07 &        NGC7052 &  19326 &   319.638 &    26.447 &  0.012000 &      51.5 &       Sb &      39.06 \\
\end{tabular}
\caption{Compiled sample of 138 sources with {\it Chandra} data available from the extended CALIFA sample. (1) Name of the source. T1 and T2 in this column are the Type-I and Type-II AGN in \citet{lacerda20} classified through optical diagnostics, which have available {\it Chandra} data. (2) Observation ID in the {\it Chandra} database. (3) Right ascension in degrees. (4) Declination in degrees. (5) redshift. (6) Distance in Mpc. (7) Hubble type. (8) Exposure time of the observation measured in kiloseconds.}
\label{tab:sample_online}
\end{table}
\newpage

\begin{table}
\begin{tabular}{p{1.3cm}rrrrrp{1.3cm}rrrrrr}
\toprule
Name &  Amp. &   b & $b/a$ &  Noise & Cont & Name &  Amp. &   b & $b/a$ &  Noise & Cont. \\
& (cts/s/px) &  ($\rm{arcsec})$ &  & (cts/s/px) & & (cts/s/px) &  ($\rm{arcsec})$ &  & (cts/s/px) \\
(1) & (2) & (3) & (4) &(5) & (6) & (1) & (2) & (3) & (4) &(5) & (6)  \\
\midrule \midrule
 
  NGC0023   &      0.7$\pm$0.1 &      0.4$\pm$0.1  &    1*            &    0.039 &  0.33        &       NGC4291   &      0.8$\pm$0.5 &      0.2*            &    1*            &    0.136 &  0.04      \\
  NGC0192   &      0.08$\pm$0.01 &      0.5*            &    1*            &    0.089 &  0.14    &      NGC4486B  &      0.3$\pm$0.1 &      0.3*            &    1*            &    0.076 &  0.29      \\
  NGC0214   &      1.7$\pm$0.3 &      0.3$\pm$0.2  &    1.1$\pm$0.2 &    0.009 &  0.04           &      NGC4676A  &      0.2$\pm$0.1 &      0.13$\pm$0.04 &    1.00$\pm$0.04 &    0.019 &  0.42        \\
  NGC0499   &      0.16$\pm$0.03 &      0.5$\pm$0.1  &    1.1$\pm$0.1 &    0.049 &  0.27         &      NGC4676B  &      1.5$\pm$0.5 &      0.2*            &    1*            &    0.046 &  0.14      \\
  NGC0507   &      0.5$\pm$0.6 &      0.2*            &    1*            &    0.058 &  0.27       &     NGC4874   &      1.3$\pm$0.2 &      0.2*            &    1*            &    0.095 &  0.17       \\
  NGC0508   &      0.07$\pm$0.01 &      0.4$\pm$0.3  &    0.9$\pm$0.3 &    0.01 &  0.032          &     NGC5198   &      0.02*         &      0.1*            &    1*            &    0.017 &  0.11    \\
  NGC0741   &      0.4$\pm$0.1 &      0.6*            &    1*            &    0.09 &  0.31       &      NGC5216   &      2.3$\pm$0.4 &      0.3$\pm$0.1  &    0.9$\pm$0.1 &    0.013 &  0.01          \\
  NGC0833   &      0.8$\pm$0.5 &      0.1*            &    1*            &    0.094 & 0.27     &        NGC5218   &      0.04$\pm$0.01 &      0.51$\pm$0.01 &    1*            &    0.005 &  0.45      \\
  NGC0835   &      6.2$\pm$7.1 &      0.4$\pm$0.2  &    0.9$\pm$0.2 &    1.636 &  0.015           &     NGC5394   &      0.4$\pm$0.1 &      0.1*            &    1*            &    0.013 &  0.36      \\
  UGC01859  &      0.3$\pm$0.1 &      0.2*            &    1*            &    0.008 &  0.27      &      NGC5395   &      1.9$\pm$0.4 &      0.3$\pm$0.2  &    0.9$\pm$0.2 &    0.037 &  0.02           \\
  NGC1060   &      0.3$\pm$0.1 &      0.5*            &    1*            &    0.019 &  0.09       &     NGC5427   &      0.12$\pm$0.02 &      0.1*            &    1*            &    0.008 &  0.02    \\
  PGC11179  &      0.02$\pm$0.01 &      0.6$\pm$0.1  &    1.00$\pm$0.02 &    0.011 &  0.11       &      NGC5532   &      7.1$\pm$5.8 &      0.4$\pm$0.1  &    0.9$\pm$0.1 &    0.285 &  0.12           \\
  NGC1167   &      0.2$\pm$0.1 &      0.1*            &    1*            &    0.023 &  0.19      &      NGC5576   &      0.19$\pm$0.03 &      0.6*            &    1*            &    0.039 &  0.09    \\
  NGC1259   &      0.4$\pm$0.1 &      0.1*            &    1*            &    0.027 &  0.23      &      NGC5614   &      2.5$\pm$0.5 &      0.4$\pm$0.1  &    0.9$\pm$0.1 &    0.017 &  0.01           \\
  NGC1277   &      2.9$\pm$1.2 &      0.6*            &    1*            &    0.397 &  0.143      &     NGC5623   &      3.5$\pm$0.7 &      0.4$\pm$0.2  &    1.1$\pm$0.2 &    1.4 &  0.056            \\
  NGC2315   &      0.04$\pm$0.01 &      0.6$\pm$0.1  &    1.0$\pm$0.1 &    0.013 &  0.08&               NGC5675   &      0.05$\pm$0.01 &      0.5*            &    1*            &    0.01 &  0.39     \\
  UGC03816  &      0.1$\pm$0.1 &      0.4*1           &    1*            &    0.052 &  0.02      &      NGC5845   &      1.6$\pm$0.4 &      0.1*            &    1*            &    0.07 &  0.32       \\
  UGC03995  &      5.3$\pm$1.2 &      0.4$\pm$0.1  &    1.0$\pm$0.1 &    0.707 &  0.06          &       NGC5929   &      3.60$\pm$4.04 &      0.4$\pm$0.3  &    1.2$\pm$0.3 &    0.398 &  0.02        \\
  NGC2445   &      1.1$\pm$0.5 &      0.2*            &    1*            &    0.086 &  0.04      &      NGC5953   &      0.8$\pm$0.3 &      0.6$\pm$0.1  &    0.9$\pm$0.1 &    0.026 &  0.15           \\
  NGC2484   &      1.6$\pm$0.3 &      0.4$\pm$0.2  &    0.9$\pm$0.2 &    0.021 &  0.11           &      ARP220    &      4.7$\pm$2.9 &      0.050*          &    1*            &    0.151 &  0.36      \\
  UGC04132  &      0.01$\pm$0.01 &      0.5$\pm$0.2  &    1.0$\pm$0.2 &    0.009 &  0.77         &      NGC6090   &      0.2$\pm$0.1 &      0.47$\pm$0.01 &    1.00$\pm$0.01 &    0.034 &  0.22        \\
  NGC2513   &      0.06*         &      0.050$\pm$0.004 &    1*            &    0.018 &  >1  &          NGC6125   &      0.02*         &      0.48$\pm$0.01 &    1*            &    0.008 &  0.22      \\
  NGC2558   &      0.05$\pm$0.05 &      0.2*            &    1*            &    0.009 & >1    &         NGC6166N1 &      2.9$\pm$0.6 &      0.360$\pm$0.002 &    1*            &    0.284 &  0.38      \\
  NGC2623   &      0.3*          &      0.090*          &    1*            &    0.033 &  0.06    &      NGC6251   &      448$\pm$268 &      0.3$\pm$0.1  &    0.9$\pm$0.1 &    9.411 &  0.03          \\
  NGC2639   &      0.04$\pm$0.06 &      0.280*          &    1*            &    0.031 &  0.83    &      NGC6285   &      1.2$\pm$0.2 &      0.3$\pm$0.2  &    0.9$\pm$0.2 &    0.015 &  0.05          \\
  NGC2787   &      12.3$\pm$3.6 &      0.330$\pm$0.130 &    1.0$\pm$0.1 &    0.27 &  0.03        &      NGC6278   &      8.3$\pm$2.6 &      0.3$\pm$0.1  &    1.0$\pm$0.1 &    0.034 &  0.02           \\
  NGC2906   &      0.1$\pm$0.1 &      0.09$\pm$0.01 &    1.00$\pm$0.01 &    0.025 &  0.07       &       NGC6338   &      2.4$\pm$0.8 &      0.240*          &    1*            &    0.092 &  0.15       \\
  NGC3396   &      0.10$\pm$0.02 &      0.3*            &    1*            &    0.008 &  0.24    &      NGC7236   &      0.10$\pm$0.05 &      0.430*          &    1*            &    0.018 &  0.26    \\
  PGC32873  &      1.3$\pm$0.2 &      0.1*            &    1*            &    0.067 &  0.14      &      UGC11958  &      1.4$\pm$0.6 &      0.170*          &    1*            &    0.027 &  >1       \\
  PGC033423 &      0.3$\pm$0.1 &      0.2*            &    1*            &    0.064 &  0.38      &      UGC12127  &      0.6$\pm$0.1 &      0.130*          &    1*            &    0.016 &  0.37\\
  NGC3842   &      0.3$\pm$0.2 &      0.15*           &    1*            &    0.051 &  >1      &        NGC7457   &      3.6$\pm$1.1 &      0.4$\pm$0.05 &    1.0$\pm$0.1 &    0.051 &  0.16          \\
  NGC3860   &      3.8$\pm$1.2 &      0.240*          &    1*            &    0.144 &  0.18      &      NGC7619   &      0.80$\pm$0.04 &      0.050$\pm$0.002 &    1.*           &    0.042 &  0.47    \\
  NGC3945   &      15.1$\pm$7.6 &      0.1*            &    1*            &    0.079 &  0.02     &      NGC7052   &      3.1$\pm$1.1 &      0.1*            &    1*            &    0.095 &  0.14      \\

\end{tabular}
\caption{Best-fit values of the Gaussian fit for the nuclear source. Col. (1) is the name of the source, (2) is the amplitude of the nuclear Gaussian, (3) is the semi-major axis of the Gaussian, in arcsec. (4) is the ratio between the semi-major and semi-minor axes, (5) is the noise and (6) is the contamination from the extended emission. Values with an $*$ symbol have errors below $0.005$, while for values without the symbol and error, these are well below the significance of the main value. }
\label{tab:Gaussian_fits_online}
\end{table}

\begin{table}
\begin{tabular}{l|ll|ll|lll|lll}
\toprule
 Name &   \multicolumn{2}{c}{PL ($\rm{M_1}$)} &  \multicolumn{2}{c}{APEC ($\rm{M_2}$)} &  \multicolumn{3}{c}{2PL ($\rm{M_3}$)} &  \multicolumn{3}{c}{A+PL ($\rm{M_4}$)}    \\

\midrule

         & $\rm{C/dof}$ & $\rm{BIC}$ & $\rm{C/dof}$ & $\rm{BIC}$ &  $\rm{C/dof}$ & $\rm{BIC}$ &  $\rm{f_{1}}$ & $\rm{C/dof}$ & $\rm{BIC}$ &  $\rm{f_{1}}$/$\rm{f_{2}}$  \\
         (1) & (2) & (3) & (4) & (5) & (6) &(7) & (8) & (9) & (10) & (11)   \\
\toprule

  NGC0023   &   95.42/46             &    83.75  & 86.01/47           &   78.23   &  95.43/45              &    79.86    &   x              &  49.21/44$\clubsuit$    &    29.75   &      $\checkmark$/$\checkmark$    \\
  NGC0192   &   65.49/44             &    53.94  & 90.86/45           &   83.16   &  44.01/43$\clubsuit$     &    28.61  &   $\checkmark$   &  64.21/42               &   44.96    &      x/$\checkmark$              \\
  NGC0499   &   58.09/44             &    46.54  & 55.41/45           &   47.71   &  58.10/43              &    42.7     &   x              &  45.83/42$\clubsuit$    &    26.57   &      $\checkmark$/$\checkmark$    \\
  NGC0507   &   99.51/45             &    87.9   & 30.08/46$\clubsuit$  &   22.33  &   99.51/44              &    84.03  &   x             &   30.08/43             &     10.72    &      $\checkmark$/x             \\
  NGC0741   &   310.20/50            &    298.29 & 91.20/51           &   83.25   &  305.01/49             &    289.12   &   $\checkmark$    &   53.22/48$\clubsuit$    &    33.37 &      $\checkmark$/$\checkmark$   \\
  NGC0833   &   315.49/52            &    303.47 & 889.70/53          &   881.68  &  58.07/51$\clubsuit$     &    42.04  &   $\checkmark$   &   80.84/50              &    60.8    &      $\checkmark$/$\checkmark$   \\
  NGC0835   &   989.69/60            &    977.26 & 3023.78/61         &   3015.5  &  144.52/59             &    127.94   &   $\checkmark$    &   87.55/58$\clubsuit$    &    66.84 &      $\checkmark$/$\checkmark$   \\
  NGC1060   &   76.47/39             &    65.26  & 62.78/40           &   55.31   &  75.80/38              &    60.85    &   x              &  41.57/37$\clubsuit$    &    22.88   &      $\checkmark$/$\checkmark$    \\
  NGC1167   &   37.94/44$\clubsuit$    &    26.39 &  50.60/45           &   42.9   &   35.71/43              &    20.31  &   x             &   35.78/42             &     16.53    &      x/$\checkmark$             \\
  NGC1277   &   63.91/56             &    51.68  & 95.28/57           &   87.12   &  63.91/55              &    47.6     &   x              &  34.83/54$\clubsuit$    &    14.44   &      $\checkmark$/$\checkmark$    \\
  UGC03816  &   48.32/44             &    36.77  & 67.81/45           &   60.11   &  45.23/43$\clubsuit$     &    29.83  &   $\checkmark$   &   44.27/42              &    25.02   &      x/$\checkmark$             \\
  UGC03995  &   194.53/47            &    182.79 & 389.03/48          &   381.2   &  84.57/46$\clubsuit$     &    68.92  &   $\checkmark$   &   116.05/45             &    96.49   &      $\checkmark$/$\checkmark$   \\
  NGC2445   &   38.06/44             &    26.51  & 51.29/45           &   43.59   &  38.06/43              &    22.66    &   x              &  29.45/42$\clubsuit$    &    10.2    &      $\checkmark$/$\checkmark$    \\
  NGC2484   &   30.46/44$\clubsuit$    &    18.91 &  55.55/45           &   47.85  &   30.10/43              &    14.7   &   x             &   29.80/42             &     10.55    &      x/$\checkmark$             \\
  NGC2623   &   62.07/43             &    50.58  & 125.72/44          &   118.06  &  59.25/42$\clubsuit$     &    43.94  &   $\checkmark$   &   61.24/41              &    42.1    &      x/$\checkmark$             \\
  NGC2639   &   66.85/42             &    55.43  & 37.91/43$\clubsuit$  &   30.29  &   63.64/41              &    48.41  &   $\checkmark$   &   33.83/40              &    14.8    &      $\checkmark$/x             \\
  NGC2787   &   53.80/50$\clubsuit$    &    41.89 &  256.35/51          &   248.41 &   53.80/49              &    37.92  &   x             &   51.21/48             &     31.36    &      x/$\checkmark$             \\
  PGC32873  &   65.35/45             &    53.73  & 76.16/46           &   68.42   &  61.08/44              &    45.6     &   $\checkmark$    &   54.10/43$\clubsuit$    &    34.74 &      $\checkmark$/$\checkmark$   \\
  PGC033423 &   56.55/46             &    44.88  & 125.55/47          &   117.77  &  52.88/45              &    37.31    &   $\checkmark$    &   42.97/44$\clubsuit$    &    23.51 &      $\checkmark$/$\checkmark$   \\
  NGC3842   &   195.95/48            &    184.16 & 55.61/49           &   47.75   &  195.95/47             &    180.22   &   x              &  40.69/46$\clubsuit$    &    21.03   &      $\checkmark$/$\checkmark$    \\
  NGC3860   &   39.04/47$\clubsuit$    &    27.3  &  235.58/48          &   227.76 &   37.08/46              &    21.43  &   x             &   37.18/45             &     17.62    &      x/$\checkmark$             \\
  NGC3945   &   67.62/49$\clubsuit$    &    55.77 &  208.02/50          &   200.11 &   67.62/48              &    51.81  &   x             &   62.41/47             &     42.65    &      x/$\checkmark$             \\
  NGC4291   &   43.56/45             &    31.95  & 87.47/46           &   79.73   &  40.65/44              &    25.17    &   $\checkmark$    &   34.58/43$\clubsuit$    &    15.22 &      $\checkmark$/$\checkmark$   \\
  NGC4486B  &   45.06/44$\clubsuit$    &    33.51 &  104.90/45          &   97.2   &   44.10/43              &    28.7   &   x             &   43.83/42             &     24.58    &      x/$\checkmark$             \\
  NGC4676B  &   36.80/44$\clubsuit$    &    25.25 &  73.43/45           &   65.73  &   36.80/43              &    21.4   &   x             &   35.36/42             &     16.11    &      x/$\checkmark$             \\
  NGC4874   &   198.99/46            &    187.31 & 91.03/47           &   83.25   &  198.99/45             &    183.42   &   x              &  74.30/44$\clubsuit$    &    54.84   &      $\checkmark$/$\checkmark$    \\
  NGC5216   &   41.56/42$\clubsuit$    &    30.14 &  45.09/43           &   37.48  &   41.56/41              &    26.33  &   x             &   37.23/40             &     18.19    &      x/x                       \\
  NGC5395   &   68.04/43$\clubsuit$    &    56.56 &  98.60/44           &   90.94  &   66.12/42              &    50.8   &   x             &   66.72/41             &     47.58    &      x/$\checkmark$             \\
  NGC5532   &   356.28/56            &    344.04 & 1125.79/57         &   1117.64 &  307.87/55             &    291.56   &   $\checkmark$    &   67.15/54$\clubsuit$    &    46.76 &      $\checkmark$/$\checkmark$   \\
  NGC5576   &   37.07/43$\clubsuit$    &    25.58 &  54.06/44           &   46.4   &   36.81/42              &    21.49  &   x             &   34.89/41             &     15.75    &      x/$\checkmark$             \\
  NGC5614   &   44.18/46$\clubsuit$    &    32.51 &  92.31/47           &   84.53  &   43.09/45              &    27.52  &   x             &   42.33/44             &     22.88    &      x/$\checkmark$             \\
  NGC5623   &   50.09/47$\clubsuit$    &    38.35 &  104.33/48          &   96.5   &   50.09/46              &    34.44  &   x             &   44.71/45             &     25.15    &      x/$\checkmark$             \\
  NGC5675   &   51.58/50             &    39.67  & 155.42/51          &   147.48  &  43.78/49$\clubsuit$     &    27.9   &   $\checkmark$    &  42.34/48$\clubsuit$     &   22.49  &      $\checkmark$/$\checkmark$    \\
  NGC5845   &   60.10/45$\clubsuit$    &    48.48 &  100.67/46          &   92.93  &   60.10/44              &    44.61  &   x             &   58.67/43             &     39.32    &      x/$\checkmark$             \\
  NGC5929   &   289.27/56            &    277.04 & 1392.54/57         &   1384.38 &  94.74/55$\clubsuit$     &    78.43  &   $\checkmark$   &   122.04/54             &    101.65  &      $\checkmark$/$\checkmark$   \\
  NGC5953   &   46.39/45             &    34.78  & 41.99/46$\clubsuit$  &   34.25  &   46.39/44              &    30.91  &   x             &   41.99/43             &     22.64    &      x/x                       \\
  ARP220    &   108.20/59            &    95.82  & 1369.20/60         &   1360.94 &  102.96/58             &    86.45    &   $\checkmark$    &   94.10/57$\clubsuit$    &    73.47 &      $\checkmark$/$\checkmark$   \\
  NGC6090   &   64.94/46             &    53.27  & 81.69/47           &   73.91   &  64.95/45              &    49.38    &   x              &  43.80/44$\clubsuit$    &    24.34   &      $\checkmark$/$\checkmark$    \\
  NGC6125   &   59.95/44             &    48.4   & 80.94/49           &   88.64   &  59.94/43              &    44.54    &   x              &  27.37/42$\clubsuit$    &    8.12    &      $\checkmark$/$\checkmark$    \\
  NGC6166N1 &   56.02/48             &    44.22  & 27.37/45$\clubsuit$  &   3.78   &   56.02/47              &    40.29  &   x             &   40.08/46             &     20.42    &      $\checkmark$/x             \\
  NGC6251   &   137.47/77            &    124.32 & 23833.12/78        &  23824.36&  137.35/76              &    119.83   &   x              &  107.38/75$\clubsuit$   &     85.47  &      $\checkmark$/$\checkmark$   \\
  NGC6278   &   77.82/48$\clubsuit$    &    66.02 &  227.09/49          &   219.23 &   76.91/47              &    61.19  &   x             &   75.62/46             &     55.96    &      x/$\checkmark$             \\
  NGC6338   &   197.92/38            &    186.78 & 80.60/39           &   73.17   &  197.94/37             &    183.09   &   x              &  55.37/36$\clubsuit$    &    36.8    &      $\checkmark$/$\checkmark$    \\
  UGC11958  &   48.76/43             &    37.27  & 227.85/44          &   220.2   &  44.43/42$\clubsuit$     &    29.11  &   $\checkmark$   &   k 45.60/41            &      26.45 &      x/$\checkmark$           \\
  UGC12127  &   60.02/44             &    48.47  & 23.93/45$\clubsuit$  &   16.23  &   60.02/43              &    44.62  &   x             &   23.93/42             &     4.68     &      $\checkmark$/x             \\
  NGC7457   &   35.48/47$\clubsuit$    &    23.75 &  97.83/48           &   90.0   &   35.48/46              &    19.83  &   x             &   35.48/45             &     15.92    &      x/$\checkmark$             \\
  NGC7619   &   283.66/49            &    271.81 & 48.94/50$\clubsuit$  &   41.04  &   283.65/48             &    267.85 &   x             &   48.94/47             &     29.19    &      $\checkmark$/x             \\
  NGC7052   &   66.85/47             &    55.12  & 146.06/48          &   138.24  &  63.54/45              &    47.89    &   $\checkmark$    &   52.51/45$\clubsuit$    &    32.95 &      $\checkmark$/$\checkmark$   \\

 \end{tabular}
\caption{Statistical results of the spectral fits for the nuclear extraction. Column (1) is the name of the source. Columns (2), (4), (6) and (9) are the $\rm{C/dof}$ all the models used in the analysis respectively, while Columns (3), (5), (7) and (10) are the Bayesian Information Criterion (BIC) values for each of the models, and Columns (8) and (11) are the f-tests values for the comparison between models $\rm{M_1-M_3}$ and $\rm{M_1-M_4}$/$\rm{M_2-M_4}$, respectively. The club-suit ($\clubsuit$) symbol in each of the $\rm{C/dof}$ columns, represents the preferred model (models) for each source.  }

\label{tab:spectral-fits_online}
\end{table}

\begin{landscape}
\tiny
\begin{longtable}{l|rrrrr|rrrrr|rrrrr}
    \caption{Emission line flux ratios. Col. (1) is the name of the source. Cols. (2), (7) and (12) is the ratio between \textbf{[$\ion{N}{II}$]} and H${\alpha}$ for central CALIFA, $R_e$ and {\it complete galaxy}, respectively. Cols. (3), (8) and (13) is the ratio between \textbf{[$\ion{O}{III}$]} and H${\beta}$ for the {\it center}, $R_e$ and {\it complete galaxy}, respectively. Cols. (4), (9) and (14) is the ratio between \textbf{[$\ion{S}{II}$]} and H${\alpha}$ for the {\it center}, $R_e$ and {\it complete galaxy}, respectively. Cols (5), (10) and (15) is the ratio between \textbf{[$\ion{O}{I}$]} and H${\alpha}$ for the {\it center}, $R_e$ and {\it complete galaxy}, while Cols. (6), (11) and (16) is the $\rm{EW(H\alpha)}$ for {\it center}, $R_e$ and {\it complete galaxy}, respectively. All measurements are in logarithmic scale. Upper limits correspond to the 3$\sigma$ limit. }\\
    \label{tab:flux_online}


%
Name & \multicolumn{5}{c}{{\it Center} } & \multicolumn{5}{c}{$R_e$} & \multicolumn{5}{c}{$All$}  \\
\hline
& \textbf{[$\ion{N}{II}$]/H${\alpha}$} & \textbf{[$\ion{O}{III}$]/H${\beta}$} & \textbf{[$\ion{S}{II}$]}/H${\alpha}$ & \textbf{[$\ion{O}{I}$]/H${\alpha}$} & EW(H${\alpha}$) &  \textbf{[$\ion{N}{II}$]/H${\alpha}$} & \textbf{[$\ion{O}{III}$]/H${\beta}$} & \textbf{[$\ion{S}{II}$]}/H${\alpha}$ & \textbf{[$\ion{O}{I}$]/H${\alpha}$} & EW(H${\alpha}$) & \textbf{[$\ion{N}{II}$]/H${\alpha}$} & \textbf{[$\ion{O}{III}$]/H${\beta}$} & \textbf{[$\ion{S}{II}$]}/H${\alpha}$ & \textbf{[$\ion{O}{I}$]/H${\alpha}$} & EW(H${\alpha}$)  \\
(1) & (2) & (3) & (4) & (5) & (6) & (7) & (8) & (9) & (10) & (11) & (12) & (13) & (14) & (15)  \\
\hline \hline

NGC7803 & $-0.311\pm0.047$ & $-0.349\pm0.137$ & $-0.375\pm0.047$ & $-2.303\pm0.164$ & $1.181\pm0.829$ & $-0.347\pm0.028$ & $-0.468\pm-0.468$ & $-0.404\pm0.049$ & $-2.265\pm0.217$ & $1.486\pm1.311$ & $-0.323\pm0.049$ & $-0.380\pm0.163$ & $-0.386\pm0.093$ & $-1.946\pm0.459$ & $1.397\pm1.306$ \\
NGC0023 & $-0.262\pm0.028$ & $-0.453\pm0.062$ & $-0.432\pm0.047$ & $-1.844\pm0.140$ & $1.610\pm0.964$ & $-0.250\pm0.045$ & $-0.346\pm-0.346$ & $-0.356\pm0.081$ & $-1.593\pm0.253$ & $1.365\pm1.093$ & $-0.260\pm0.065$ & $-0.294\pm0.178$ & $-0.350\pm0.097$ & $-1.506\pm0.302$ & $1.315\pm1.237$ \\
NGC0192 & $-0.291\pm0.040$ & $0.027\pm0.241$ & $-0.496\pm0.044$ & $-1.802\pm0.139$ & $1.289\pm0.477$ & $-0.183\pm0.086$ & $0.136\pm0.136$ & $-0.319\pm0.139$ & $-1.151\pm0.245$ & $0.581\pm0.269$ & $-0.278\pm0.133$ & $-0.114\pm0.405$ & $-0.390\pm0.173$ & $-1.199\pm0.424$ & $0.820\pm0.754$ \\
NGC0197 & $-0.413\pm0.115$ & $-0.181\pm0.269$ & $-0.438\pm0.068$ & $<-1.240$ & $0.995\pm0.779$ & $-0.559\pm0.239$ & $-0.047\pm-0.047$ & $-0.197\pm0.225$ & $-0.447\pm0.331$ & $0.808\pm0.852$ & $-0.529\pm0.219$ & $-0.100\pm0.404$ & $-0.222\pm0.220$ & $-0.490\pm0.400$ & $0.502\pm0.837$ \\
NGC0214 & $0.047\pm0.061$ & $0.462\pm0.206$ & $-0.309\pm0.093$ & $-1.447\pm0.267$ & $0.547\pm-0.231$ & $-0.442\pm0.056$ & $-0.647\pm-0.647$ & $-0.698\pm0.123$ & $-1.697\pm0.295$ & $1.335\pm0.950$ & $-0.388\pm0.138$ & $-0.358\pm0.397$ & $-0.523\pm0.239$ & $-1.317\pm0.438$ & $1.224\pm1.027$ \\
NGC0384 & $-0.025\pm0.127$ & $0.150\pm0.281$ & $-0.429\pm0.177$ & $-1.141\pm0.288$ & $0.026\pm-0.567$ & $-0.363\pm0.226$ & $-0.185\pm-0.185$ & $-0.365\pm0.287$ & $-0.329\pm0.329$ & $-0.419\pm-0.525$ & $-0.302\pm0.304$ & $-0.122\pm0.344$ & $-0.394\pm0.245$ & $-0.489\pm0.353$ & $-0.296\pm-0.116$ \\
NGC0495 & $-0.401\pm0.202$ & $-0.390\pm0.504$ & $-0.732\pm0.061$ & $-0.677\pm0.327$ & $-0.352\pm-0.760$ & $-0.562\pm0.293$ & $-0.245\pm-0.245$ & $-0.504\pm0.134$ & $-0.731\pm0.255$ & $-0.383\pm-0.236$ & $-0.423\pm0.255$ & $-0.241\pm0.382$ & $-0.317\pm0.255$ & $-0.499\pm0.287$ & $-0.334\pm-0.119$ \\
NGC0499 & $0.016\pm0.101$ & $0.077\pm0.256$ & $-0.280\pm0.114$ & $-1.214\pm0.404$ & $-0.011\pm-0.648$ & $-0.143\pm0.226$ & $-0.211\pm-0.211$ & $-0.362\pm0.231$ & $-0.515\pm0.283$ & $-0.368\pm-0.671$ & $-0.200\pm0.273$ & $-0.131\pm0.352$ & $-0.380\pm0.234$ & $-0.560\pm0.343$ & $-0.331\pm-0.457$ \\
NGC0496 & $-0.467\pm0.015$ & $-0.781\pm0.139$ & $-0.569\pm0.025$ & $-1.900\pm0.194$ & $1.474\pm0.515$ & $-0.526\pm0.087$ & $-0.427\pm-0.427$ & $-0.432\pm0.149$ & $-1.462\pm0.399$ & $1.370\pm1.228$ & $-0.608\pm0.192$ & $-0.275\pm0.329$ & $-0.387\pm0.216$ & $-1.184\pm0.538$ & $1.262\pm1.128$ \\
NGC0504 & $-0.239\pm0.156$ & $0.174\pm0.153$ & $-0.488\pm0.065$ & $-0.353\pm0.185$ & $-0.598\pm-0.930$ & $-0.267\pm0.147$ & $-0.505\pm-0.505$ & $-0.138\pm0.222$ & $-0.402\pm0.225$ & $-0.429\pm-0.519$ & $-0.283\pm0.235$ & $-0.316\pm0.462$ & $-0.216\pm0.256$ & $-0.414\pm0.298$ & $-0.485\pm-0.496$ \\
NGC0507 & $-0.143\pm0.168$ & $-0.502\pm0.184$ & $-0.455\pm0.090$ & $-0.754\pm0.233$ & $-0.169\pm-0.331$ & $-0.250\pm0.201$ & $-0.400\pm-0.400$ & $<-0.253$ & $-0.636\pm0.241$ & $-0.309\pm-0.595$ & $-0.229\pm0.219$ & $-0.430\pm0.380$ & $-0.336\pm0.381$ & $-0.642\pm0.220$ & $-0.307\pm-0.586$ \\
NGC0508 & $-0.355\pm0.074$ & $-0.850\pm0.178$ & $-0.770\pm0.101$ & $<-1.597$ & $0.522\pm-0.269$ & $-0.417\pm0.076$ & $-0.691\pm-0.691$ & $-0.709\pm0.162$ & $-1.518\pm0.489$ & $0.478\pm0.022$ & $-0.359\pm0.160$ & $-0.518\pm0.350$ & $-0.650\pm0.214$ & $-1.051\pm0.495$ & $0.300\pm0.044$ \\
NGC0548 & $-0.651\pm0.225$ & $-0.013\pm0.502$ & $<-0.234$ & $-0.563\pm0.277$ & $-0.519\pm-0.441$ & $-0.380\pm0.366$ & $-0.217\pm-0.217$ & $0.009\pm0.447$ & $-0.289\pm0.283$ & $-0.416\pm-0.327$ & $-0.421\pm0.262$ & $-0.192\pm0.372$ & $-0.195\pm0.397$ & $-0.357\pm0.276$ & $-0.081\pm0.161$ \\
NGC0741 & $0.001\pm0.163$ & $0.165\pm0.475$ & $-0.620\pm0.164$ & $<-1.310$ & $-0.171\pm-0.742$ & $-0.445\pm0.282$ & $-0.219\pm-0.219$ & $-0.225\pm0.276$ & $-0.455\pm0.279$ & $-0.483\pm-0.500$ & $-0.336\pm0.276$ & $-0.106\pm0.345$ & $-0.255\pm0.335$ & $-0.450\pm0.299$ & $-0.341\pm0.549$ \\
MCG-02-06-016 & $-1.032\pm0.206$ & $0.290\pm0.256$ & $-0.396\pm0.137$ & $-0.964\pm0.321$ & $1.132\pm0.689$ & $-0.883\pm0.322$ & $0.044\pm0.044$ & $-0.455\pm0.204$ & $-1.148\pm0.499$ & $1.484\pm1.505$ & $-0.811\pm0.335$ & $0.102\pm0.333$ & $-0.426\pm0.257$ & $-0.968\pm0.473$ & $1.349\pm1.605$ \\
NGC0833 & $0.031\pm0.033$ & $0.299\pm0.117$ & $0.054\pm0.028$ & $-1.135\pm0.111$ & $0.832\pm0.233$ & $<0.004$ & $0.374\pm0.374$ & $<0.037$ & $<-1.186$ & $<0.966$ & $0.031\pm0.032$ & $0.308\pm0.087$ & $0.053\pm0.042$ & $-1.087\pm0.116$ & $0.765\pm0.367$ \\
NGC0835 & $-0.251\pm0.134$ & $-0.341\pm0.358$ & $-0.511\pm0.138$ & $-1.798\pm0.222$ & $1.249\pm1.004$ & $-0.335\pm0.049$ & $-0.576\pm-0.576$ & $-0.566\pm0.059$ & $-1.880\pm0.147$ & $1.449\pm0.957$ & $-0.284\pm0.084$ & $-0.476\pm0.201$ & $-0.478\pm0.128$ & $-1.744\pm0.290$ & $1.285\pm1.082$ \\
PGC008502 & $-0.117\pm0.115$ & $-0.616\pm0.472$ & $-0.450\pm0.291$ & $-1.457\pm0.287$ & $0.398\pm0.015$ & $-0.277\pm0.157$ & $-0.572\pm-0.572$ & $-0.355\pm0.220$ & $-1.043\pm0.389$ & $0.471\pm0.364$ & $-0.345\pm0.149$ & $-0.435\pm0.333$ & $-0.369\pm0.191$ & $-1.192\pm0.455$ & $0.921\pm0.972$ \\
NGC0890 & $-0.168\pm0.150$ & $-0.860\pm0.337$ & $<0.042$ & $-0.287\pm0.259$ & $-0.756\pm-0.834$ & $-0.275\pm0.233$ & $0.044\pm0.044$ & $-0.328\pm0.229$ & $-0.313\pm0.253$ & $-0.544\pm-0.619$ & $-0.268\pm0.246$ & $-0.007\pm0.383$ & $-0.299\pm0.238$ & $-0.368\pm0.263$ & $-0.497\pm-0.468$ \\
UGC01859 & $0.011\pm0.033$ & $0.387\pm0.174$ & $-0.048\pm0.048$ & $<-1.109$ & $0.650\pm0.172$ & $-0.182\pm0.164$ & $0.212\pm0.212$ & $-0.148\pm0.207$ & $-0.521\pm0.424$ & $0.026\pm0.048$ & $-0.207\pm0.215$ & $0.157\pm0.332$ & $-0.135\pm0.245$ & $-0.643\pm0.429$ & $0.144\pm0.240$ \\
IC0225 & $-0.694\pm0.040$ & $-0.008\pm0.040$ & $-0.533\pm0.103$ & $-2.341\pm0.329$ & $1.960\pm1.678$ & $-0.561\pm0.157$ & $-0.016\pm-0.016$ & $-0.366\pm0.117$ & $-1.202\pm0.327$ & $0.970\pm0.648$ & $-0.541\pm0.235$ & $-0.077\pm0.295$ & $-0.368\pm0.216$ & $-1.031\pm0.574$ & $0.994\pm1.284$ \\
NGC0991 & $-0.284\pm0.071$ & $-0.350\pm0.141$ & $-0.163\pm0.077$ & $-1.212\pm0.235$ & $0.414\pm-0.114$ & $-0.609\pm0.151$ & $-0.337\pm-0.337$ & $-0.287\pm0.180$ & $-1.296\pm0.463$ & $1.253\pm1.301$ & $-0.609\pm0.181$ & $-0.316\pm0.294$ & $-0.284\pm0.184$ & $-1.228\pm0.445$ & $1.188\pm1.434$ \\
NGC1060 & $-0.066\pm0.087$ & $-0.035\pm0.215$ & $<-0.889$ & $-1.006\pm0.223$ & $-0.046\pm-0.682$ & $-0.184\pm0.276$ & $-0.275\pm-0.275$ & $-0.613\pm0.133$ & $-0.578\pm0.252$ & $-0.277\pm-0.484$ & $-0.187\pm0.238$ & $-0.161\pm0.295$ & $-0.457\pm0.269$ & $-0.611\pm0.291$ & $-0.309\pm-0.296$ \\
NGC1132 & $-0.256\pm0.282$ & $-0.457\pm0.350$ & $0.699\pm0.120$ & $-0.680\pm0.282$ & $-0.310\pm-0.647$ & $-0.748\pm0.165$ & $-0.387\pm-0.387$ & $0.105\pm0.307$ & $-0.254\pm0.241$ & $-0.375\pm-0.340$ & $-0.426\pm0.274$ & $-0.325\pm0.380$ & $0.311\pm0.380$ & $-0.444\pm0.337$ & $-0.257\pm-0.070$ \\
NGC1129 & $-0.192\pm0.196$ & $0.012\pm0.481$ & $<-0.234$ & $<-1.310$ & $-0.424\pm-0.786$ & $-0.352\pm0.238$ & $-0.205\pm-0.205$ & $-0.346\pm0.272$ & $-0.451\pm0.232$ & $-0.544\pm-0.480$ & $-0.315\pm0.254$ & $-0.170\pm0.370$ & $-0.349\pm0.339$ & $-0.548\pm0.295$ & $-0.455\pm-0.225$ \\
PGC11179 & $<-0.183$ & $<0.029n$ & $<-0.234$ & $<-1.310$ & $-0.968\pm-0.270$ & $<-0.330$ & $-0.154\pm-0.154$ & $<-0.253$ & $<-0.922$ & $-0.557\pm-0.236$ & $<7.979$ & $<-0.126$ & $<-0.244$ & $<-0.880$ & $-0.028\pm0.792$ \\
NGC1167 & $0.089\pm0.028$ & $0.512\pm0.067$ & $0.086\pm0.031$ & $-1.595\pm0.188$ & $1.095\pm0.806$ & $0.004\pm0.170$ & $0.072\pm0.072$ & $-0.272\pm0.305$ & $-0.676\pm0.264$ & $0.042\pm-0.218$ & $-0.098\pm0.237$ & $0.038\pm0.334$ & $-0.190\pm0.315$ & $-0.675\pm0.365$ & $0.306\pm0.773$ \\
NGC1259 & $-0.156\pm0.133$ & $-0.409\pm0.394$ & $-0.657\pm0.137$ & $-1.096\pm0.420$ & $0.094\pm-0.229$ & $-0.429\pm0.196$ & $-0.346\pm-0.346$ & $-0.108\pm0.298$ & $-0.291\pm0.441$ & $-0.151\pm0.283$ & $-0.287\pm0.221$ & $-0.273\pm0.398$ & $-0.362\pm0.335$ & $-0.477\pm0.393$ & $0.392\pm1.200$ \\
NGC1270 & $<-0.183$ & $<0.029n$ & $<-0.234$ & $<-1.310$ & $<0.581$ & $<-0.330$ & $-0.154\pm-0.154$ & $<-0.253$ & $<-0.922$ & $<0.385$ & $<-0.335$ & $<-0.126$ & $<-0.244$ & $<-0.880$ & $-3.499\pm-1.983$ \\
NGC1271 & $<-0.183$ & $<0.029n$ & $<-0.234$ & $<-1.310$ & $<0.581$ & $<-0.330$ & $-0.154\pm-0.154$ & $<-0.253$ & $<-0.922$ & $-1.549\pm-0.765$ & $<-0.335$ & $<-0.126$ & $<-0.244$ & $<-0.880$ & $-0.731\pm0.687$ \\
NGC1277 & $<-0.183$ & $<0.029n$ & $<-0.234$ & $<-1.310$ & $<0.581$ & $<-0.330$ & $-0.154\pm-0.154$ & $<-0.253$ & $<-0.922$ & $<0.385$ & $<-0.618$ & $<-0.126$ & $<-0.244$ & $<-0.880$ & $-0.547\pm0.517$ \\
NGC1281 & $<-0.183$ & $<0.029n$ & $<-0.234$ & $<-1.310$ & $<0.581$ & $<-0.330$ & $-0.154\pm-0.154$ & $<-0.253$ & $<-0.922$ & $<0.385$ & $<-0.335$ & $<-0.126$ & $<-0.244$ & $<-0.880$ & $<0.435$ \\
PGC012562 & $<-0.183$ & $<0.029n$ & $<-0.234$ & $<-1.310$ & $0.581$ & $<-0.330$ & $-0.154\pm-0.154$ & $<-0.253$ & $<-0.922$ & $<0.385$ & $<-0.335$ & $<-0.126$ & $<-0.244$ & $<-0.880$ & $<0.435$ \\
UGC02698 & $0.005\pm0.070$ & $0.115\pm0.372$ & $-0.417\pm0.295$ & $<-1.310$ & $-0.004\pm-0.639$ & $-0.396\pm0.288$ & $-0.305\pm-0.305$ & $-0.278\pm0.238$ & $-0.516\pm0.255$ & $0.135\pm0.018$ & $-0.338\pm0.290$ & $-0.292\pm0.369$ & $-0.240\pm0.295$ & $-0.510\pm0.285$ & $0.404\pm0.997$ \\
NGC2315 & $-0.157\pm0.079$ & $-0.290\pm0.165$ & $-0.525\pm0.066$ & $<-1.310$ & $0.640\pm-0.041$ & $-0.318\pm0.129$ & $-0.514\pm-0.514$ & $-0.505\pm0.231$ & $-1.201\pm0.374$ & $0.521\pm0.304$ & $-0.309\pm0.203$ & $-0.405\pm0.354$ & $-0.416\pm0.286$ & $-0.856\pm0.476$ & $0.229\pm0.279$ \\
UGC03816 & $0.009\pm0.079$ & $0.130\pm0.133$ & $-0.363\pm0.161$ & $-1.050\pm0.272$ & $0.314\pm-0.176$ & $-0.139\pm0.182$ & $0.145\pm0.145$ & $-0.323\pm0.215$ & $-0.548\pm0.242$ & $-0.049\pm-0.430$ & $-0.279\pm0.280$ & $0.129\pm0.292$ & $-0.357\pm0.252$ & $-0.661\pm0.288$ & $-0.041\pm-0.176$ \\
UGC03995 & $-0.043\pm0.053$ & $0.883\pm0.047$ & $-0.231\pm0.051$ & $-1.685\pm0.204$ & $0.962\pm0.456$ & $-0.275\pm0.173$ & $-0.051\pm-0.051$ & $-0.402\pm0.232$ & $-1.163\pm0.444$ & $0.777\pm0.825$ & $-0.346\pm0.192$ & $-0.165\pm0.461$ & $-0.376\pm0.209$ & $-1.147\pm0.415$ & $0.941\pm1.002$ \\
NGC2445 & $-0.339\pm0.024$ & $-0.670\pm0.120$ & $-0.601\pm0.051$ & $-2.295\pm0.294$ & $1.729\pm1.213$ & $-0.343\pm0.189$ & $-0.055\pm-0.055$ & $-0.146\pm0.188$ & $-1.069\pm0.355$ & $1.108\pm1.194$ & $-0.401\pm0.219$ & $-0.091\pm0.290$ & $-0.174\pm0.210$ & $-1.242\pm0.469$ & $1.430\pm1.711$ \\
NGC2484 & $0.139\pm0.042$ & $0.346\pm0.197$ & $-0.235\pm0.128$ & $-1.338\pm0.544$ & $0.298\pm0.079$ & $-0.468\pm0.179$ & $-0.517\pm-0.517$ & $-0.386\pm0.185$ & $-0.563\pm0.335$ & $-0.333\pm-0.186$ & $-0.216\pm0.412$ & $0.077\pm0.402$ & $-0.293\pm0.175$ & $-0.642\pm0.447$ & $-0.201\pm0.107$ \\
UGC04132 & $-0.403\pm0.031$ & $-0.496\pm0.092$ & $-0.617\pm0.047$ & $-1.892\pm0.253$ & $1.244\pm0.391$ & $-0.385\pm0.047$ & $-0.507\pm-0.507$ & $-0.503\pm0.087$ & $-1.886\pm0.263$ & $1.372\pm0.927$ & $-0.403\pm0.059$ & $-0.290\pm0.225$ & $-0.441\pm0.111$ & $-1.799\pm0.341$ & $1.442\pm1.259$ \\
NGC2513 & $0.018\pm0.053$ & $-0.132\pm0.435$ & $<-0.234$ & $-0.996\pm0.218$ & $-0.175\pm-0.827$ & $-0.392\pm0.290$ & $-0.269\pm-0.269$ & $-0.255\pm0.261$ & $-0.520\pm0.285$ & $-0.441\pm-0.514$ & $-0.248\pm0.277$ & $-0.278\pm0.361$ & $-0.358\pm0.280$ & $-0.544\pm0.294$ & $-0.418\pm-0.344$ \\
NGC2553 & $-0.767\pm0.273$ & $-0.300\pm0.474$ & $-0.347\pm0.189$ & $-0.464\pm0.286$ & $-0.612\pm-0.859$ & $-0.430\pm0.315$ & $-0.400\pm-0.400$ & $-0.292\pm0.199$ & $-0.122\pm0.192$ & $-0.706\pm-0.686$ & $-0.533\pm0.328$ & $-0.279\pm0.454$ & $-0.332\pm0.211$ & $-0.342\pm0.309$ & $-0.479\pm-0.363$ \\
NGC2558 & $0.129\pm0.045$ & $0.234\pm0.220$ & $-0.017\pm0.071$ & $-1.485\pm0.243$ & $0.321\pm-0.382$ & $-0.347\pm0.194$ & $-0.234\pm-0.234$ & $-0.475\pm0.265$ & $-0.909\pm0.325$ & $0.775\pm0.532$ & $-0.343\pm0.220$ & $-0.250\pm0.348$ & $-0.441\pm0.258$ & $-0.882\pm0.378$ & $0.926\pm1.048$ \\
IC2341 & $0.087\pm0.041$ & $0.210\pm0.278$ & $0.059\pm0.198$ & $-1.065\pm0.203$ & $0.081\pm-0.597$ & $-0.052\pm0.167$ & $0.009\pm0.009$ & $-0.162\pm0.257$ & $-0.895\pm0.250$ & $-0.014\pm-0.631$ & $-0.115\pm0.239$ & $-0.071\pm0.348$ & $-0.194\pm0.275$ & $-0.762\pm0.289$ & $0.013\pm-0.314$ \\
UGC04414 & $-0.012\pm0.063$ & $0.194\pm0.377$ & $0.307\pm0.083$ & $-1.044\pm0.265$ & $0.080\pm-0.650$ & $-0.283\pm0.252$ & $-0.140\pm-0.140$ & $0.103\pm0.203$ & $-0.543\pm0.294$ & $-0.018\pm-0.057$ & $-0.372\pm0.239$ & $-0.204\pm0.341$ & $-0.137\pm0.299$ & $-0.806\pm0.426$ & $0.638\pm0.840$ \\
NGC2595 & $-0.377\pm0.090$ & $-0.578\pm0.208$ & $-0.559\pm0.101$ & $-2.045\pm0.181$ & $1.157\pm0.708$ & $-0.343\pm0.197$ & $-0.303\pm-0.303$ & $-0.290\pm0.232$ & $-0.814\pm0.423$ & $0.601\pm0.555$ & $-0.390\pm0.173$ & $-0.333\pm0.320$ & $-0.364\pm0.235$ & $-1.095\pm0.521$ & $0.982\pm1.066$ \\
UGC04461 & $-0.428\pm0.011$ & $-0.553\pm0.048$ & $-0.505\pm0.031$ & $-1.964\pm0.202$ & $1.459\pm0.472$ & $-0.507\pm0.058$ & $-0.191\pm-0.191$ & $-0.388\pm0.120$ & $-1.682\pm0.288$ & $1.439\pm1.197$ & $-0.671\pm0.209$ & $0.019\pm0.272$ & $-0.445\pm0.156$ & $-1.552\pm0.399$ & $1.505\pm1.474$ \\
NGC2623 & $-0.079\pm0.039$ & $0.338\pm0.219$ & $-0.234\pm0.079$ & $-1.361\pm0.177$ & $1.112\pm0.896$ & $-0.101\pm0.163$ & $0.063\pm0.063$ & $-0.185\pm0.220$ & $-0.794\pm0.335$ & $0.473\pm0.261$ & $-0.195\pm0.221$ & $-0.006\pm0.359$ & $-0.219\pm0.270$ & $-0.897\pm0.417$ & $0.579\pm0.673$ \\
NGC2639 & $0.096\pm0.027$ & $0.721\pm0.255$ & $-0.033\pm0.037$ & $-1.645\pm0.184$ & $1.047\pm0.702$ & $-0.314\pm0.105$ & $-0.233\pm-0.233$ & $-0.504\pm0.119$ & $-1.491\pm0.265$ & $0.678\pm0.114$ & $-0.338\pm0.155$ & $-0.339\pm0.380$ & $-0.438\pm0.174$ & $-1.388\pm0.320$ & $0.757\pm0.430$ \\
NGC2780 & $-0.506\pm0.039$ & $-0.786\pm0.120$ & $-0.540\pm0.049$ & $-2.025\pm0.209$ & $1.478\pm1.073$ & $-0.480\pm0.056$ & $-0.823\pm-0.823$ & $-0.482\pm0.097$ & $-1.600\pm0.312$ & $1.143\pm0.720$ & $-0.453\pm0.141$ & $-0.583\pm0.354$ & $-0.387\pm0.183$ & $-1.280\pm0.524$ & $0.971\pm0.813$ \\
NGC2748 & $-0.458\pm0.019$ & $-0.387\pm0.103$ & $-0.493\pm0.052$ & $-1.967\pm0.288$ & $1.254\pm0.619$ & $-0.475\pm0.026$ & $-0.398\pm-0.398$ & $-0.434\pm0.059$ & $-1.877\pm0.334$ & $1.457\pm0.959$ & $-0.519\pm0.091$ & $-0.205\pm0.233$ & $-0.391\pm0.103$ & $-1.730\pm0.373$ & $1.455\pm1.191$ \\
NGC2787 & $-0.030\pm0.046$ & $0.201\pm0.051$ & $0.025\pm0.044$ & $-1.607\pm0.240$ & $0.406\pm-0.081$ & $-0.398\pm0.207$ & $0.143\pm0.143$ & $-0.112\pm0.209$ & $-0.717\pm0.280$ & $-0.067\pm-0.523$ & $-0.426\pm0.264$ & $0.086\pm0.277$ & $-0.167\pm0.241$ & $-0.724\pm0.319$ & $-0.061\pm-0.325$ \\
NGC2805 & $-0.433\pm0.042$ & $-0.530\pm0.148$ & $-0.382\pm0.031$ & $-1.584\pm0.263$ & $1.019\pm0.468$ & $-0.474\pm0.228$ & $-0.436\pm-0.436$ & $-0.326\pm0.235$ & $-0.947\pm0.431$ & $0.906\pm1.104$ & $-0.516\pm0.204$ & $-0.440\pm0.374$ & $-0.384\pm0.199$ & $-1.173\pm0.500$ & $1.115\pm1.366$ \\
NGC2906 & $-0.039\pm0.041$ & $0.306\pm0.269$ & $-0.170\pm0.069$ & $-1.290\pm0.203$ & $0.232\pm-0.732$ & $-0.370\pm0.057$ & $-0.601\pm-0.601$ & $-0.482\pm0.091$ & $-1.680\pm0.285$ & $1.021\pm0.635$ & $-0.339\pm0.081$ & $-0.511\pm0.269$ & $-0.449\pm0.108$ & $-1.765\pm0.360$ & $1.275\pm1.184$ \\
UGC05187 & $-0.547\pm0.040$ & $-0.234\pm0.072$ & $-0.356\pm0.077$ & $-1.807\pm0.271$ & $1.268\pm0.924$ & $-0.655\pm0.104$ & $-0.113\pm-0.113$ & $-0.320\pm0.112$ & $-1.560\pm0.374$ & $1.401\pm1.270$ & $-0.653\pm0.206$ & $-0.090\pm0.278$ & $-0.323\pm0.178$ & $-1.228\pm0.519$ & $1.224\pm1.239$ \\
MCG+08-19-17 & $-0.525\pm0.044$ & $-0.183\pm0.048$ & $-0.437\pm0.060$ & $-2.008\pm0.276$ & $1.520\pm1.154$ & $-0.604\pm0.305$ & $0.071\pm0.071$ & $-0.308\pm0.255$ & $-0.753\pm0.348$ & $1.412\pm1.576$ & $-0.605\pm0.191$ & $0.029\pm0.237$ & $-0.389\pm0.204$ & $-1.316\pm0.623$ & $1.117\pm1.490$ \\
NGC3353 & $-0.980\pm0.049$ & $0.490\pm0.027$ & $-0.775\pm0.083$ & $-2.269\pm0.095$ & $2.406\pm2.086$ & $-0.935\pm0.072$ & $0.456\pm0.456$ & $-0.639\pm0.123$ & $-2.090\pm0.237$ & $2.100\pm2.053$ & $-0.901\pm0.075$ & $0.430\pm0.071$ & $-0.552\pm0.124$ & $-1.932\pm0.267$ & $1.968\pm1.946$ \\
IC2604 & $-0.743\pm0.049$ & $0.139\pm0.082$ & $-0.449\pm0.071$ & $-1.410\pm0.253$ & $1.437\pm1.211$ & $-0.973\pm0.191$ & $0.245\pm0.245$ & $-0.467\pm0.144$ & $-1.488\pm0.415$ & $1.737\pm1.636$ & $-0.918\pm0.235$ & $0.197\pm0.287$ & $-0.472\pm0.190$ & $-1.263\pm0.533$ & $1.594\pm1.751$ \\
NGC3395 & $-0.545\pm0.032$ & $-0.053\pm0.062$ & $-0.408\pm0.038$ & $-1.946\pm0.086$ & $1.574\pm0.883$ & $-0.675\pm0.077$ & $0.073\pm0.073$ & $-0.352\pm0.088$ & $-1.730\pm0.263$ & $1.765\pm1.809$ & $-0.695\pm0.122$ & $0.073\pm0.132$ & $-0.333\pm0.127$ & $-1.685\pm0.360$ & $1.724\pm1.762$ \\
NGC3396 & $-0.628\pm0.038$ & $0.266\pm0.056$ & $-0.630\pm0.084$ & $-2.123\pm0.170$ & $2.132\pm1.866$ & $-0.629\pm0.095$ & $0.042\pm0.042$ & $-0.276\pm0.113$ & $-1.485\pm0.350$ & $1.511\pm1.546$ & $-0.639\pm0.170$ & $0.060\pm0.227$ & $-0.284\pm0.182$ & $-1.395\pm0.518$ & $1.500\pm1.759$ \\
PGC32873 & $<-0.183$ & $<0.029n$ & $<-0.234$ & $<-1.310$ & $<0.581$ & $<-0.330$ & $-0.154\pm-0.154$ & $<-0.253$ & $<-0.922$ & $<0.385$ & $<-0.335$ & $<-0.126$ & $<-0.244$ & $<-0.880$ & $<0.435$ \\
PGC033423 & $-0.402\pm0.035$ & $-0.308\pm0.142$ & $-0.319\pm0.094$ & $-1.753\pm0.443$ & $1.145\pm0.879$ & $-0.366\pm0.090$ & $-0.185\pm-0.185$ & $-0.235\pm0.134$ & $-1.547\pm0.325$ & $1.391\pm1.243$ & $-0.359\pm0.153$ & $-0.119\pm0.237$ & $-0.202\pm0.164$ & $-1.366\pm0.446$ & $1.575\pm1.502$ \\
NGC3600 & $-0.861\pm0.026$ & $0.296\pm0.018$ & $-0.512\pm0.062$ & $-1.998\pm0.142$ & $1.673\pm1.288$ & $-0.892\pm0.099$ & $0.222\pm0.222$ & $-0.390\pm0.109$ & $-1.739\pm0.253$ & $1.486\pm1.303$ & $-0.890\pm0.142$ & $0.140\pm0.182$ & $-0.359\pm0.119$ & $-1.562\pm0.368$ & $1.382\pm1.352$ \\
NGC3605 & $-0.482\pm0.250$ & $-0.358\pm0.362$ & $-0.230\pm0.363$ & $-0.537\pm0.268$ & $-0.588\pm-0.780$ & $-0.418\pm0.201$ & $-0.420\pm-0.420$ & $-0.337\pm0.183$ & $-0.409\pm0.296$ & $-0.342\pm-0.596$ & $-0.481\pm0.231$ & $-0.356\pm0.307$ & $-0.293\pm0.273$ & $-0.440\pm0.271$ & $-0.410\pm-0.521$ \\
NGC3773 & $-0.773\pm0.042$ & $0.158\pm0.038$ & $-0.623\pm0.103$ & $-2.498\pm0.243$ & $2.089\pm1.678$ & $-0.734\pm0.049$ & $0.090\pm0.090$ & $-0.498\pm0.135$ & $-2.368\pm0.365$ & $2.078\pm1.912$ & $-0.718\pm0.069$ & $0.072\pm0.084$ & $-0.428\pm0.110$ & $-1.927\pm0.454$ & $1.799\pm1.744$ \\
NGC3842 & $-0.450\pm0.162$ & $-0.502\pm0.450$ & $-0.038\pm0.243$ & $<-0.644$ & $-0.231\pm-0.689$ & $-0.422\pm0.234$ & $-0.195\pm-0.195$ & $-0.106\pm0.386$ & $-0.309\pm0.340$ & $-0.331\pm-0.268$ & $-0.488\pm0.279$ & $-0.324\pm0.337$ & $-0.211\pm0.327$ & $-0.476\pm0.325$ & $-0.078\pm0.226$ \\
NGC3860 & $-0.278\pm0.088$ & $-0.094\pm0.168$ & $-0.614\pm0.105$ & $-1.968\pm0.228$ & $1.069\pm0.258$ & $-0.435\pm0.236$ & $-0.164\pm-0.164$ & $-0.274\pm0.347$ & $-0.465\pm0.327$ & $-0.198\pm-0.109$ & $-0.374\pm0.162$ & $-0.323\pm0.339$ & $-0.579\pm0.262$ & $-1.254\pm0.602$ & $0.137\pm0.519$ \\
NGC3861 & $0.083\pm0.040$ & $0.285\pm0.205$ & $0.003\pm0.080$ & $<-1.310$ & $0.511\pm-0.460$ & $-0.213\pm0.122$ & $-0.350\pm-0.350$ & $-0.440\pm0.226$ & $-1.080\pm0.334$ & $0.496\pm0.219$ & $-0.318\pm0.159$ & $-0.429\pm0.328$ & $-0.446\pm0.202$ & $-1.327\pm0.364$ & $1.029\pm0.920$ \\
NGC3893 & $-0.414\pm0.046$ & $-0.646\pm0.086$ & $-0.635\pm0.042$ & $-2.258\pm0.264$ & $0.883\pm0.788$ & $-0.446\pm0.056$ & $-0.585\pm-0.585$ & $-0.481\pm0.088$ & $-2.020\pm0.335$ & $1.533\pm1.316$ & $-0.432\pm0.071$ & $-0.545\pm0.199$ & $-0.441\pm0.130$ & $-1.875\pm0.399$ & $1.362\pm1.366$ \\
NGC3896 & $-0.786\pm0.027$ & $0.291\pm0.060$ & $-0.408\pm0.048$ & $-2.007\pm0.284$ & $1.373\pm0.625$ & $-0.753\pm0.102$ & $0.161\pm0.161$ & $-0.284\pm0.120$ & $-1.583\pm0.330$ & $1.311\pm1.026$ & $-0.650\pm0.254$ & $0.045\pm0.305$ & $-0.162\pm0.220$ & $-1.269\pm0.537$ & $0.937\pm0.976$ \\
NGC3945 & $0.004\pm0.109$ & $-0.175\pm0.529$ & $-0.231\pm0.117$ & $-<1.339$ & $0.057\pm-0.462$ & $-0.002\pm0.099$ & $-0.259\pm-0.259$ & $-0.191\pm0.201$ & $-1.134\pm0.238$ & $0.128\pm-0.374$ & $-0.108\pm0.238$ & $-0.159\pm0.363$ & $-0.226\pm0.251$ & $-0.822\pm0.376$ & $-0.072\pm-0.314$ \\
NGC4059 & $0.010\pm0.051$ & $0.277\pm0.247$ & $0.304\pm0.058$ & $-1.345\pm0.291$ & $0.300\pm-0.619$ & $-0.114\pm0.166$ & $0.063\pm0.063$ & $0.361\pm0.158$ & $-0.623\pm0.254$ & $0.125\pm-0.267$ & $-0.203\pm0.223$ & $-0.013\pm0.326$ & $0.232\pm0.246$ & $-0.673\pm0.313$ & $0.172\pm0.182$ \\
IC3065 & $-0.248\pm0.200$ & $-0.387\pm0.357$ & $<-0.779$ & $-0.444\pm0.122$ & $-0.575\pm-0.706$ & $-0.580\pm0.427$ & $-0.135\pm-0.135$ & $-0.196\pm0.252$ & $-0.487\pm0.101$ & $-0.635\pm-0.534$ & $-0.470\pm0.367$ & $-0.209\pm0.386$ & $-0.293\pm0.325$ & $-0.580\pm0.292$ & $-0.417\pm-0.165$ \\
NGC4291 & $-0.085\pm0.196$ & $-0.349\pm0.295$ & $-0.351\pm0.239$ & $-0.456\pm0.302$ & $-0.377\pm-0.702$ & $-0.435\pm0.236$ & $-0.262\pm-0.262$ & $-0.377\pm0.218$ & $-0.394\pm0.245$ & $-0.278\pm-0.328$ & $-0.297\pm0.242$ & $-0.449\pm0.364$ & $-0.383\pm0.274$ & $-0.415\pm0.243$ & $-0.121\pm0.112$ \\
NGC4390 & $-0.415\pm0.101$ & $-0.357\pm0.186$ & $-0.108\pm0.061$ & $-0.871\pm0.196$ & $0.401\pm-0.224$ & $-0.665\pm0.100$ & $-0.144\pm-0.144$ & $-0.283\pm0.140$ & $-1.476\pm0.402$ & $1.491\pm1.469$ & $-0.689\pm0.164$ & $-0.126\pm0.238$ & $-0.271\pm0.166$ & $-1.331\pm0.472$ & $1.422\pm1.556$ \\
PGC040616 & $-0.563\pm0.233$ & $-0.584\pm0.364$ & $-0.415\pm0.262$ & $-0.637\pm0.280$ & $-0.288\pm-0.673$ & $-0.419\pm0.259$ & $-0.297\pm-0.297$ & $-0.318\pm0.322$ & $-0.491\pm0.286$ & $-0.306\pm-0.248$ & $-0.548\pm0.238$ & $-0.397\pm0.337$ & $-0.296\pm0.258$ & $-0.468\pm0.246$ & $-0.223\pm0.023$ \\
NGC4470 & $-0.523\pm0.025$ & $-0.358\pm0.044$ & $-0.414\pm0.034$ & $-2.076\pm0.155$ & $1.498\pm0.712$ & $-0.586\pm0.025$ & $-0.247\pm-0.247$ & $-0.361\pm0.054$ & $-1.893\pm0.216$ & $1.598\pm1.338$ & $-0.597\pm0.094$ & $-0.207\pm0.188$ & $-0.296\pm0.121$ & $-1.640\pm0.443$ & $1.391\pm1.337$ \\
NGC4479 & $-0.292\pm0.232$ & $-0.759\pm0.238$ & $-0.383\pm0.329$ & $-0.745\pm0.280$ & $-0.477\pm-0.683$ & $-0.426\pm0.229$ & $-0.267\pm-0.267$ & $-0.507\pm0.243$ & $-0.297\pm0.289$ & $-0.462\pm-0.517$ & $-0.439\pm0.252$ & $-0.355\pm0.375$ & $-0.404\pm0.267$ & $-0.370\pm0.285$ & $-0.266\pm-0.145$ \\
NGC4486B & $-0.599\pm0.283$ & $-0.653\pm0.359$ & $-0.470\pm0.342$ & $-1.018\pm0.385$ & $-0.209\pm-0.309$ & $-0.440\pm0.237$ & $-0.410\pm-0.410$ & $-0.455\pm0.336$ & $-0.530\pm0.310$ & $-0.142\pm-0.144$ & $-0.522\pm0.262$ & $-0.587\pm0.342$ & $-0.431\pm0.289$ & $-0.693\pm0.348$ & $0.373\pm1.048$ \\
IC3586 & $-0.346\pm0.286$ & $<0.029n$ & $-0.582\pm0.280$ & $-0.406\pm0.165$ & $-0.277\pm-0.326$ & $-0.443\pm0.116$ & $-0.248\pm-0.248$ & $-0.170\pm0.146$ & $-0.603\pm0.247$ & $-0.004\pm0.206$ & $-0.466\pm0.276$ & $-0.237\pm0.236$ & $-0.274\pm0.311$ & $-0.474\pm0.264$ & $0.222\pm0.646$ \\
IC3652 & $<-0.183$ & $<0.364$ & $-0.327\pm0.064$ & $0.063\pm0.114$ & $-0.608\pm-0.772$ & $-0.650\pm0.217$ & $0.036\pm0.036$ & $-0.312\pm0.267$ & $-0.105\pm0.214$ & $0.021\pm0.007$ & $-0.604\pm0.274$ & $-0.056\pm0.424$ & $-0.343\pm0.254$ & $-0.186\pm0.282$ & $0.177\pm0.664$ \\
NGC4676A & $-0.195\pm0.099$ & $0.008\pm0.183$ & $-0.164\pm0.098$ & $-1.487\pm0.224$ & $1.229\pm0.828$ & $-0.275\pm0.211$ & $0.080\pm0.080$ & $-0.114\pm0.226$ & $-0.888\pm0.402$ & $0.690\pm0.409$ & $-0.263\pm0.199$ & $0.076\pm0.346$ & $-0.096\pm0.239$ & $-0.962\pm0.490$ & $0.727\pm0.762$ \\
NGC4676B & $-0.142\pm0.062$ & $-0.078\pm0.064$ & $0.110\pm0.050$ & $-1.400\pm0.178$ & $1.145\pm0.443$ & $-0.252\pm0.058$ & $-0.027\pm-0.027$ & $-0.010\pm0.084$ & $-1.406\pm0.161$ & $1.228\pm0.733$ & $-0.276\pm0.096$ & $0.010\pm0.180$ & $-0.060\pm0.152$ & $-1.409\pm0.279$ & $1.161\pm0.983$ \\
PGC092948 & $-0.185\pm0.142$ & $-0.474\pm0.309$ & $-0.334\pm0.183$ & $-0.892\pm0.369$ & $0.949\pm1.004$ & $-0.356\pm0.292$ & $-0.256\pm-0.256$ & $-0.443\pm0.230$ & $-0.791\pm0.208$ & $1.002\pm1.242$ & $-0.347\pm0.251$ & $-0.251\pm0.387$ & $-0.359\pm0.247$ & $-0.893\pm0.336$ & $1.150\pm1.608$ \\
NGC4841A & $-0.143\pm0.118$ & $-0.225\pm0.567$ & $0.460\pm0.178$ & $-1.066\pm0.409$ & $-0.225\pm-0.704$ & $-0.454\pm0.347$ & $-0.463\pm-0.463$ & $-0.022\pm0.320$ & $-0.442\pm0.257$ & $-0.343\pm-0.288$ & $-0.333\pm0.314$ & $-0.306\pm0.423$ & $0.198\pm0.297$ & $-0.498\pm0.306$ & $-0.287\pm-0.063$ \\
NGC4861 & $<-0.183$ & $0.749\pm0.081$ & $-0.932\pm0.054$ & $-2.559\pm0.124$ & $<2.667$ & $<-0.330$ & $0.785\pm0.785$ & $<-0.253$ & $-2.594\pm0.055$ & $2.678\pm1.242$ & $<-0.335$ & $0.781\pm0.041$ & $-0.984\pm0.009$ & $-2.592\pm0.103$ & $2.560\pm2.056$ \\
NGC4874 & $-0.358\pm0.264$ & $-0.107\pm0.194$ & $0.083\pm0.167$ & $-0.686\pm0.309$ & $-0.336\pm-0.776$ & $-0.460\pm0.193$ & $-0.130\pm-0.130$ & $-0.164\pm0.268$ & $-0.486\pm0.294$ & $-0.272\pm-0.243$ & $-0.380\pm0.242$ & $-0.142\pm0.341$ & $-0.004\pm0.277$ & $-0.424\pm0.280$ & $-0.321\pm-0.111$ \\
NGC5198 & $-0.057\pm0.077$ & $0.253\pm0.188$ & $-0.129\pm0.078$ & $-0.915\pm0.237$ & $0.128\pm-0.383$ & $-0.422\pm0.251$ & $-0.196\pm-0.196$ & $-0.122\pm0.246$ & $-0.470\pm0.261$ & $-0.363\pm-0.495$ & $-0.336\pm0.254$ & $-0.144\pm0.333$ & $-0.189\pm0.278$ & $-0.458\pm0.277$ & $-0.321\pm-0.341$ \\
NGC5216 & $-0.121\pm0.035$ & $0.319\pm0.066$ & $0.141\pm0.026$ & $-1.114\pm0.103$ & $0.547\pm-0.237$ & $-0.263\pm0.168$ & $0.008\pm0.008$ & $-0.015\pm0.183$ & $-0.765\pm0.278$ & $0.254\pm-0.215$ & $-0.277\pm0.202$ & $0.043\pm0.334$ & $-0.044\pm0.223$ & $-0.714\pm0.325$ & $0.188\pm-0.013$ \\
NGC5218 & $-0.188\pm0.080$ & $-0.208\pm0.118$ & $-0.350\pm0.063$ & $-1.605\pm0.092$ & $1.075\pm0.672$ & $-0.172\pm0.131$ & $-0.100\pm-0.100$ & $-0.251\pm0.166$ & $-1.167\pm0.384$ & $0.837\pm1.037$ & $-0.186\pm0.179$ & $-0.091\pm0.367$ & $-0.258\pm0.209$ & $-1.025\pm0.423$ & $0.669\pm0.857$ \\
NGC5358 & $-0.504\pm0.310$ & $-0.621\pm0.191$ & $-0.566\pm0.262$ & $-0.558\pm0.217$ & $-0.775\pm-0.954$ & $<-0.712$ & $-0.652\pm-0.652$ & $-0.568\pm0.019$ & $-0.597\pm0.232$ & $-0.636\pm-0.781$ & $-0.510\pm0.301$ & $-0.485\pm0.342$ & $-0.479\pm0.264$ & $-0.544\pm0.203$ & $-0.526\pm-0.480$ \\
NGC5394 & $-0.237\pm0.019$ & $-0.531\pm0.021$ & $-0.495\pm0.016$ & $-1.882\pm0.073$ & $1.709\pm0.711$ & $-0.266\pm0.041$ & $-0.503\pm-0.503$ & $-0.421\pm0.067$ & $-1.758\pm0.220$ & $1.412\pm1.048$ & $-0.246\pm0.093$ & $-0.345\pm0.317$ & $-0.360\pm0.145$ & $-1.493\pm0.431$ & $1.226\pm1.174$ \\
NGC5395 & $0.036\pm0.035$ & $0.463\pm0.212$ & $-0.071\pm0.025$ & $-0.851\pm0.132$ & $0.260\pm-0.432$ & $-0.232\pm0.124$ & $-0.199\pm-0.199$ & $-0.270\pm0.161$ & $-1.263\pm0.399$ & $0.820\pm0.755$ & $-0.280\pm0.136$ & $-0.255\pm0.320$ & $-0.321\pm0.158$ & $-1.437\pm0.466$ & $1.132\pm1.127$ \\
NGC5426 & $-0.359\pm0.057$ & $-0.400\pm0.150$ & $-0.492\pm0.068$ & $-1.502\pm0.164$ & $0.764\pm0.126$ & $-0.464\pm0.047$ & $-0.547\pm-0.547$ & $-0.455\pm0.075$ & $-1.715\pm0.355$ & $1.476\pm1.154$ & $-0.517\pm0.116$ & $-0.327\pm0.323$ & $-0.428\pm0.132$ & $-1.616\pm0.422$ & $1.540\pm1.557$ \\
NGC5427 & $-0.051\pm0.131$ & $0.785\pm0.263$ & $-0.315\pm0.119$ & $-1.583\pm0.097$ & $1.301\pm0.756$ & $-0.370\pm0.102$ & $-0.433\pm-0.433$ & $-0.392\pm0.140$ & $-1.618\pm0.429$ & $1.281\pm1.252$ & $-0.344\pm0.135$ & $-0.322\pm0.338$ & $-0.339\pm0.180$ & $-1.477\pm0.491$ & $1.110\pm1.225$ \\
NGC5473 & $-0.318\pm0.126$ & $-0.025\pm0.474$ & $-0.422\pm0.223$ & $-0.650\pm0.258$ & $-0.506\pm-1.111$ & $-0.276\pm0.238$ & $-0.157\pm-0.157$ & $-0.273\pm0.282$ & $-0.467\pm0.309$ & $-0.648\pm-0.697$ & $-0.379\pm0.251$ & $0.063\pm0.351$ & $-0.312\pm0.260$ & $-0.442\pm0.267$ & $-0.638\pm-0.578$ \\
NGC5485 & $-0.063\pm0.063$ & $0.339\pm0.376$ & $-0.019\pm0.082$ & $-0.956\pm0.257$ & $-0.086\pm-0.930$ & $-0.204\pm0.329$ & $-0.078\pm-0.078$ & $-0.187\pm0.251$ & $-0.597\pm0.349$ & $-0.264\pm-0.277$ & $-0.184\pm0.300$ & $-0.073\pm0.368$ & $-0.160\pm0.223$ & $-0.653\pm0.357$ & $-0.339\pm-0.310$ \\
NGC5532 & $0.066\pm0.121$ & $0.068\pm0.389$ & $0.412\pm0.122$ & $-0.996\pm0.461$ & $0.129\pm-0.206$ & $-0.455\pm0.380$ & $-0.309\pm-0.309$ & $0.351\pm0.173$ & $-0.503\pm0.201$ & $-0.535\pm-0.539$ & $-0.241\pm0.303$ & $-0.128\pm0.397$ & $0.406\pm0.239$ & $-0.554\pm0.352$ & $-0.418\pm-0.282$ \\
NGC5546 & $0.086\pm0.033$ & $0.114\pm0.225$ & $0.164\pm0.048$ & $-1.829\pm0.296$ & $0.581\pm-0.129$ & $-0.156\pm0.215$ & $-0.143\pm-0.143$ & $0.232\pm0.194$ & $-0.490\pm0.237$ & $-0.151\pm-0.336$ & $-0.100\pm0.260$ & $-0.067\pm0.352$ & $0.215\pm0.186$ & $-0.704\pm0.447$ & $-0.152\pm-0.054$ \\
NGC5557 & $-0.167\pm0.232$ & $0.093\pm0.428$ & $-0.407\pm0.230$ & $-0.279\pm0.206$ & $-0.609\pm-0.890$ & $-0.394\pm0.319$ & $0.227\pm0.227$ & $-0.327\pm0.268$ & $-0.281\pm0.151$ & $-0.722\pm-0.682$ & $-0.367\pm0.257$ & $0.082\pm0.339$ & $-0.226\pm0.270$ & $-0.308\pm0.231$ & $-0.528\pm-0.376$ \\
NGC5576 & $-0.496\pm0.258$ & $-0.403\pm0.508$ & $-0.683\pm0.211$ & $-0.535\pm0.271$ & $-0.248\pm-0.566$ & $-0.367\pm0.259$ & $-0.611\pm-0.611$ & $-0.504\pm0.251$ & $-0.529\pm0.296$ & $-0.230\pm-0.409$ & $-0.392\pm0.277$ & $-0.453\pm0.418$ & $-0.475\pm0.274$ & $-0.512\pm0.311$ & $-0.193\pm-0.406$ \\
NGC5614 & $0.132\pm0.021$ & $0.690\pm0.232$ & $0.047\pm0.047$ & $-1.283\pm0.185$ & $0.410\pm-0.380$ & $-0.149\pm0.103$ & $0.034\pm0.034$ & $-0.292\pm0.163$ & $-1.111\pm0.305$ & $0.455\pm0.093$ & $-0.196\pm0.168$ & $-0.076\pm0.418$ & $-0.307\pm0.229$ & $-1.026\pm0.333$ & $0.494\pm0.364$ \\
NGC5631 & $0.061\pm0.041$ & $0.410\pm0.219$ & $0.152\pm0.028$ & $-1.234\pm0.264$ & $0.166\pm-0.631$ & $-0.128\pm0.131$ & $0.164\pm0.164$ & $-0.027\pm0.164$ & $-0.770\pm0.291$ & $-0.011\pm-0.386$ & $-0.193\pm0.215$ & $0.181\pm0.323$ & $-0.060\pm0.231$ & $-0.710\pm0.327$ & $-0.042\pm-0.332$ \\
NGC5623 & $0.013\pm0.042$ & $0.086\pm0.095$ & $-0.008\pm0.034$ & $-1.344\pm0.225$ & $0.410\pm-0.165$ & $-0.259\pm0.218$ & $-0.103\pm-0.103$ & $-0.207\pm0.235$ & $-0.619\pm0.321$ & $-0.018\pm-0.132$ & $-0.262\pm0.234$ & $-0.111\pm0.315$ & $-0.209\pm0.246$ & $-0.646\pm0.351$ & $-0.105\pm-0.095$ \\
NGC5656 & $-0.303\pm0.086$ & $-0.308\pm0.202$ & $-0.427\pm0.059$ & $-1.339\pm0.202$ & $0.528\pm0.262$ & $-0.482\pm0.043$ & $-0.655\pm-0.655$ & $-0.481\pm0.053$ & $-1.851\pm0.249$ & $1.361\pm0.881$ & $-0.471\pm0.062$ & $-0.418\pm0.229$ & $-0.389\pm0.110$ & $-1.687\pm0.379$ & $1.367\pm1.351$ \\
NGC5675 & $0.033\pm0.026$ & $0.466\pm0.185$ & $0.113\pm0.030$ & $-1.191\pm0.121$ & $1.055\pm0.718$ & $-0.093\pm0.128$ & $0.142\pm0.142$ & $-0.138\pm0.164$ & $-0.882\pm0.348$ & $0.355\pm0.261$ & $-0.162\pm0.195$ & $0.025\pm0.381$ & $-0.152\pm0.228$ & $-0.947\pm0.477$ & $0.805\pm1.326$ \\
UGC9562 & $-0.929\pm0.073$ & $0.353\pm0.094$ & $-0.500\pm0.090$ & $-1.892\pm0.264$ & $1.483\pm1.140$ & $-0.825\pm0.211$ & $0.173\pm0.173$ & $-0.310\pm0.168$ & $-1.257\pm0.502$ & $1.382\pm1.494$ & $-0.840\pm0.265$ & $0.163\pm0.274$ & $-0.378\pm0.220$ & $-1.284\pm0.549$ & $1.467\pm1.628$ \\
NGC5794 & $-0.082\pm0.171$ & $0.029\pm0.505$ & $-0.478\pm0.152$ & $-0.723\pm0.262$ & $-0.241\pm-0.758$ & $-0.266\pm0.230$ & $-0.120\pm-0.120$ & $-0.375\pm0.306$ & $-0.468\pm0.254$ & $-0.287\pm-0.638$ & $-0.259\pm0.241$ & $-0.091\pm0.378$ & $-0.351\pm0.274$ & $-0.539\pm0.287$ & $-0.205\pm-0.357$ \\
NGC5797 & $-0.414\pm0.390$ & $-0.080\pm0.431$ & $<-0.378$ & $-0.087\pm0.252$ & $-0.631\pm-0.644$ & $-0.366\pm0.369$ & $-0.445\pm-0.445$ & $-0.326\pm0.334$ & $-0.177\pm0.217$ & $-0.570\pm-0.694$ & $-0.412\pm0.339$ & $-0.272\pm0.404$ & $-0.424\pm0.217$ & $-0.318\pm0.308$ & $-0.459\pm-0.439$ \\
UGC9661 & $-0.556\pm0.024$ & $-0.451\pm0.090$ & $-0.331\pm0.054$ & $-1.880\pm0.226$ & $1.423\pm0.907$ & $-0.643\pm0.115$ & $-0.345\pm-0.345$ & $-0.337\pm0.134$ & $-1.411\pm0.426$ & $1.406\pm1.268$ & $-0.642\pm0.175$ & $-0.315\pm0.271$ & $-0.272\pm0.175$ & $-1.172\pm0.520$ & $1.188\pm1.204$ \\
NGC5845 & $-0.111\pm0.076$ & $-0.438\pm0.442$ & $-0.642\pm0.228$ & $-0.694\pm0.273$ & $-0.020\pm-0.160$ & $-0.164\pm0.098$ & $-0.736\pm-0.736$ & $-0.660\pm0.213$ & $-0.852\pm0.266$ & $0.041\pm-0.041$ & $-0.224\pm0.198$ & $-0.410\pm0.453$ & $-0.511\pm0.267$ & $-0.614\pm0.268$ & $-0.163\pm-0.325$ \\
NGC5929 & $-0.202\pm0.029$ & $0.516\pm0.027$ & $-0.038\pm0.030$ & $-1.138\pm0.066$ & $1.596\pm1.046$ & $-0.251\pm0.062$ & $0.302\pm0.302$ & $-0.034\pm0.097$ & $-1.222\pm0.266$ & $0.890\pm0.475$ & $-0.259\pm0.138$ & $0.205\pm0.260$ & $-0.081\pm0.159$ & $-1.085\pm0.323$ & $0.810\pm0.762$ \\
NGC5930 & $-0.249\pm0.024$ & $-0.459\pm0.079$ & $-0.382\pm0.039$ & $-1.877\pm0.098$ & $1.748\pm1.115$ & $-0.295\pm0.062$ & $-0.172\pm-0.172$ & $-0.180\pm0.122$ & $-1.406\pm0.366$ & $0.954\pm0.873$ & $-0.273\pm0.155$ & $-0.102\pm0.320$ & $-0.145\pm0.187$ & $-1.102\pm0.451$ & $0.748\pm0.891$ \\
NGC5953 & $-0.096\pm0.089$ & $0.229\pm0.208$ & $-0.298\pm0.082$ & $-1.714\pm0.092$ & $1.495\pm0.807$ & $-0.361\pm0.046$ & $-0.367\pm-0.367$ & $-0.511\pm0.043$ & $-2.058\pm0.142$ & $1.823\pm1.320$ & $-0.240\pm0.100$ & $-0.057\pm0.280$ & $-0.364\pm0.112$ & $-1.684\pm0.295$ & $1.455\pm1.378$ \\
NGC5954 & $-0.436\pm0.029$ & $-0.695\pm0.072$ & $-0.589\pm0.033$ & $-2.147\pm0.215$ & $1.856\pm1.239$ & $-0.447\pm0.047$ & $-0.371\pm-0.371$ & $-0.477\pm0.094$ & $-2.092\pm0.292$ & $1.785\pm1.855$ & $-0.423\pm0.123$ & $-0.288\pm0.273$ & $-0.415\pm0.151$ & $-1.599\pm0.508$ & $1.456\pm1.565$ \\
ARP220 & $0.050\pm0.028$ & $0.638\pm0.251$ & $-0.185\pm0.056$ & $-1.646\pm0.191$ & $1.457\pm1.216$ & $0.122\pm0.116$ & $0.185\pm0.185$ & $-0.145\pm0.229$ & $-0.908\pm0.350$ & $0.791\pm0.756$ & $0.052\pm0.173$ & $0.162\pm0.393$ & $-0.148\pm0.235$ & $-0.937\pm0.448$ & $0.814\pm0.858$ \\
UGC10205 & $-0.251\pm0.109$ & $-0.136\pm0.133$ & $-0.159\pm0.112$ & $-1.636\pm0.279$ & $0.858\pm0.516$ & $-0.386\pm0.160$ & $-0.018\pm-0.018$ & $-0.140\pm0.247$ & $-0.820\pm0.400$ & $0.672\pm0.817$ & $-0.368\pm0.174$ & $-0.086\pm0.263$ & $-0.158\pm0.248$ & $-1.013\pm0.486$ & $0.435\pm0.742$ \\
NGC6090 & $-0.372\pm0.018$ & $-0.348\pm0.063$ & $-0.610\pm0.066$ & $-2.146\pm0.076$ & $2.240\pm1.561$ & $-0.414\pm0.033$ & $-0.160\pm-0.160$ & $-0.454\pm0.062$ & $-1.869\pm0.203$ & $1.867\pm1.477$ & $-0.422\pm0.096$ & $-0.187\pm0.218$ & $-0.451\pm0.136$ & $-1.629\pm0.408$ & $1.653\pm1.622$ \\
NGC6125 & $-0.074\pm0.131$ & $0.040\pm0.207$ & $<-0.234$ & $-0.486\pm0.085$ & $-0.401\pm-1.042$ & $-0.321\pm0.279$ & $-0.110\pm-0.110$ & $-0.197\pm0.269$ & $-0.289\pm0.215$ & $-0.639\pm-0.752$ & $-0.287\pm0.288$ & $-0.063\pm0.328$ & $-0.355\pm0.201$ & $-0.398\pm0.253$ & $-0.533\pm-0.473$ \\
NGC6166NED01 & $0.143\pm0.094$ & $-0.324\pm0.244$ & $-0.004\pm0.094$ & $<-1.310$ & $0.727\pm0.606$ & $-0.450\pm0.265$ & $-0.398\pm-0.398$ & $-0.130\pm0.273$ & $-0.673\pm0.281$ & $0.228\pm-0.035$ & $-0.415\pm0.303$ & $-0.374\pm0.303$ & $-0.130\pm0.264$ & $-0.658\pm0.304$ & $0.225\pm0.267$ \\
NGC6251 & $0.034\pm0.025$ & $0.754\pm0.222$ & $0.088\pm0.056$ & $-1.699\pm0.151$ & $0.808\pm0.295$ & $-0.345\pm0.359$ & $-0.182\pm-0.182$ & $0.210\pm0.204$ & $-0.437\pm0.218$ & $-0.170\pm-0.235$ & $-0.160\pm0.308$ & $0.362\pm0.485$ & $0.242\pm0.251$ & $-0.776\pm0.534$ & $-0.137\pm0.062$ \\
PGC2172338 & $-0.034\pm0.048$ & $0.044\pm0.196$ & $-0.020\pm0.096$ & $-1.185\pm0.329$ & $0.537\pm0.046$ & $-0.347\pm0.221$ & $-0.150\pm-0.150$ & $-0.040\pm0.377$ & $-0.416\pm0.304$ & $0.521\pm0.729$ & $-0.175\pm0.238$ & $-0.150\pm0.367$ & $-0.006\pm0.227$ & $-0.779\pm0.439$ & $0.722\pm1.340$ \\
NGC6285 & $-0.407\pm0.019$ & $-0.447\pm0.078$ & $-0.441\pm0.024$ & $-2.079\pm0.143$ & $1.707\pm1.219$ & $<-0.330$ & $-0.154\pm-0.154$ & $<-0.253$ & $<-0.922$ & $<0.385$ & $-0.413\pm0.018$ & $-0.433\pm0.100$ & $-0.441\pm0.028$ & $-2.032\pm0.128$ & $1.735\pm1.187$ \\
NGC6286 & $-0.356\pm0.089$ & $-0.374\pm0.108$ & $-0.433\pm0.094$ & $-1.737\pm0.231$ & $1.476\pm0.466$ & $-0.247\pm0.111$ & $-0.281\pm-0.281$ & $-0.334\pm0.128$ & $-1.594\pm0.237$ & $1.349\pm0.825$ & $-0.203\pm0.143$ & $-0.188\pm0.168$ & $-0.321\pm0.144$ & $-1.539\pm0.293$ & $1.308\pm0.981$ \\
NGC6278 & $0.048\pm0.070$ & $0.241\pm0.312$ & $-0.006\pm0.098$ & $-0.735\pm0.222$ & $0.068\pm-0.375$ & $-0.397\pm0.274$ & $0.099\pm0.099$ & $-0.137\pm0.237$ & $-0.387\pm0.261$ & $-0.290\pm-0.686$ & $-0.339\pm0.259$ & $-0.035\pm0.374$ & $-0.161\pm0.270$ & $-0.488\pm0.293$ & $-0.184\pm-0.385$ \\
NGC6338 & $0.103\pm0.033$ & $-0.018\pm0.181$ & $0.188\pm0.051$ & $-1.721\pm0.160$ & $0.577\pm0.099$ & $-0.091\pm0.267$ & $-0.093\pm-0.093$ & $0.252\pm0.168$ & $-0.616\pm0.432$ & $-0.027\pm-0.190$ & $-0.057\pm0.239$ & $-0.112\pm0.338$ & $0.189\pm0.209$ & $-0.641\pm0.404$ & $-0.128\pm-0.075$ \\
NGC7236 & $-0.069\pm0.085$ & $-0.446\pm0.555$ & $0.389\pm0.143$ & $-0.596\pm0.305$ & $-0.276\pm-0.687$ & $-0.119\pm0.166$ & $-0.430\pm-0.430$ & $0.379\pm0.156$ & $-0.667\pm0.310$ & $-0.301\pm-0.499$ & $-0.183\pm0.244$ & $-0.414\pm0.451$ & $0.335\pm0.223$ & $-0.599\pm0.285$ & $-0.452\pm-0.507$ \\
UGC11958 & $0.119\pm0.031$ & $0.319\pm0.099$ & $0.242\pm0.032$ & $-1.441\pm0.266$ & $0.615\pm-0.024$ & $-0.089\pm0.338$ & $-0.121\pm-0.121$ & $0.216\pm0.217$ & $-0.559\pm0.217$ & $-0.210\pm-0.257$ & $-0.089\pm0.277$ & $0.009\pm0.372$ & $0.152\pm0.238$ & $-0.744\pm0.459$ & $-0.197\pm-0.054$ \\
UGC12127 & $-0.045\pm0.153$ & $0.015\pm0.287$ & $0.250\pm0.156$ & $-0.779\pm0.299$ & $-0.089\pm-0.492$ & $-0.270\pm0.239$ & $-0.380\pm-0.380$ & $-0.059\pm0.148$ & $-0.303\pm0.281$ & $-0.327\pm-0.232$ & $-0.257\pm0.259$ & $-0.141\pm0.403$ & $0.042\pm0.277$ & $-0.439\pm0.337$ & $-0.314\pm0.056$ \\
NGC7457 & $-0.395\pm0.229$ & $<0.029n$ & $-0.274\pm0.160$ & $<-0.655$ & $-1.092\pm-0.995$ & $-0.363\pm0.248$ & $0.215\pm0.215$ & $-0.450\pm0.283$ & $-0.523\pm0.194$ & $-0.586\pm-0.528$ & $-0.410\pm0.260$ & $0.258\pm0.430$ & $-0.320\pm0.301$ & $-0.451\pm0.270$ & $-0.560\pm0.439$ \\
IC5309 & $-0.448\pm0.056$ & $-0.503\pm0.109$ & $-0.592\pm0.050$ & $-1.764\pm0.304$ & $1.028\pm0.407$ & $-0.514\pm0.078$ & $-0.491\pm-0.491$ & $-0.464\pm0.104$ & $-1.490\pm0.324$ & $1.145\pm0.753$ & $-0.548\pm0.167$ & $-0.365\pm0.331$ & $-0.433\pm0.185$ & $-1.293\pm0.478$ & $0.992\pm1.056$ \\
NGC7611 & $-0.087\pm0.170$ & $0.069\pm0.576$ & $-0.284\pm0.270$ & $-0.707\pm0.389$ & $-0.470\pm-0.729$ & $-0.478\pm0.298$ & $-0.296\pm-0.296$ & $-0.299\pm0.287$ & $-0.422\pm0.345$ & $-0.748\pm-0.780$ & $-0.326\pm0.307$ & $-0.323\pm0.496$ & $-0.296\pm0.275$ & $-0.484\pm0.339$ & $-0.876\pm-0.743$ \\
NGC7619 & $-0.129\pm0.063$ & $0.175\pm0.259$ & $-0.579\pm0.250$ & $-0.793\pm0.277$ & $-0.287\pm-0.942$ & $-0.246\pm0.240$ & $-0.087\pm-0.087$ & $-0.399\pm0.237$ & $-0.417\pm0.279$ & $-0.432\pm-0.633$ & $-0.239\pm0.247$ & $-0.084\pm0.341$ & $-0.333\pm0.234$ & $-0.440\pm0.286$ & $-0.405\pm-0.381$ \\
NGC7623 & $-0.684\pm0.439$ & $<0.077$ & $-0.325\pm0.037$ & $-0.374\pm0.025$ & $-0.872\pm-0.900$ & $-0.514\pm0.119$ & $-0.319\pm-0.319$ & $-0.143\pm0.340$ & $-0.542\pm0.251$ & $-0.504\pm-0.464$ & $-0.512\pm0.310$ & $-0.277\pm0.311$ & $-0.252\pm0.259$ & $-0.474\pm0.259$ & $-0.427\pm-0.318$ \\
NGC7684 & $-0.008\pm0.108$ & $0.198\pm0.487$ & $-0.525\pm0.236$ & $<-1.310$ & $-0.177\pm-0.587$ & $-0.180\pm0.184$ & $0.433\pm0.433$ & $-0.517\pm0.240$ & $<-1.336$ & $-0.188\pm-0.557$ & $-0.209\pm0.250$ & $0.247\pm0.400$ & $-0.452\pm0.256$ & $-0.779\pm0.266$ & $-0.093\pm-0.301$ \\
NGC7716 & $-0.064\pm0.055$ & $0.437\pm0.426$ & $0.091\pm0.063$ & $-0.746\pm0.184$ & $-0.027\pm-0.838$ & $-0.419\pm0.079$ & $-0.238\pm-0.238$ & $-0.335\pm0.119$ & $-1.404\pm0.356$ & $0.915\pm0.719$ & $-0.428\pm0.117$ & $-0.189\pm0.255$ & $-0.334\pm0.156$ & $-1.290\pm0.412$ & $1.134\pm1.058$ \\
NGC7052 & $0.018\pm0.044$ & $-0.064\pm0.154$ & $-0.469\pm0.128$ & $-1.605\pm0.252$ & $0.530\pm-0.034$ & $-0.219\pm0.222$ & $-0.043\pm-0.043$ & $-0.380\pm0.127$ & $-0.430\pm0.227$ & $-0.477\pm-0.545$ & $-0.175\pm0.233$ & $-0.090\pm0.305$ & $-0.426\pm0.263$ & $-0.647\pm0.395$ & $-0.487\pm-0.286$ \\

\bottomrule
\end{longtable}
\end{landscape}

\end{document}